\documentclass[10pt,a4paper]{article} 
\usepackage{graphicx}
\usepackage{epstopdf}
\usepackage{epsfig}
\usepackage{amsmath,amssymb,bm,amsfonts,bbm}
\usepackage{style}
\usepackage{yfonts}

\newcommand{\be}{\begin{equation}}
\newcommand{\ee}{\end{equation}}
\newcommand{\bea}{\begin{eqnarray}}

\newcommand{\eea}{\end{eqnarray}}

\newcommand{\mpl}{M_{\rm Pl}}

\def\Hc{{\cal H}}
\def\Rc{{\cal R}}
\def\Vc{{\cal V}}
\def\Vct{\tilde{\cal V}}
\def\Rct{\tilde{\cal R}}
\def\Rt{\tilde{R}}

\newcommand{\kt}{{\tilde{k}}}
\newcommand{\vt}{{\tilde{v}}}

\newcommand{\dN}{{{\delta N}}}

\newcommand{\dphi}{\delta \varphi}

\def\exx{\varepsilon}

\def \lta {\mathrel{\vcenter
     {\hbox{$<$}\nointerlineskip\hbox{$\sim$}}}}

\DeclareSymbolFont{mathscrUC}{U}{rsfs}{m}{n}  
\DeclareSymbolFont{mathscrLC}{OT1}{pzc}{m}{n} 

\makeatletter

\DeclareRobustCommand*{\mathscr}[1]{\gdef\F@ntPrefix{mathscr@char@}%
  \@EachCharacter #1\@EndEachCharacter}
\long\def\DoLongFutureLet #1#2#3#4{%
   \def\@FutureLetDecide{#1#2\@FutureLetToken
      \def\@FutureLetNext{#3}\else
      \def\@FutureLetNext{#4}\fi\@FutureLetNext}
   \futurelet\@FutureLetToken\@FutureLetDecide}
\def\DoFutureLet #1#2#3#4{\DoLongFutureLet{#1}{#2}{#3}{#4}}
\def\@EachCharacter{\DoFutureLet{\ifx}{\@EndEachCharacter}%
   {\@EachCharacterDone}{\@PickUpTheCharacter}}
\def\m@keCharacter#1{\csname\F@ntPrefix#1\endcsname}
\def\@PickUpTheCharacter#1{\m@keCharacter{#1}\@EachCharacter}
\def\@EachCharacterDone \@EndEachCharacter{}

\DeclareMathSymbol{\mathscr@char@A}{\mathord}{mathscrUC}{`A}
\DeclareMathSymbol{\mathscr@char@B}{\mathord}{mathscrUC}{`B}
\DeclareMathSymbol{\mathscr@char@C}{\mathord}{mathscrUC}{`C}
\DeclareMathSymbol{\mathscr@char@D}{\mathord}{mathscrUC}{`D}
\DeclareMathSymbol{\mathscr@char@E}{\mathord}{mathscrUC}{`E}
\DeclareMathSymbol{\mathscr@char@F}{\mathord}{mathscrUC}{`F}
\DeclareMathSymbol{\mathscr@char@G}{\mathord}{mathscrUC}{`G}
\DeclareMathSymbol{\mathscr@char@H}{\mathord}{mathscrUC}{`H}
\DeclareMathSymbol{\mathscr@char@I}{\mathord}{mathscrUC}{`I}
\DeclareMathSymbol{\mathscr@char@J}{\mathord}{mathscrUC}{`J}
\DeclareMathSymbol{\mathscr@char@K}{\mathord}{mathscrUC}{`K}
\DeclareMathSymbol{\mathscr@char@L}{\mathord}{mathscrUC}{`L}
\DeclareMathSymbol{\mathscr@char@M}{\mathord}{mathscrUC}{`M}
\DeclareMathSymbol{\mathscr@char@N}{\mathord}{mathscrUC}{`N}
\DeclareMathSymbol{\mathscr@char@O}{\mathord}{mathscrUC}{`O}
\DeclareMathSymbol{\mathscr@char@P}{\mathord}{mathscrUC}{`P}
\DeclareMathSymbol{\mathscr@char@Q}{\mathord}{mathscrUC}{`Q}
\DeclareMathSymbol{\mathscr@char@R}{\mathord}{mathscrUC}{`R}
\DeclareMathSymbol{\mathscr@char@S}{\mathord}{mathscrUC}{`S}
\DeclareMathSymbol{\mathscr@char@T}{\mathord}{mathscrUC}{`T}
\DeclareMathSymbol{\mathscr@char@U}{\mathord}{mathscrUC}{`U}
\DeclareMathSymbol{\mathscr@char@V}{\mathord}{mathscrUC}{`V}
\DeclareMathSymbol{\mathscr@char@W}{\mathord}{mathscrUC}{`W}
\DeclareMathSymbol{\mathscr@char@X}{\mathord}{mathscrUC}{`X}
\DeclareMathSymbol{\mathscr@char@Y}{\mathord}{mathscrUC}{`Y}
\DeclareMathSymbol{\mathscr@char@Z}{\mathord}{mathscrUC}{`Z}
\DeclareMathSymbol{\mathscr@char@a}{\mathord}{mathscrLC}{`a}
\DeclareMathSymbol{\mathscr@char@b}{\mathord}{mathscrLC}{`b}
\DeclareMathSymbol{\mathscr@char@c}{\mathord}{mathscrLC}{`c}
\DeclareMathSymbol{\mathscr@char@d}{\mathord}{mathscrLC}{`d}
\DeclareMathSymbol{\mathscr@char@e}{\mathord}{mathscrLC}{`e}
\DeclareMathSymbol{\mathscr@char@f}{\mathord}{mathscrLC}{`f}
\DeclareMathSymbol{\mathscr@char@g}{\mathord}{mathscrLC}{`g}
\DeclareMathSymbol{\mathscr@char@h}{\mathord}{mathscrLC}{`h}
\DeclareMathSymbol{\mathscr@char@i}{\mathord}{mathscrLC}{`i}
\DeclareMathSymbol{\mathscr@char@j}{\mathord}{mathscrLC}{`j}
\DeclareMathSymbol{\mathscr@char@k}{\mathord}{mathscrLC}{`k}
\DeclareMathSymbol{\mathscr@char@l}{\mathord}{mathscrLC}{`l}
\DeclareMathSymbol{\mathscr@char@m}{\mathord}{mathscrLC}{`m}
\DeclareMathSymbol{\mathscr@char@n}{\mathord}{mathscrLC}{`n}
\DeclareMathSymbol{\mathscr@char@o}{\mathord}{mathscrLC}{`o}
\DeclareMathSymbol{\mathscr@char@p}{\mathord}{mathscrLC}{`p}
\DeclareMathSymbol{\mathscr@char@q}{\mathord}{mathscrLC}{`q}
\DeclareMathSymbol{\mathscr@char@r}{\mathord}{mathscrLC}{`r}
\DeclareMathSymbol{\mathscr@char@s}{\mathord}{mathscrLC}{`s}
\DeclareMathSymbol{\mathscr@char@t}{\mathord}{mathscrLC}{`t}
\DeclareMathSymbol{\mathscr@char@u}{\mathord}{mathscrLC}{`u}
\DeclareMathSymbol{\mathscr@char@v}{\mathord}{mathscrLC}{`v}
\DeclareMathSymbol{\mathscr@char@w}{\mathord}{mathscrLC}{`w}
\DeclareMathSymbol{\mathscr@char@x}{\mathord}{mathscrLC}{`x}
\DeclareMathSymbol{\mathscr@char@y}{\mathord}{mathscrLC}{`y}
\DeclareMathSymbol{\mathscr@char@z}{\mathord}{mathscrLC}{`z}

\makeatother

\relax

\title{Features of the inflaton potential and the power spectrum of 
cosmological perturbations}

\author{K. Kefala, G.P. Kodaxis, I.D. Stamou and N. Tetradis}

\affiliation{Department of Physics, University of Athens, University Campus, Zographou 157 84, Greece}

\emailAdd{kyrkfl@phys.uoa.gr, gekontax@phys.uoa.gr, joanstam@phys.uoa.gr, ntetrad@phys.uoa.gr}

\abstract{
We discuss features of the inflaton potential that can lead to a strong
enhancement of the power spectrum of curvature perturbations. 
We show that a steep decrease of the potential 
induces an enhancement of the spectrum by several orders of magnitude, 
which may lead to the production
of primordial black holes. The same feature can also create a distinctive 
oscillatory pattern in the spectrum of
gravitational waves generated through the scalar perturbations
at second order.
We study the additive effect of several such features. 
We analyse a simplified potential, but also discuss the possible application 
to supergravity models.}

\begin{document}

\maketitle

\section{Introduction} \label{intro}

The detection of gravitational waves emitted during black-hole mergers  
\cite{Virgo} has led to the realization that black holes are quite common in the 
Universe and
has generated a lot of interest in the question of their precise abundance
and possilbe role as dark matter. 
It was suggested a long time ago \cite{pbh1} that primordial black holes (PBHs), 
produced during
the very early stages of the evolution of the Universe, may survive until today in
significant numbers in order to be detectable. This possibility has been analysed 
in great detail in recent years. (For reviews with extensive lists of references, see
\cite{reviews}.) The production of PBHs requires the presence of strong density
perturbations.
For this to occur, 
the primordial power spectrum must be larger by several orders of magnitude than 
the value favored by the cosmic microwave background (CMB). Such an enhancement is
phenomenologically viable only in the range of length scales for which the recent 
evolution is highly nonlinear and the primordial spectrum is unconstrained by observations. 
Typically, this is the case for wavenumbers larger than approximately 1 Mpc$^{-1}$.

The enhancement of the power spectrum generated by inflationary dynamics requires the
presence of a strong feature in the inflaton potential, so that the standard
slow-roll conditions are violated \cite{slowroll}. 
Several proposals have been put forward for 
achieving this goal \cite{inflectionorig,stepinflation1,pbhdarkmatter,inflection,multifield,efstathiou}.
The most popular method introduces an inflection point in the inflaton potential 
\cite{slowroll,inflection}, which results in the slowing down of the rate of change of the
inflaton background. The slow-roll parameter $\exx$ becomes very small in the vicinity
of the inflection point, but the large increase of the parameter $\eta$ leads to the
violation of the slow-roll conditions. The calculation of the spectrum through
the solution of the Mukhanov-Sasaki equation \cite{mukhsas} shows that the
power spectrum can be enhanced by several orders of magnitude for the
scales exiting the horizon when the background field takes values in the vicinity of 
the inflection point. The necessary enhancement for black-hole production depends
on many factors, such as the asymmetry or angular velocity of the collapsing 
configuration. It also depends crucially on the equation of state of the Universe,
with matter domination requiring a significantly milder enhancement \cite{matterdom}.
A drawback of the inflection-point scenario
is that generating a large PBH abundance requires a
precise fine tuning of the inflaton potential. 
Considering a two- or multi-field inflaton sector is another framework within 
which the background evolution can be modified so as to enhance the spectrum \cite{multifield}.
The presence of entropy modes, which can backreact strongly on the adiabatic mode of interest, makes the analysis of such models more complicated.

We are interested in exploring features of the potential, other than an inflection point,  
in single-field inflation,
which can lead to a large enhancement of the power spectrum of 
perturbations. It has been observed \cite{stepinflation1} that a fast 
decrease of the potential can have such an effect, with relevance for
PBH creation. On the other hand, the emphasis in models with two inflationary stages, separated by a non-inflationary period, has been put on the oscillatory form of the 
resulting spectra \cite{stepinflation2}. Our aim here is to analyse carefully 
the conditions under which such a feature results in 
an enhancement of the spectrum by several orders of magnitude within
a certain range of short-distance scales.

The points at which the vacuum energy changes 
value abruptly may correspond to values of the inflaton field associated 
with the decoupling of modes whose quantum fluctuations contribute 
to the vacuum energy. The decoupling becomes apparent when the effective 
potential is regularized in a mass-sensitive scheme. In 
the Wilsonian approach to the renormalization group, the coarse-grained effective 
potential $U_k$ obeys an 
equation with the schematic form \cite{wetterich,review}
\be
\frac{\partial U_k(\varphi)}{\partial \ln k}=
\frac{k^4}{16\pi^2} \sum_i\, L\left(\frac{m^2_i(k,\varphi)}{k^2}\right).
\label{exactpotential} \ee
The sum extends over all fields whose masses depend
on the background field $\varphi$, with the contributions from 
bosons and fermions having opposite signs. 
The potential $U_k$ is obtained by integrating this equation,
starting with the bare potential, 
defined at some initial high scale $\Lambda$ that
can be identified with the UV cutoff of the theory, and
terminating at a physical IR scale that can be taken to zero.  
The potential incorporates quantum corrections from modes with 
momenta above the running scale $k$. 
The function $L(w)$, characterized as ``threshold function",
decays quickly for $w\gg 1$.
As a result, only modes with a running mass $m_i(k,\varphi) \lta k$
contribute to the renormalization of the potential. 
The decoupling of a given mode does not take place simultaneously for all 
values of the background field, because of the dependence 
of the mass $m_i$ on $\varphi$. This implies that the corresponding contributions 
to the vacuum energy may depend on $\varphi$ as well.
Despite the intuitive form of eq. (\ref{exactpotential}), its solution
displays a strong dependence on the UV cutoff $\Lambda$, 
which makes the precise determination of
the decoupling effects on the vacuum energy difficult.

\begin{figure}[t!]
\centering
\includegraphics[width=0.8\textwidth]{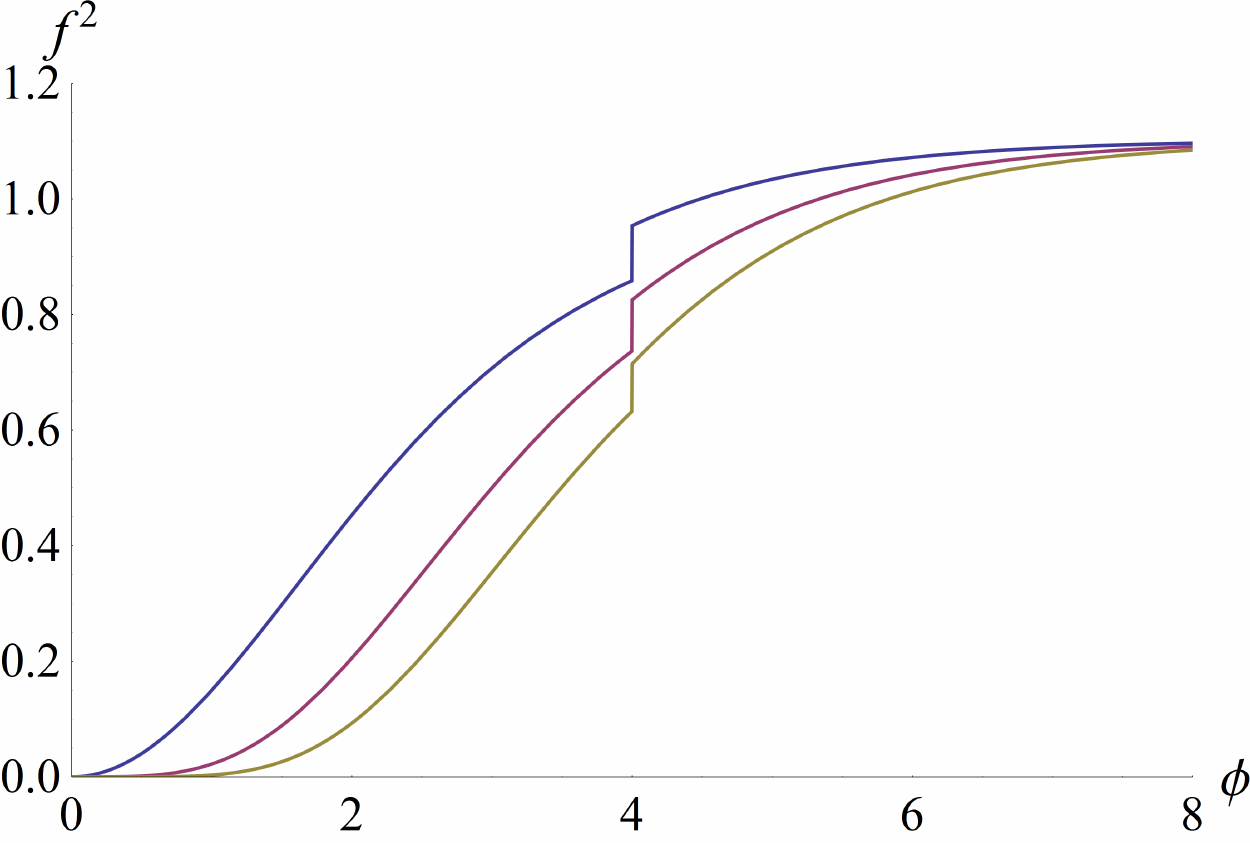}
\caption{
The square of the
function $f(\tanh[\varphi/\sqrt{6\alpha}])$ defined in eqs. (\ref{fx}) 
and (\ref{lalph}),
for $\alpha=1$, $A=0.05$, $x_0=\tanh[\varphi_0/\sqrt{6\alpha}]$ with $\varphi_0=4$, 
and $n=1,2,3$ (from top to bottom).
}
\label{alphafig}
\end{figure}

A different perspective on this issue can be obtained by considering 
the role of underlying symmetries. A specific framework is provided by the 
models associated with $\alpha$-attractors in supergravity \cite{alpha,alpha0}.
A toy model that can serve as a starting point is described by the Lagrangian \cite{alpha0}
\be
{\cal L}=\sqrt{-g}\left[ 
\frac{1}{2}\partial_\mu \chi \partial^\mu \chi +\frac{1}{12}\chi^2 R(g)
-\frac{1}{2}\partial_\mu \phi \partial^\mu \phi -\frac{1}{12}\phi^2 R(g)
-\frac{1}{36}f^2(\phi/\chi) \left(\chi^2-\phi^2 \right)^2.
\right]
\label{lagralpha} \ee
The model is invariant under the conformal transformation 
\be
g_{\mu\nu}\to e^{-2\sigma(x)}g_{\mu\nu},~~~~~~\phi\to e^{\sigma(x)}\phi,~~~~~~\chi\to e^{\sigma(x)}\chi.
\label{conformal} \ee
An interesting point is that, for constant $f(\phi/\chi)$, the model possesses a global 
$SO(1,1)$ symmetry that leaves $\chi^2-\phi^2$ invariant. 
The field $\chi$ does not have any physical degrees of freedom and
can be eliminated by imposing the gauge-fixing condition $\chi^2-\phi^2=6$. Following ref. \cite{alpha0}, we parametrize the
fields as  $\chi=\sqrt{6}\cosh(\varphi/\sqrt{6})$, $\phi=\sqrt{6}\sinh(\varphi/\sqrt{6})$.
The Lagrangian becomes
\be
{\cal L}=\sqrt{-g}\left[ 
 \frac{1}{2} R(g)-\frac{1}{2}\partial_\mu \varphi \partial^\mu \varphi
-f^2\left(\tanh \frac{\varphi}{\sqrt{6}} \right) \right].
\label{lagralphagauge} \ee
It is apparent that a constant function $f(x)$ corresponds to a cosmological constant. 
However, its value is not specified by the $SO(1,1)$ symmetry.
A possible deformation of the symmetry is obtained by assuming that $f(x)$ takes fixed values 
over two continuous ranges of $x$, with a rapid transition at a point $x_0$ in between. A stronger
deformation, which has been used extensively in the literature, 
assumes that $f(x)$ has a polynomial form.
A schematic form of $f(x)$, that displays a steep step and can also lead to power spectrum 
consistent with the cosmological constraints, is 
\be
f(x)=x^n+ A\, \Theta(x-x_0).
\label{fx} \ee
In practice, the step-function can be replaced by a smooth function. 
In the more general framework of the $\alpha$-attractors \cite{alpha,alpha0}, the 
Lagrangian takes the form 
\be
{\cal L}=\sqrt{-g}\left[ 
 \frac{1}{2} R(g)-\frac{1}{2}\partial_\mu \varphi \partial^\mu \varphi
-f^2\left(\tanh \frac{\varphi}{\sqrt{6\alpha}} \right) \right],
\label{lalph} \ee
with $\alpha$ a free parameter.
In fig. \ref{alphafig} we depict the square of the
function  $f(\tanh[\varphi/\sqrt{6\alpha}])$ defined according to eq. (\ref{fx}),
for $\alpha=1$, $A=0.05$, $x_0=\tanh[\varphi_0/\sqrt{6\alpha}]$ with $\varphi_0=4$, 
and $n=1,2,3$ (from top to bottom).

Our aim is to analyse the power spectra resulting from inflaton potentials
with the step feature displayed in fig. \ref{alphafig}.
We do not consider a specific model, but keep only a minimal
number of terms in the inflaton potential. 
The first term corresponds to vacuum energy,
for which we make the crucial assumption that it can have one or more 
transition points at which it jumps 
from one constant value to another. We also include a 
linear term, because it is the only term in a field expansion that is 
indispensable for our discussion. We
neglect the effect of higher powers of the inflaton field that would make
the analysis model dependent. Adjusting the free parameters can lead to 
the appearance of either an inflection point or a sharp drop in the potential, thus 
allowing the comparison of the effects of the two features.  
The unavoidable drawback of this simplified setup is that the potential is not flexible enough to generate the correct amplitude and tilt of the spectrum in the CMB range, as well as a sufficient number of efoldings.
This can be achieved for a potential that includes higher powers of the field.
As an example, we analyse a potential inspired by the 
Starobinsky model \cite{starobinsky}. However, we do 
not engage in detailed model building here, deferring
such an investigation to future work.

In the following section we summarize the basic formalism related to the Mukhanov-Sasaki equation. For the numerical analysis it is most convenient to express this
equation using the number of efoldings as independent variable. In section \ref{results}
we present the results of a numerical calculation of the spectum, as well as 
an analytical discussion of the appearing features. The final section includes 
a summary of our findings.

\section{The Mukhanov-Sasaki equation} \label{notes}

In this section we introduce the relevant quantities and collect the 
corresponding dynamical equations for the study of the 
curvature perturbations and their spectrum. 

The most general scalar metric perturbation 
around the Friedmann-Robertson-Walker 
(FRW) background takes the form \cite{physrep}
 \be
 ds^2=a^2(\tau)\left\{ (1+2\phi)d\tau^2-2B_{,i}\,dx^i d\tau
 -\left((1-2\psi)\delta_{ij}+2E_{,ij} \right)dx^i dx^j
 \right\},
 \label{metric} \ee
 with $B_{,i}=\partial_i B$, $E_{,ij}=\partial_i\partial_jE$.
On this background, one can parametrize the inflaton field as
$\varphi(\tau)+\delta\varphi(\tau,x)$ and define a gauge-invariant perturbation as
 \be
 v
 =a\left( \dphi+\frac{\varphi'}{\Hc}\psi \right),
 \label{vv} \ee
which satisfies the Mukhanov-Sasaki equation \cite{mukhsas}
\be
v''-\nabla^2 v-\frac{z''}{z}\, v=0,
\label{eomv} \ee
with $z=a\varphi'/\Hc$. The primes and the Hubble parameter 
correspond to derivatives with respect to conformal time.
The Fourier modes of $v$ satisfy
\be
v_k''(\tau)+\left( k^2-\frac{z''}{z} \right) v_k(\tau)=0.
\label{eomvfourier} \ee
The standard assumption, which we adopt, is that at early times the field
is in the Bunch-Davies vacuum. The strong features of the potential have not become
relevant yet, so that the background field is in the slow-roll regime.
All the modes that are phenomenologically interesting today were deeply subhorizon
at such early times. They are described by the standard expression  
$v_k=e^{-ik\tau}/\sqrt{2k}$,
which we use in order to set the initial conditions for their subsequent evolution. 
The spectrum of perturbations becomes more transparent through the use of the 
gauge-invariant comoving curvature perturbation $R=-v/z$, which satisfies
\be
R_k''+2\frac{z'}{z}R_k'+k^2 R_k=0
\label{geqqf} \ee
in Fourier space.

As we are mainly interested in the amplitude of the complex 
variables $v$ and $R$, 
we introduce polar coordinates, such that
$v_k(\tau)=\Vc_k(\tau) \exp (-i \theta_k(\tau))$,
with $\Vc_k$ and $\theta_k$ real.
From eq. (\ref{eomvfourier}) we obtain
\begin{eqnarray}
\Vc_k''+\left(k^2-\frac{z''}{z}-\theta_k'^2 \right) \Vc_k&=&0
\label{ampl} \\
\frac{\theta_k''}{\theta_k'}+2\frac{\Vc_k'}{\Vc_k}&=&0.
\label{argu} 
\end{eqnarray}
The second equation can be integrated, with the solution 
$\theta_k'\,\Vc_k^2={\rm constant}$
At early times we have
$\Vc_k=1/\sqrt{2k}$ and 
$\theta_k=k\tau$. This
fixes the constant of integration to 1/2, so that we can set
\be
\theta_k'=\frac{1}{2\Vc_k^2}
\label{thetasol} \ee
in eq. (\ref{ampl}). In this way we obtain
\be
\Vc_k''+\left(k^2-\frac{z''}{z}-\frac{1}{4\Vc_k^4} \right) \Vc_k= 0,
\label{ampln} \ee
which must be solved with initial conditions
$\Vc_k\to 1/\sqrt{2k}$, $\Vc_k'\to 0$, for $\tau\to -\infty$.
The curvature perturbation is parametrized as 
$R_k(\tau)=-\Rc_k(\tau) \exp(-i\theta_k(\tau))$, with $\Rc_k=\Vc_k/z$.
Its amplitude satisfies
\be
\Rc_k''+2\frac{z'}{z}\Rc_k'+ \left(k^2-\frac{1}{4z^4\Rc_k^4} \right) \Rc_k= 0.
\label{Rceq} \ee

It is convenient for the numerical analysis to use the number of
efoldings $N$ as the independent variable for the evolution of the
perturbations. 
The Hamilton-Jacobi slow-roll parameters are defined through the relations
\begin{eqnarray}
H^2&=&\frac{V(\varphi)}{3\mpl^2-\frac{1}{2}\varphi_{,N}^2}
\label{HH22} \\
\exx_H&=&
-\frac{d\ln H}{dN}
=\frac{ \varphi^2_{,N}}{2 \mpl^2}
\label{b1} \\
\eta_H&=&\exx_H-\frac{1}{2}\frac{d \ln \exx_H}{dN}=
\frac{ \varphi^2_{,N}}{2 \mpl^2}-\frac{\varphi_{,NN}}{\varphi_{,N}},
\label{b2} \end{eqnarray}
where $H=e^{-N}\Hc$ is the Hubble parameter defined through cosmic time,
and subscipts denote derivatives with respect to $N$.
The parameter $z$ is given by 
\be
z= e^N\, \varphi_{,N}
\label{zN}, \ee 
while the 
effective equation of state for the background is 
$w=-1+2\exx_H/3$.

The evolution of the background field is governed by the equation
\be
\varphi_{,NN}+3\varphi_{,N}-\frac{1}{2\mpl^2}\varphi_{,N}^3+
\left(3\mpl^2-\frac{1}{2}\varphi_{,N}^2 \right)\frac{V_{,\varphi}}{V}=0,
\label{eomH} \ee
with $V(\varphi)$ the inflaton potential.
The inflaton fluctuation obeys the equation
\be
v_{k,NN}+(1-\exx_H)v_{k,N}+\left(\frac{k^2}{e^{2N} H^2}
+(1+\exx_H-\eta_H)(\eta_H-2)-(\exx_H-\eta_H)_{,N}    \right) v_k =0,
\label{eovH} \ee
and its amplitude
\be
\Vc_{k,NN}+(1-\exx_H)\Vc_{k,N}+\left(\frac{k^2}{e^{2N} H^2}
\left(1-\frac{1}{4k^2\Vc_k^4} \right)
+(1+\exx_H-\eta_H)(\eta_H-2)-(\exx_H-\eta_H)_{,N}    \right) \Vc_k =0.
\label{eovHampl} \ee
In the above differential equations we can express the coefficients as
\begin{eqnarray}
1-\exx_H&=&1-\frac{\varphi^2_{,N}}{2\mpl^2}
\label{epsa} \\
(1+\exx_H-\eta_H)(\eta_H-2)-(\exx_H-\eta_H)_{,N} &=&
-2-3\frac{\varphi_{,NN}}{\varphi_{,N}}
-\frac{\varphi_{,NNN}}{\varphi_{,N}}
+\frac{\varphi_{,N}^2}{2\mpl^2}+\frac{\varphi_{,N}\,\varphi_{,NN}}{2\mpl^2}.
\label{etaa} 
\end{eqnarray}
We can also write equivalent equations for the curvature perturbation,
which take the form
\be
R_{k,NN}+\left(
3+\frac{2\varphi_{,NN}}{\varphi_{,N}}-\frac{\varphi_{,N}^2}{2\mpl^2} \right)
R_{k,N}
+\frac{k^2}{e^{2N} H^2} R_{k}=0
\label{RN} \ee
and
\be
\Rc_{k,NN}+\left(
3+\frac{2\varphi_{,NN}}{\varphi_{,N}}-\frac{\varphi_{,N}^2}{2\mpl^2} \right)
\Rc_{k,N}
+\frac{k^2}{e^{2N} H^2}
\left(1-\frac{1}{4 k^2 e^{4N}\varphi_{,N}^4\Rc_k^4} \right)\Rc_{k}=0,
\label{RcN} \ee
for the perturbation and its amplitude, respectively.

The spectrum of curvature perturbations is 
\be
\Delta_R^2=\frac{k^3}{2\pi^2}\frac{\Vc_k^2}{e^{2N}\varphi_{,N}^2}=
\frac{k^3}{2\pi^2}\Rc_k^2.
\label{result} \ee
The normalization of the spectrum can be set in terms of a 
pivot scale $k_*$ and the number of efoldings $N_*$ at which it crosses
the horizon: $k_*=e^{N_*} H_*$.
By defining dimensionless variables 
$\kt=k/k_*$, $\vt_k=\sqrt{k_*}\, v_k $, $\Vct_k=\sqrt{k_*}\, \Vc_k $, 
 $\Rt_k=\sqrt{k_*}\, R_k $, $\Rct_k=\sqrt{k_*}\, \Rc_k $, as well
as $\dN=N-N_*$, we obtain
\be
\Delta_R^2= A_s \frac{\kt^3\,2\Vct_k^2}{e^{2\,\dN}} \frac{\varphi_{,N*}^2}{\varphi_{,N}^2}.
\label{resultpiv} \ee
where 
\be
A_s= \frac{1}{4\pi^2}\frac{H^2_*}{\varphi^2_{,N*}} 
\label{As} \ee
 sets the scale for the amplitude.

For a given inflaton potential, one can integrate numerically eq. (\ref{eomH})
in order to derive the inflaton background, and then integrate one of 
eqs. (\ref{eovH}), (\ref{eovHampl}), (\ref{RN}), (\ref{RcN}) for the field or 
curvature perturbation, in order to deduce the spectrum.
The real and imaginary parts of $v_k$ and $R_k$ oscillate very rapidly
for subhorizon perturbations, as can be deduced from eqs.
(\ref{eovH}), (\ref{RN}). This makes the numerical integration of these
equations more demanding. On the other hand, the amplitudes
$\Vc_k$ and $\Rc_k$ have a smoother evolution. It is possible for
these quantities to become oscillatory also, as we shall see in
the following. However, the presence of the terms $\sim \Vc_k^{-4}$  
in eq. (\ref{eovHampl}) and $\sim \Rc_k^{-4}$ in eq. (\ref{RcN}) 
guarantees that these amplitudes remain always positive.
For the numerical analysis of the following sections, we 
solve the evolution equations both for the field perturbation $v_k$ and its amplitude $\Vc_k$ 
in order to cross-check the results. 

A quantity that plays a crucial role in determining the qualitative 
behaviour of the solutions is the one in the first parenthesis of 
eqs. (\ref{RN}), (\ref{RcN}), which we denote by
\be
f(N)=3+\frac{2\varphi_{,NN}}{\varphi_{,N}}-\frac{\varphi_{,N}^2}{2\mpl^2}
\label{AN} \ee
as a function of $N$. In the slow-roll regime, 
this quantity acts as a generalized friction term. However, for the more general
evolution that we are considering, it may become negative and lead to
a dramatic enhancement of the perturbations. 
We also define the function
\be
g(N)=1-\frac{1}{4 k^2 e^{4N}\varphi_{,N}^4\Rc_k^4},
\label{BN} \ee
appearing in the second parenthesis,
evaluated on a given solution for the perturbation. This function
diverges whenever the amplitude $\Rc_k$ approaches zero, thus preventing it from 
turning negative. An alternative way to view this point is to notice that 
eq. (\ref{RcN}) is equivalent to eq. (\ref{RN}), while the amplitude 
of $R_k$ cannot turn negative. The fact that $\Rc_k$ can approach zero at certain times during the 
later stages of the evolution, as we shall
see in the following, indicates
that during these stages the real and the imaginary part of $R_k$ are in phase and
can cross zero almost simultaneously.

\section{Features of the inflaton potential}\label{results}

We would like to explore features of the inflaton potential that can 
result in an amplification of the spectrum of curvature perturbations
by several orders 
of magnitude. Our underlying motivation is to determine the appropriate
conditions for the  
creation of primordial black holes. This is possible 
in a range of scales in which
perturbations become of order one. 
Significant deviations from the scale-invariant spectrum can occur
only at small length scales (large wavenumbers), for which the 
evolution of the spectrum is highly nonlinear, so that current
observations do not constrain its form severely. Such scales 
correspond to comoving wavenumbers larger than ${\cal O}(1)$ in 
units of Mpc$^{-1}$.

\subsection{Minimal framework}

\begin{figure}[t!]
\centering
\includegraphics[width=0.24\textwidth]{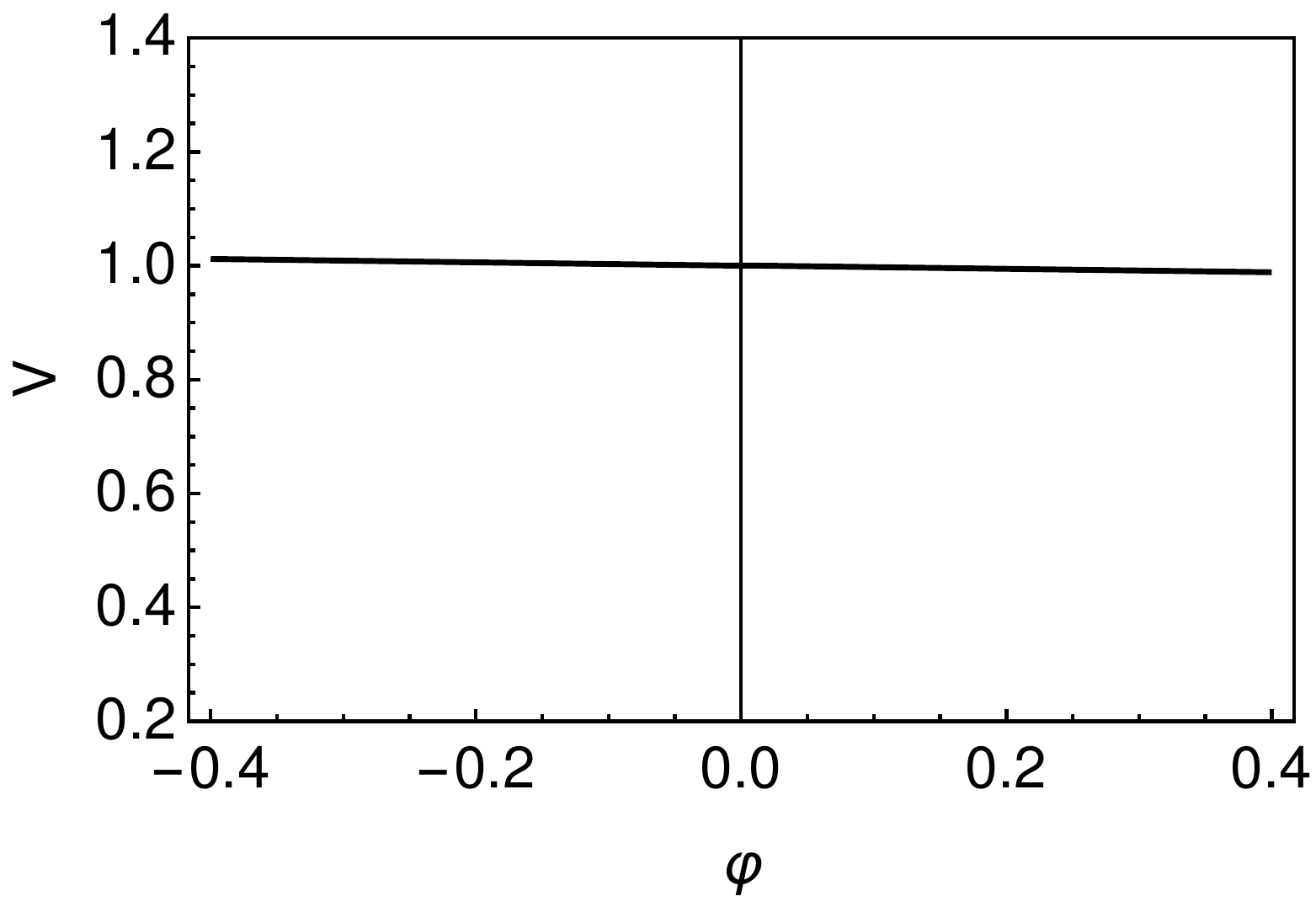} \includegraphics[width=0.24\textwidth]{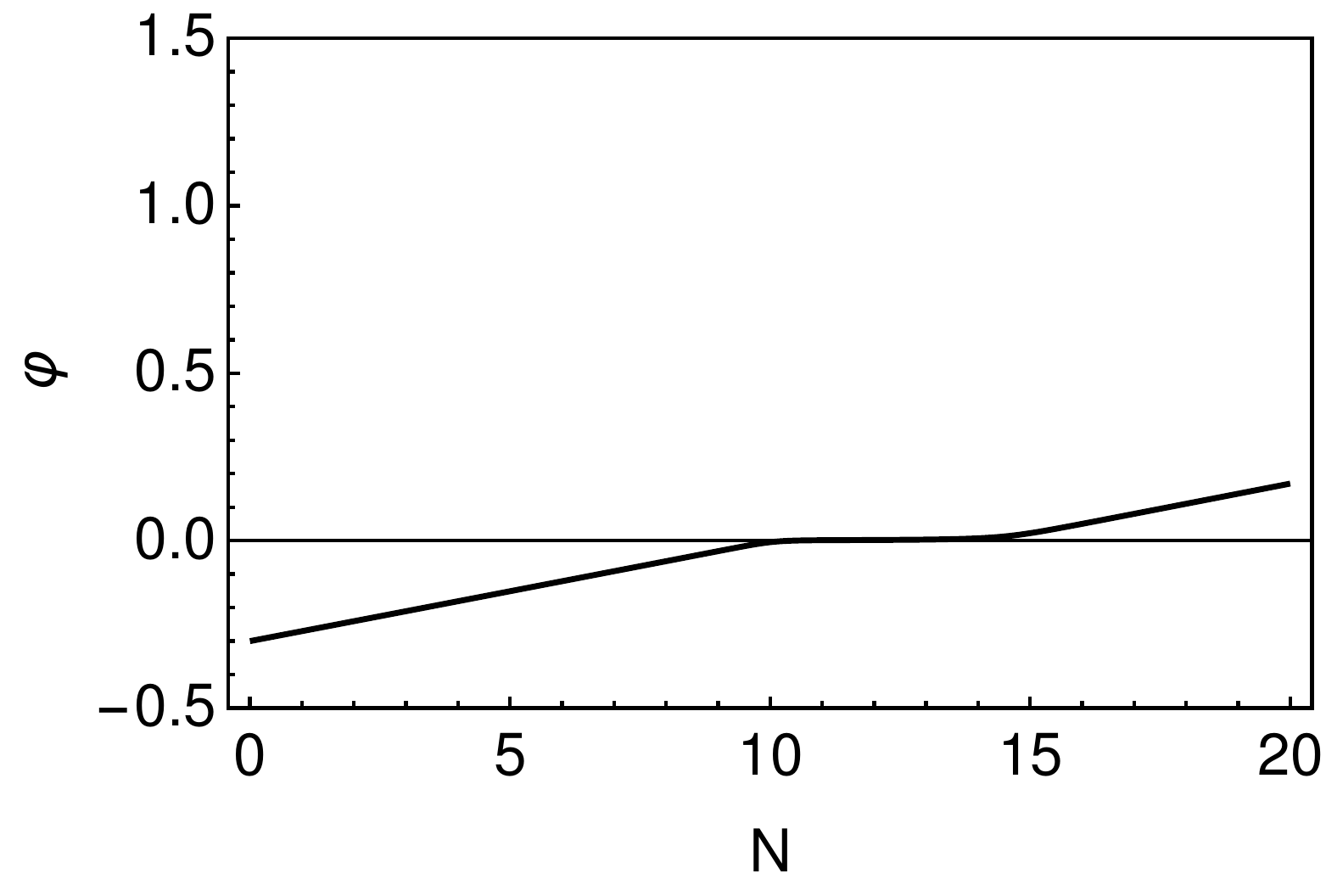} \includegraphics[width=0.24\textwidth]{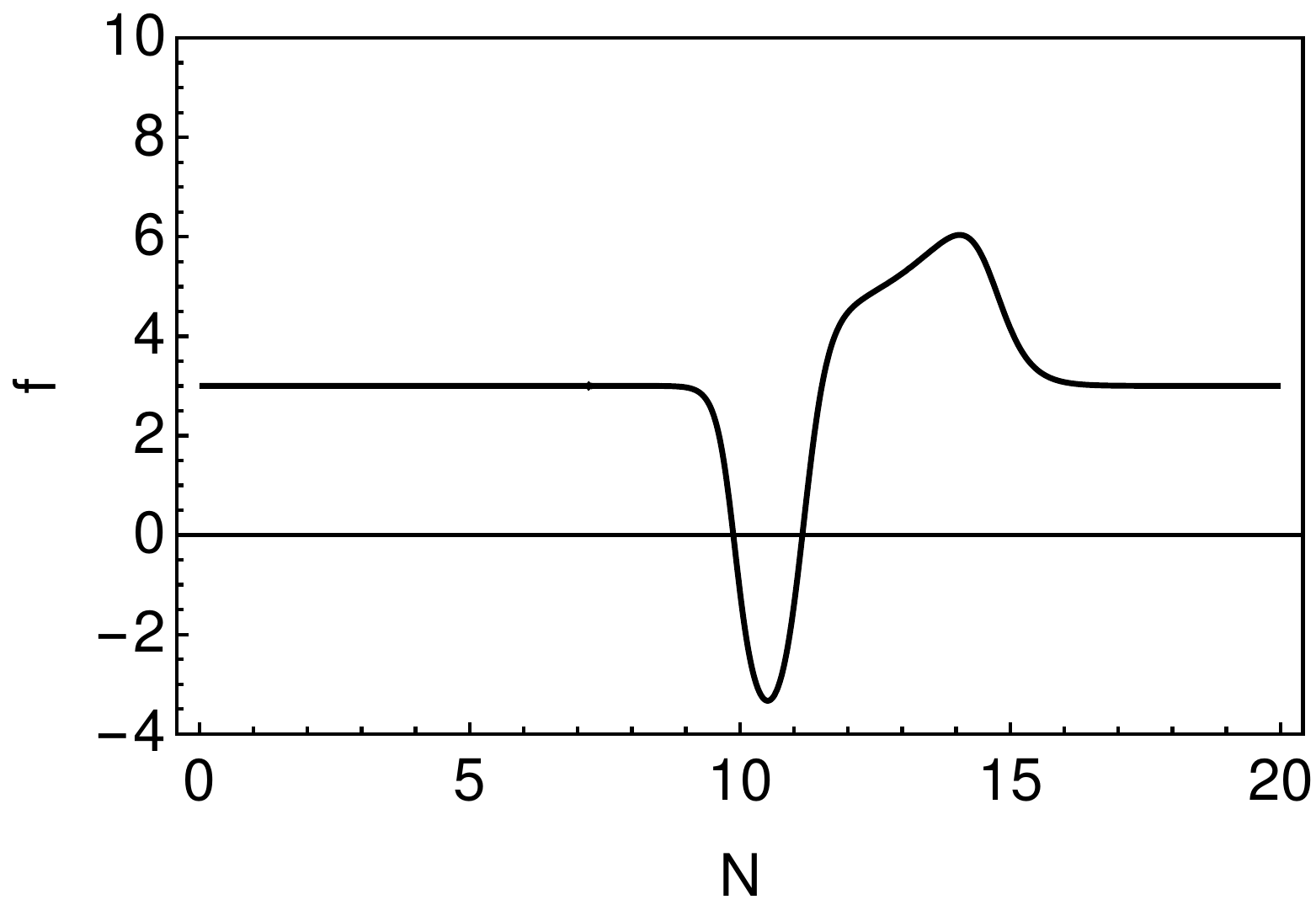} \includegraphics[width=0.25\textwidth,height= 28mm]{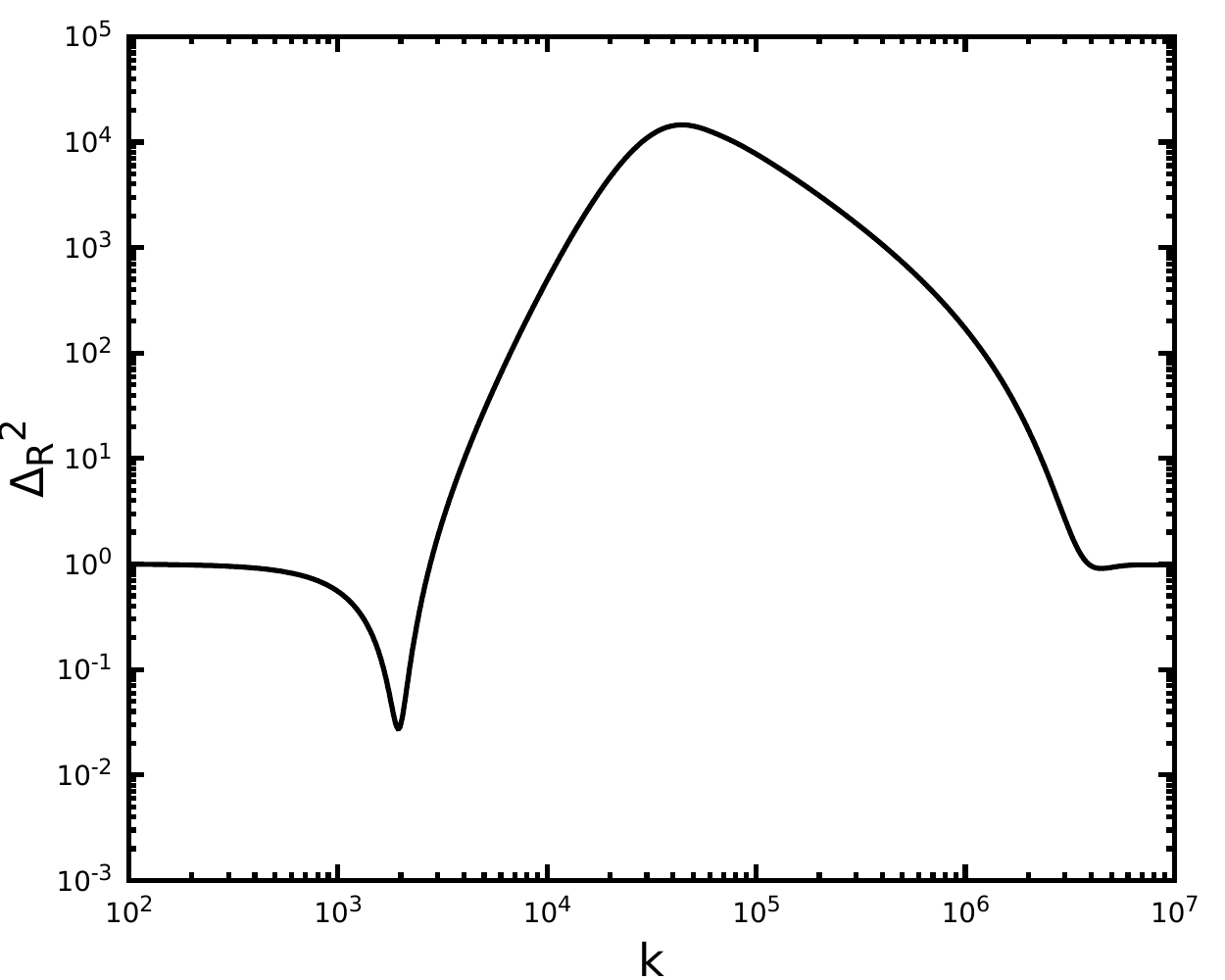}
\includegraphics[width=0.24\textwidth]{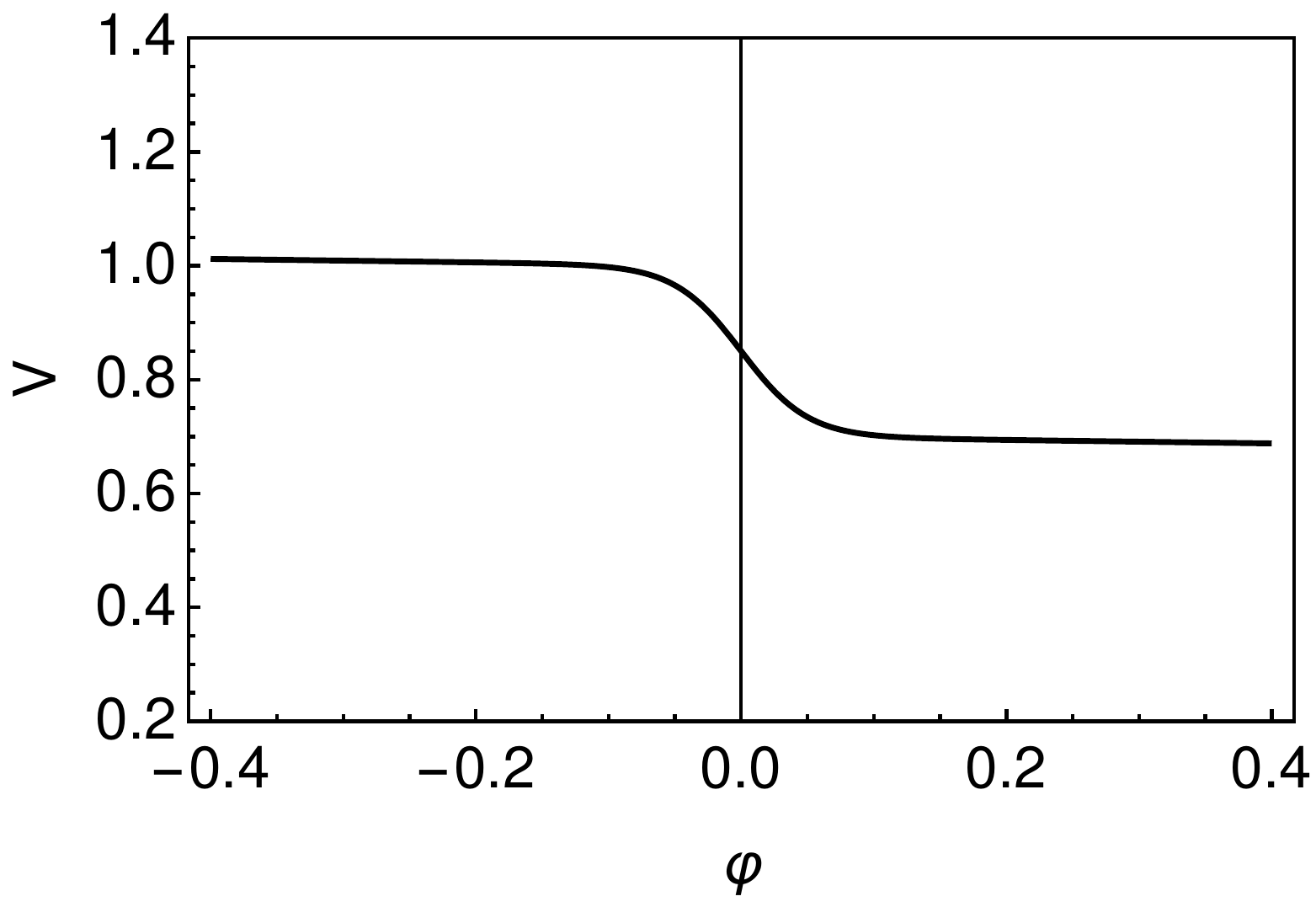} \includegraphics[width=0.24\textwidth]{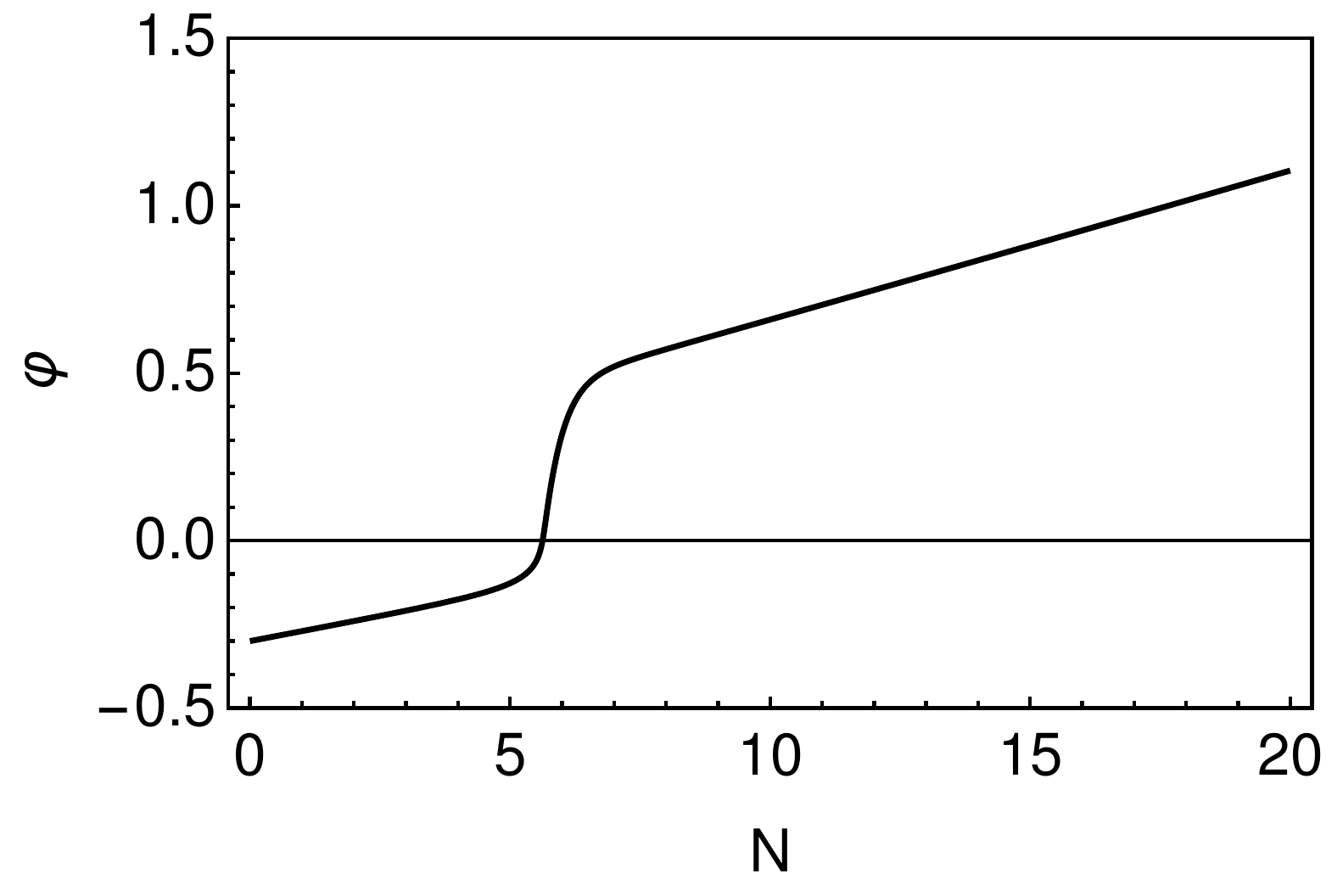}
\includegraphics[width=0.24\textwidth]{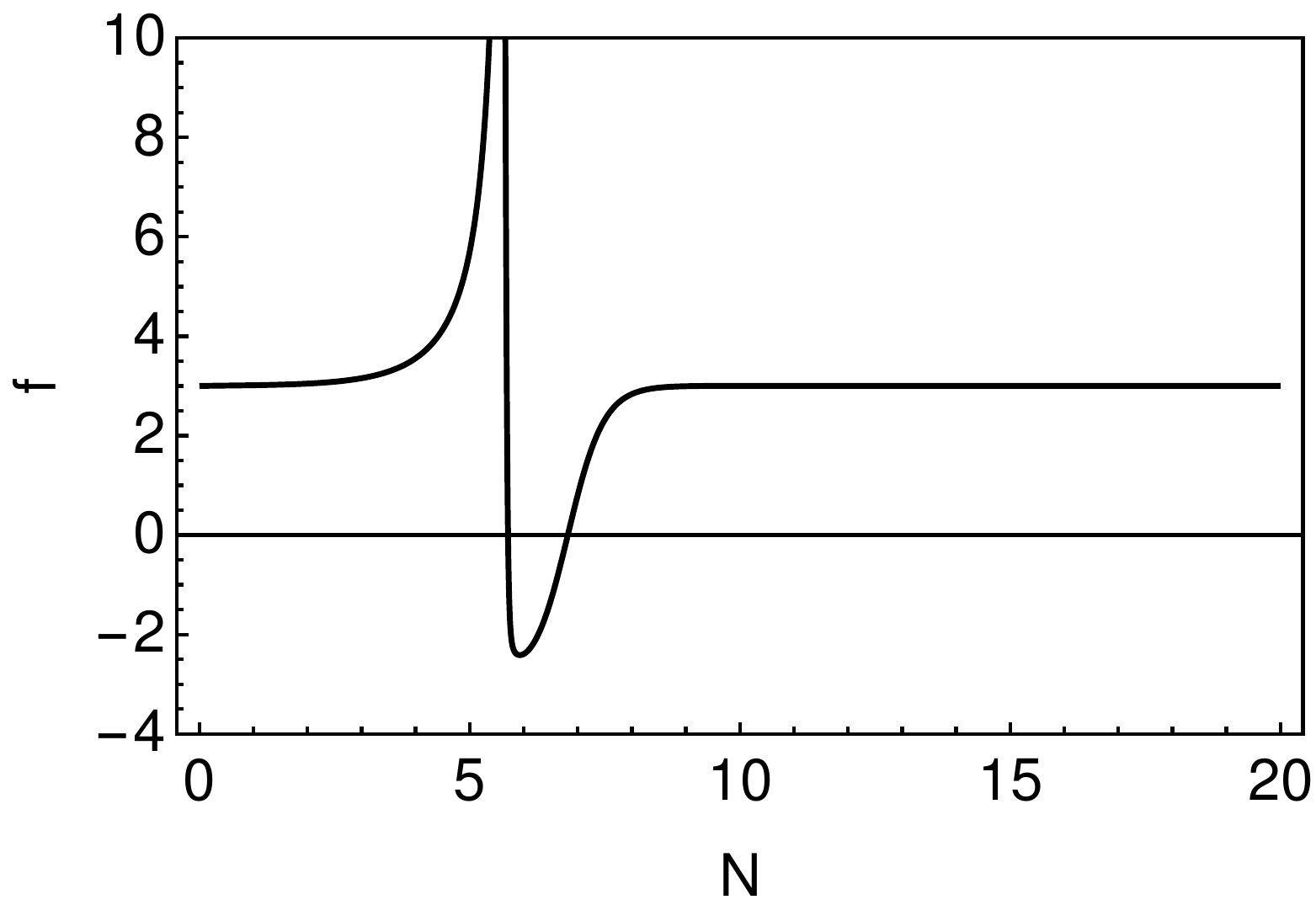} \includegraphics[width=0.25\textwidth,height= 28mm]{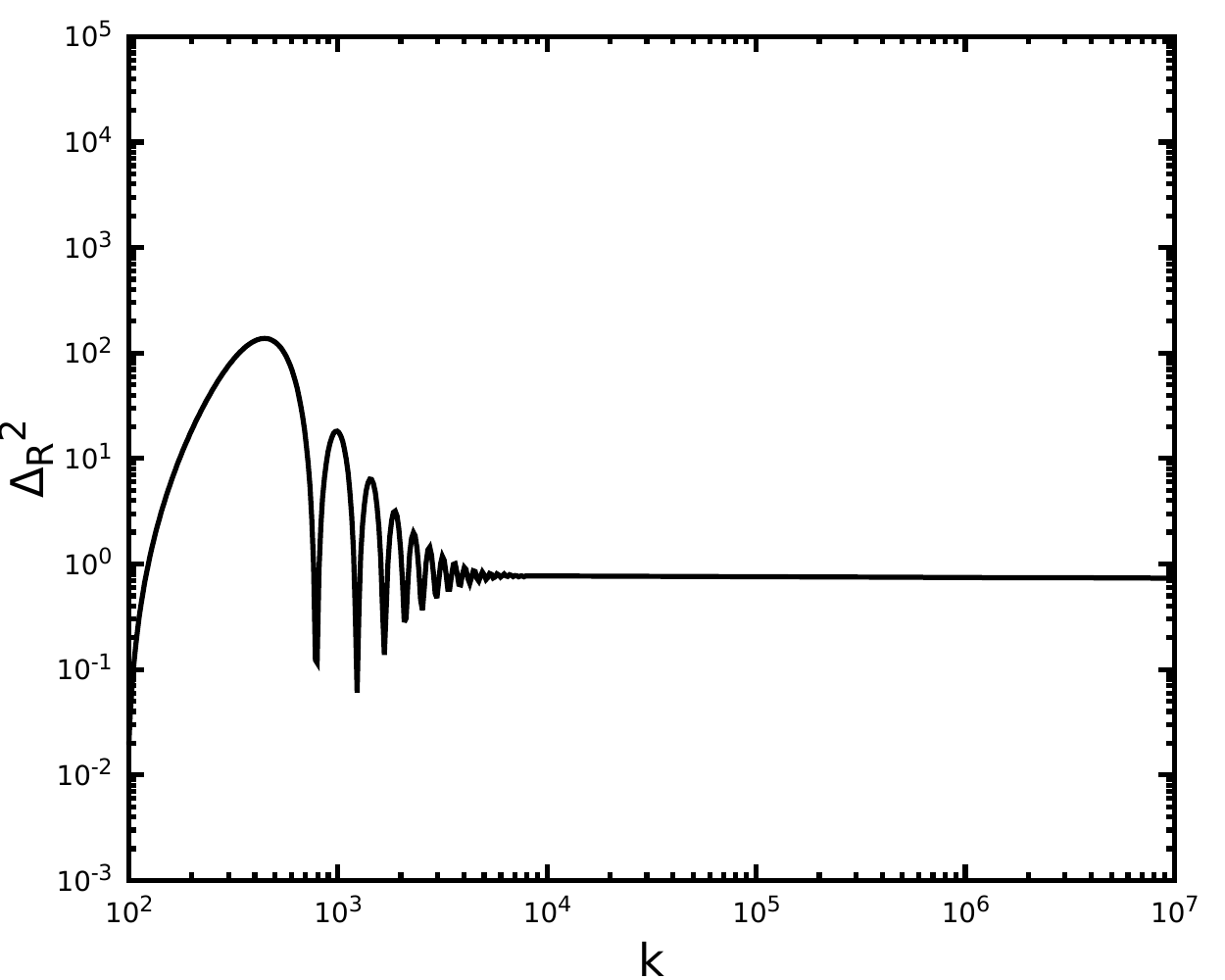}
\includegraphics[width=0.24\textwidth]{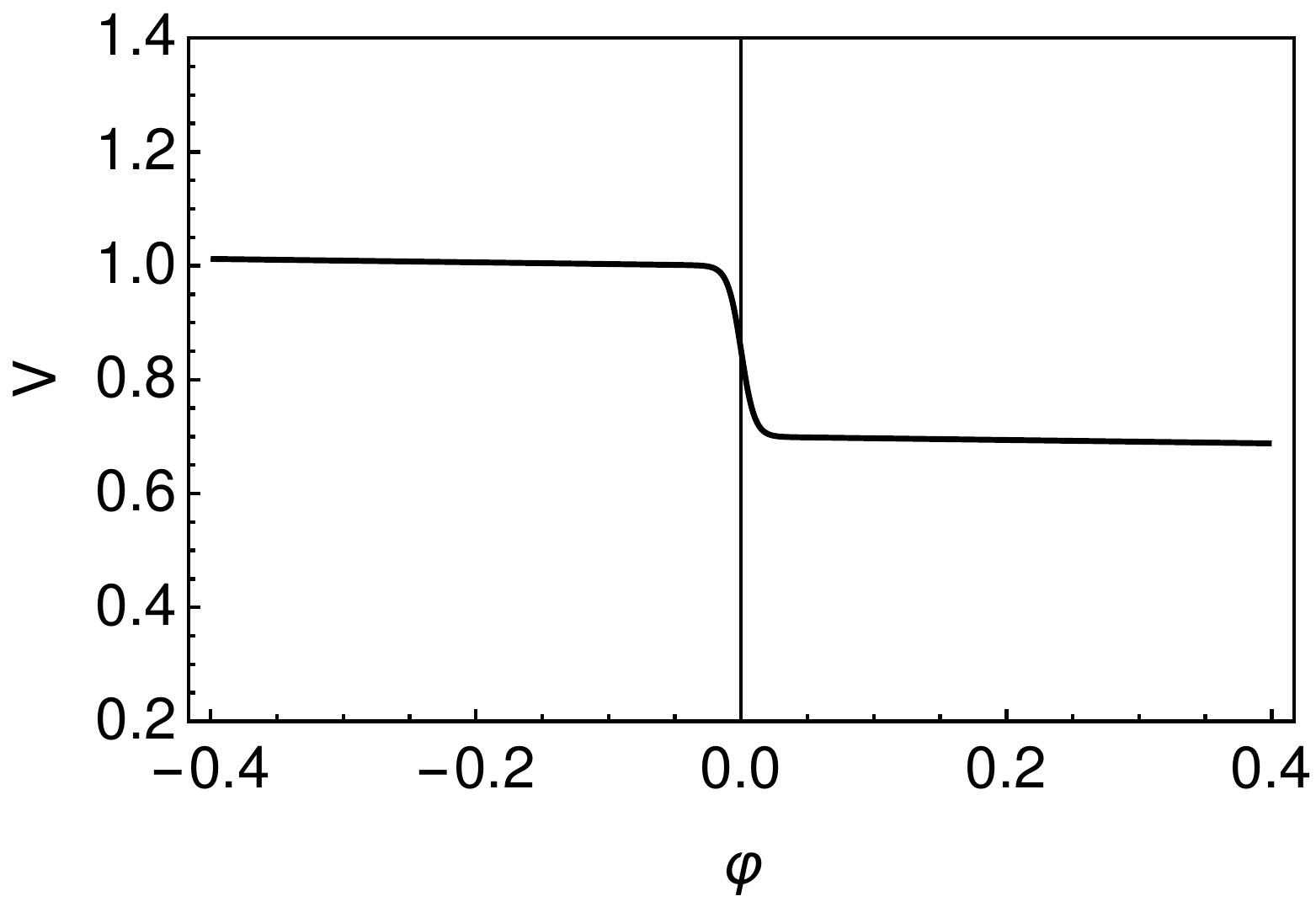} \includegraphics[width=0.24\textwidth]{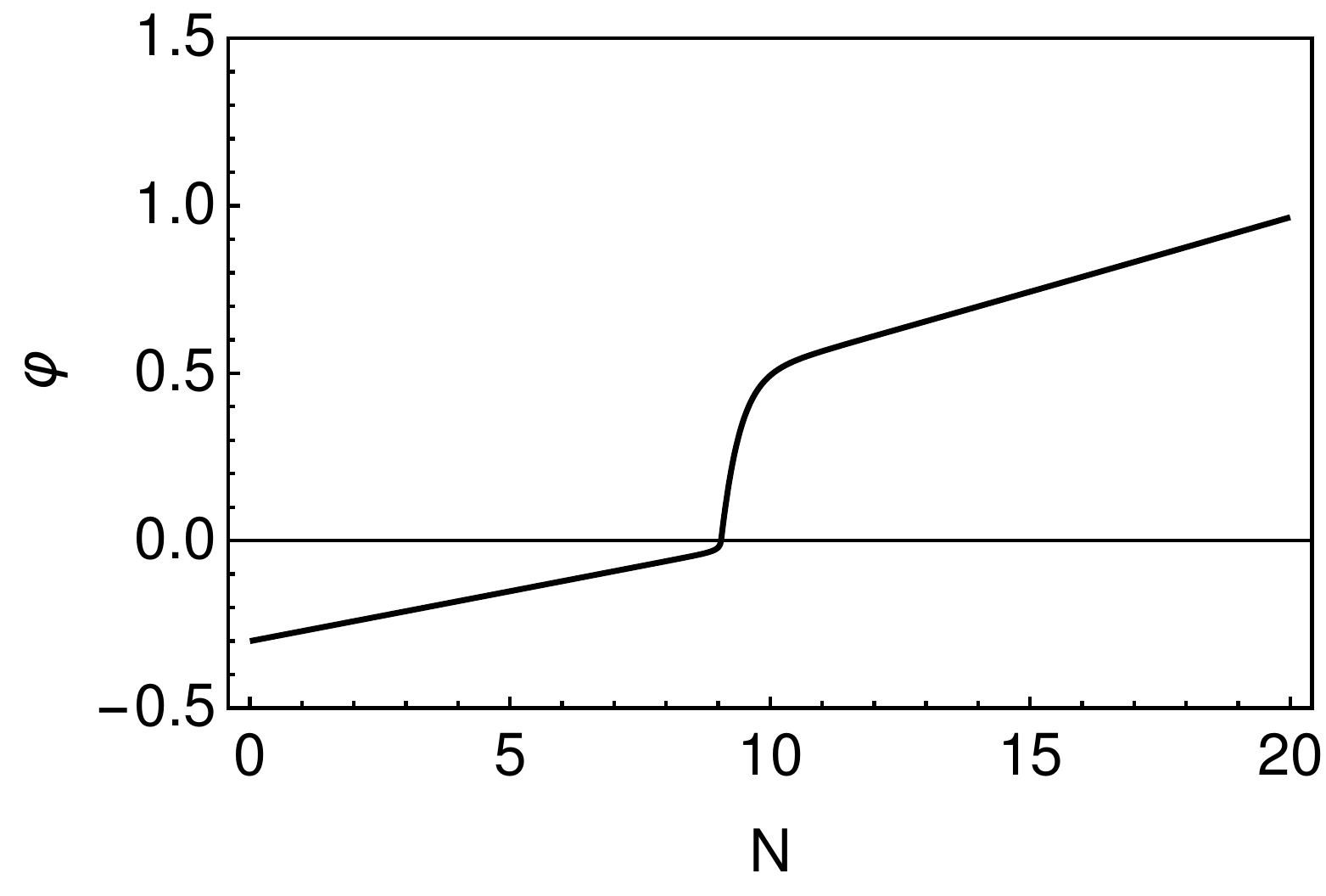}
\includegraphics[width=0.24\textwidth]{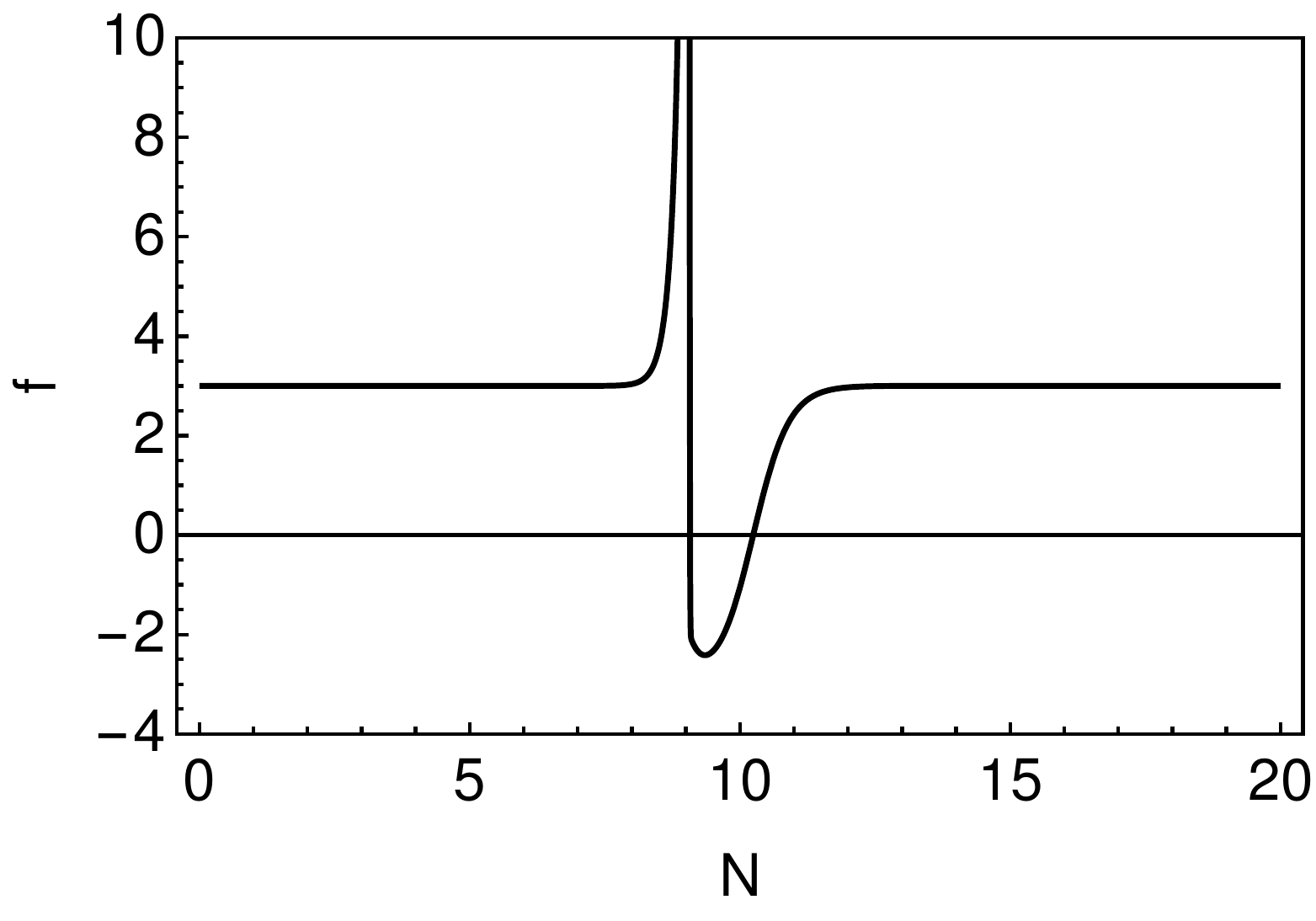} \includegraphics[width=0.25\textwidth,height= 28mm]{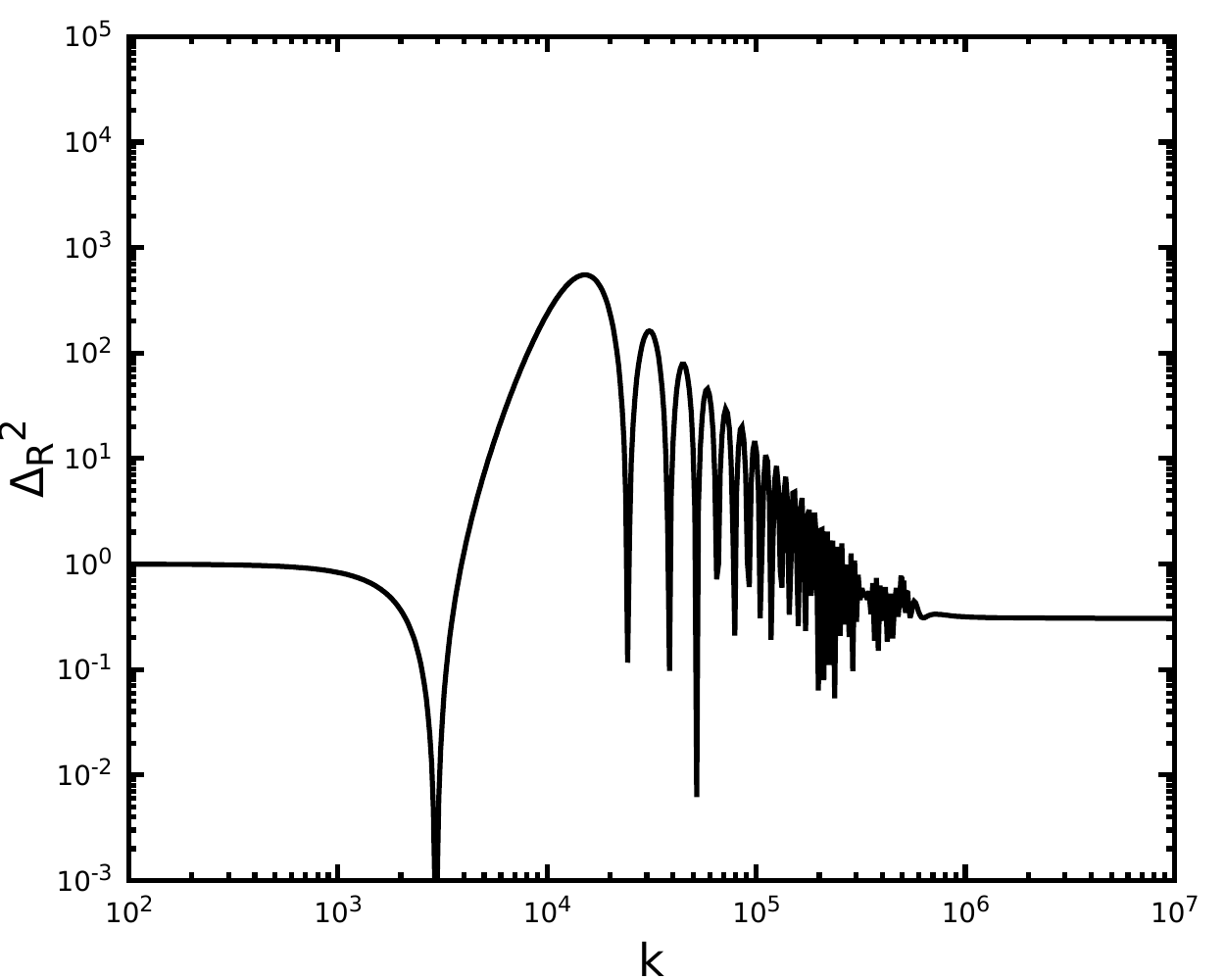}
\includegraphics[width=0.24\textwidth]{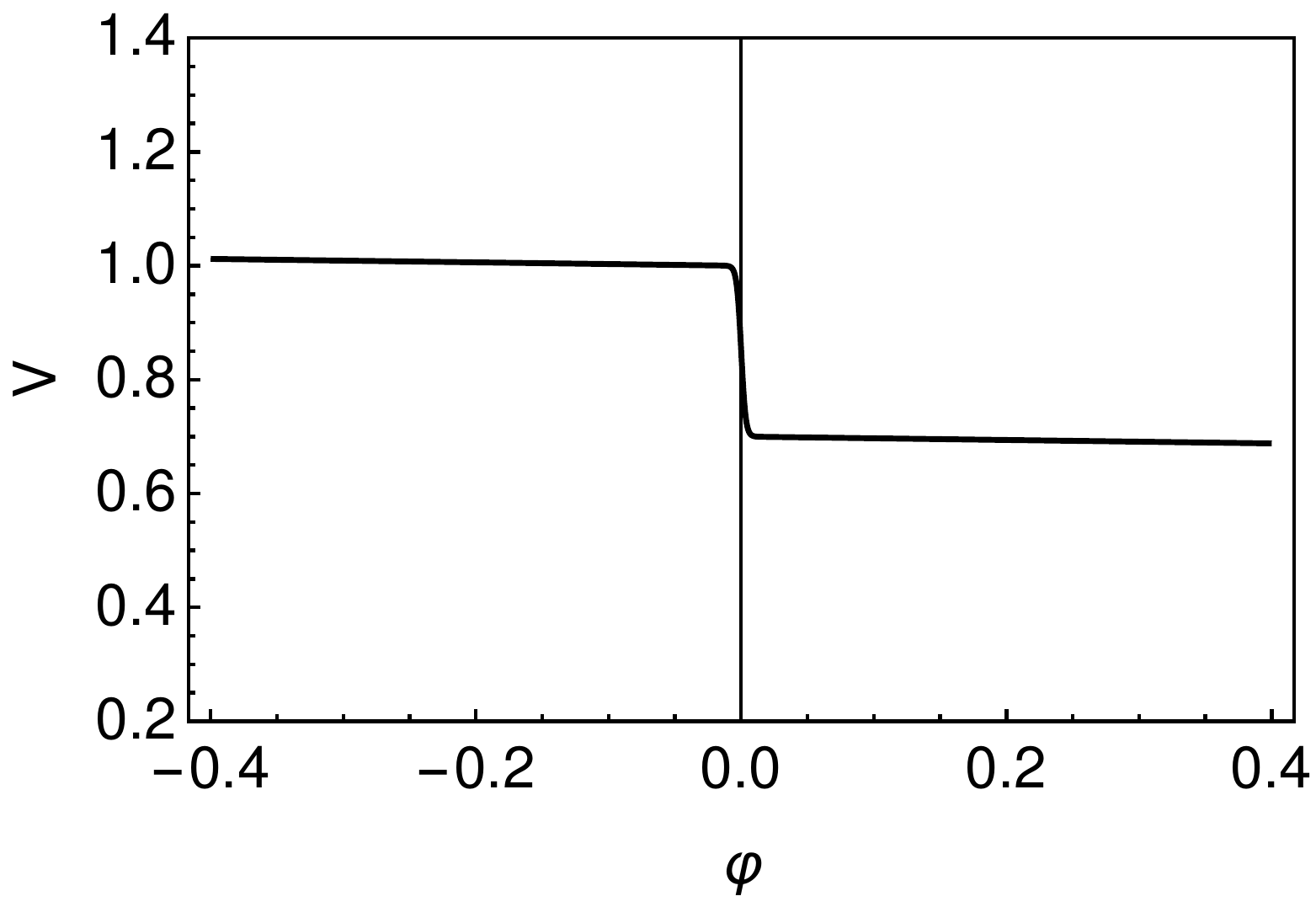} \includegraphics[width=0.24\textwidth]{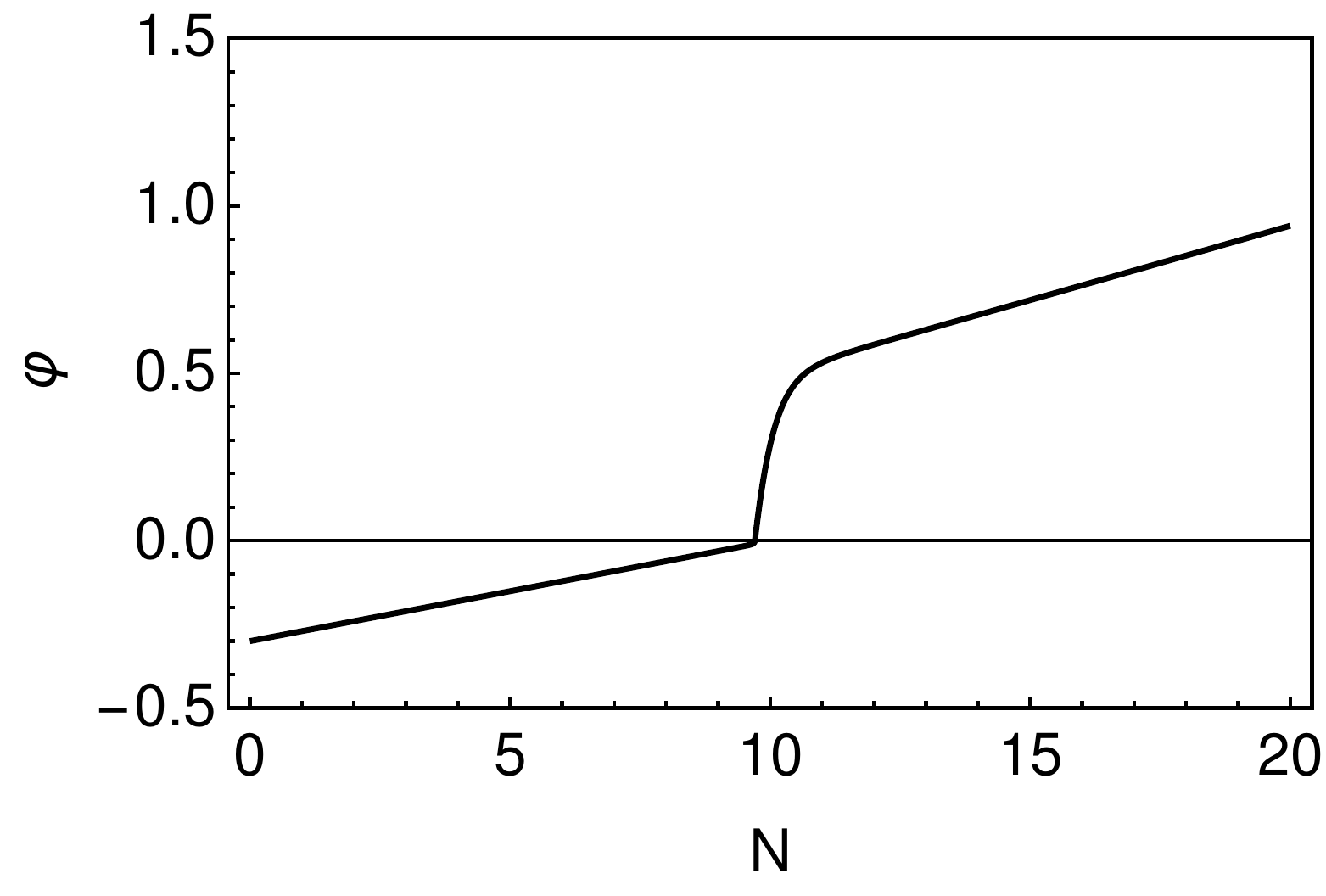}
\includegraphics[width=0.24\textwidth]{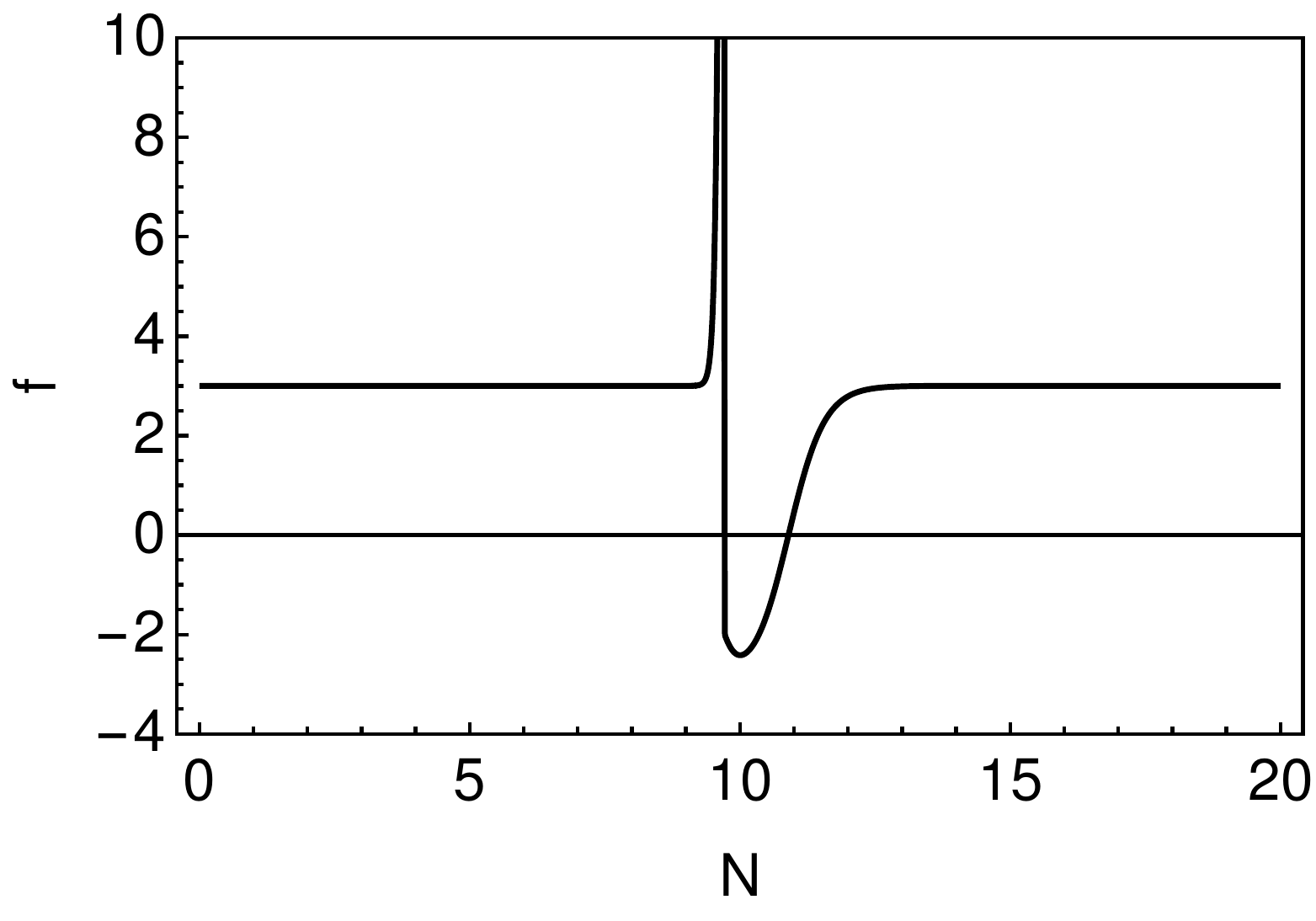} \includegraphics[width=0.25\textwidth,height= 28mm]{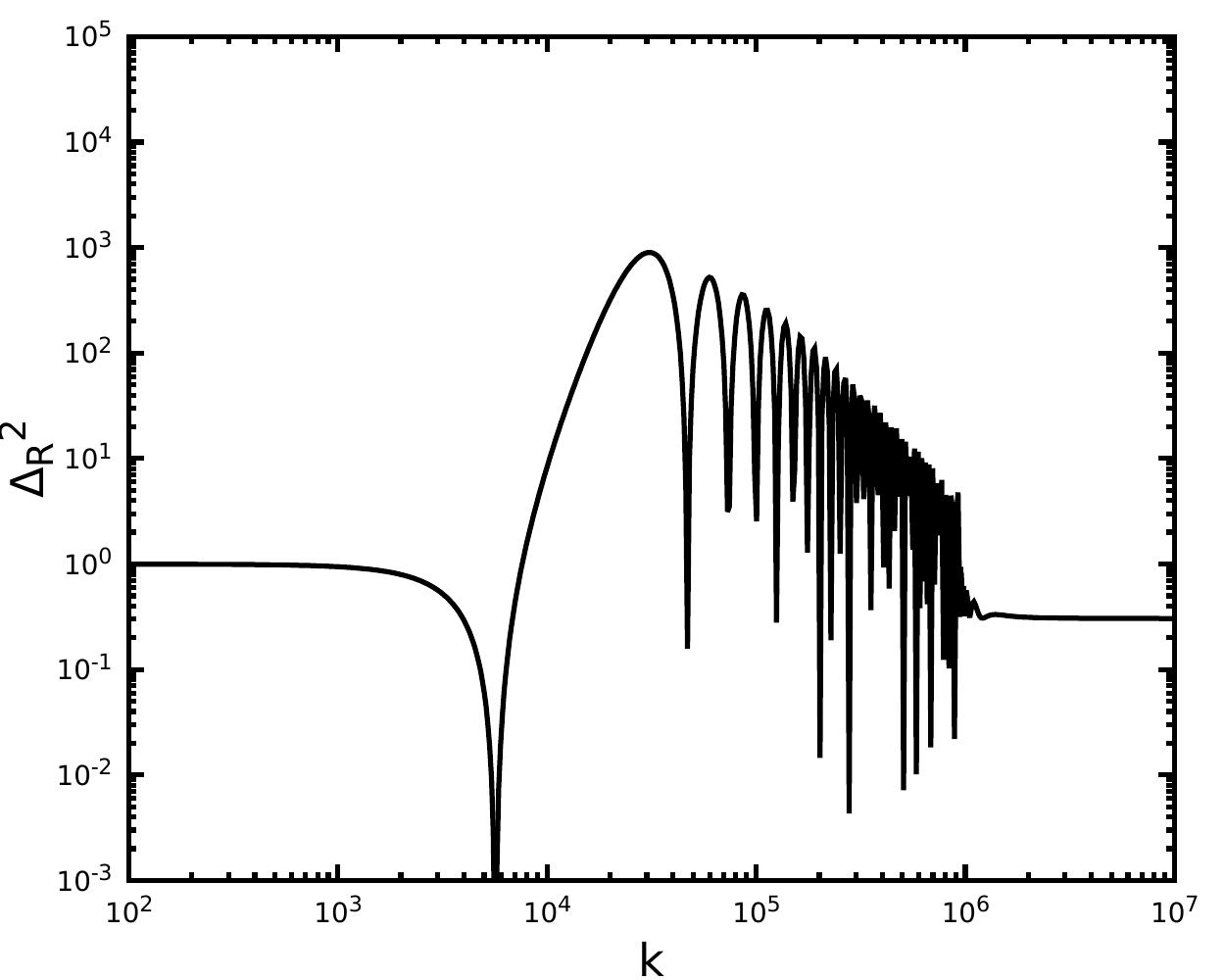}
\caption{
The inflaton potential $V(\varphi)$ of eq. (\ref{pote}), the evolution of the inflaton
$\varphi$, the function
$f(N)$ defined in eq. (\ref{AN}), and the power spectrum of curvature perturbations 
with wavenumber $k$, 
for various choices of the parameters of the potential: 
First row: $A_1=0.000605$, $c_i=100$, $B=-0.03$.~
Second row:  $A_1=-0.3$, $c_i=20$, $B=-0.03$.~
Third row:  $A_1=-0.3$, $c_i=100$, $B=-0.03$.~
Fourth row:  $A_1=-0.3$, $c_i=300$, $B=-0.03$.
The scales of $k$ and $V$ are arbitrary.
}
\label{figone}
\end{figure}

Instead of considering a specific model, we keep only the minimal
number of elements required for addressing the problem. 
We focus on only a limited range of scales, and 
the corresponding values of the inflaton background when these exit
the horizon. We approximate the inflaton potential by the smallest number of
relevant terms. The features of interest are:
\begin{enumerate}
\item
an inflection point, at which the first and second derivatives of
the potential vanish,
\item
one or more points at which the potential decreases sharply.
\end{enumerate}
Both these features can appear in a potential with the simple
parameterization
\be
V(\varphi)=V_0\left( 1+\frac{1}{2}\sum_i A_i 
\left( 1+\tanh(c_i (\varphi-\varphi_i)) \right)+B\varphi \right),
\label{pote} \ee
where $i$ is a positive integer counting certain special field values.
The first terms in the parenthesis can be identified with the vacuum energy 
that drives inflation. The crucial assumption that we have made is that
the vacuum energy can have one or more transition points at which it jumps 
from one constant value to another. As we discussed in the introduction, 
one could speculate that these
points correspond to values of the inflaton background associated with some
kind of decoupling of modes whose quantum fluctuations contribute 
to the vacuum energy. However, such a speculation cannot be put easily on formal
ground because of our lack of understanding of the nature of the 
cosmological constant. Sharp changes in the vacuum energy can also occur
during transitions from one region of a multi-field potential to another.
The analysis of such a system would require the inclusion of 
entropy perturbations. The current work is a simplified first step towards 
understanding the features that could appear in the spectrum of 
curvature perturbations for a multi-field system. The linear term in the
potential (\ref{pote}) is the only term in a field expansion that is 
indispensable for our discussion. In this subsection we
neglect the effect of higher powers of the inflaton field that would make
the analysis model dependent. 
We assume, without loss of generality, that $B<0$.
An inflection point can appear at
$\phi_1=0$ if $A_1=-2B/c$, $A_i=0$ for $i>1$. 
Negative values of $A_i$ result in a series of steps in the potential.

\begin{figure}[t!]
\centering
\includegraphics[width=0.24\textwidth]{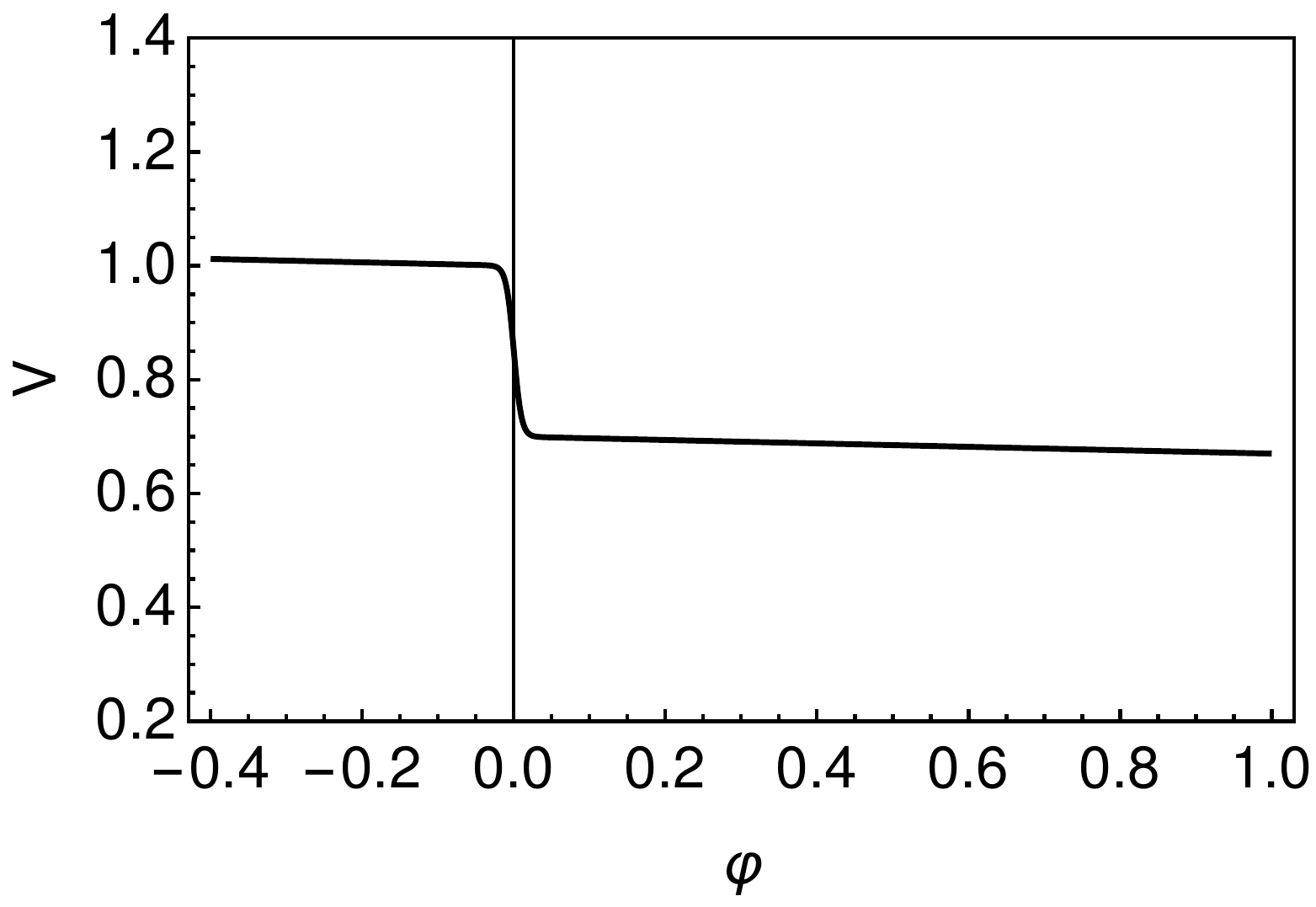} \includegraphics[width=0.24\textwidth]{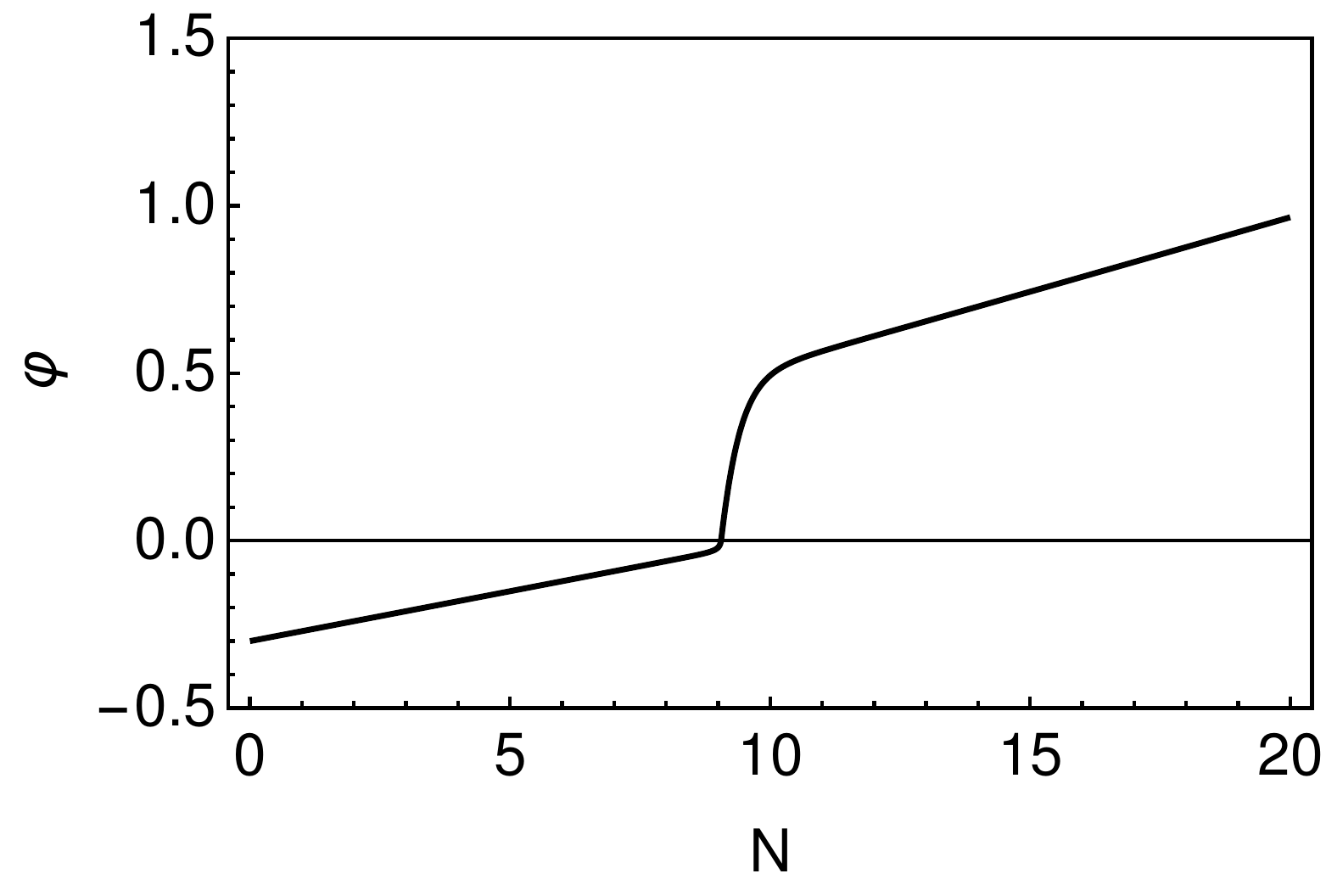} \includegraphics[width=0.24\textwidth]{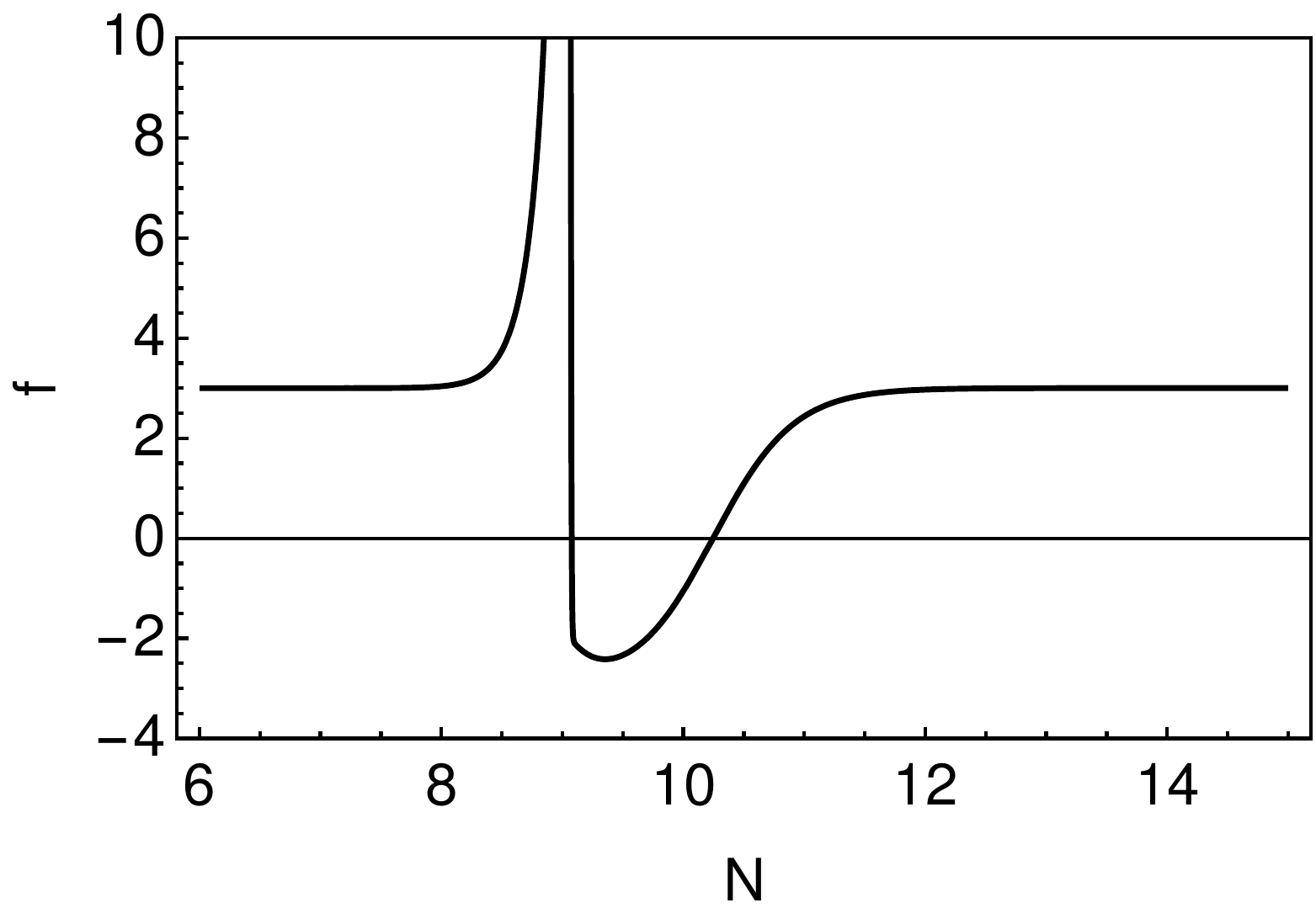} \includegraphics[width=0.25\textwidth,height= 28mm]{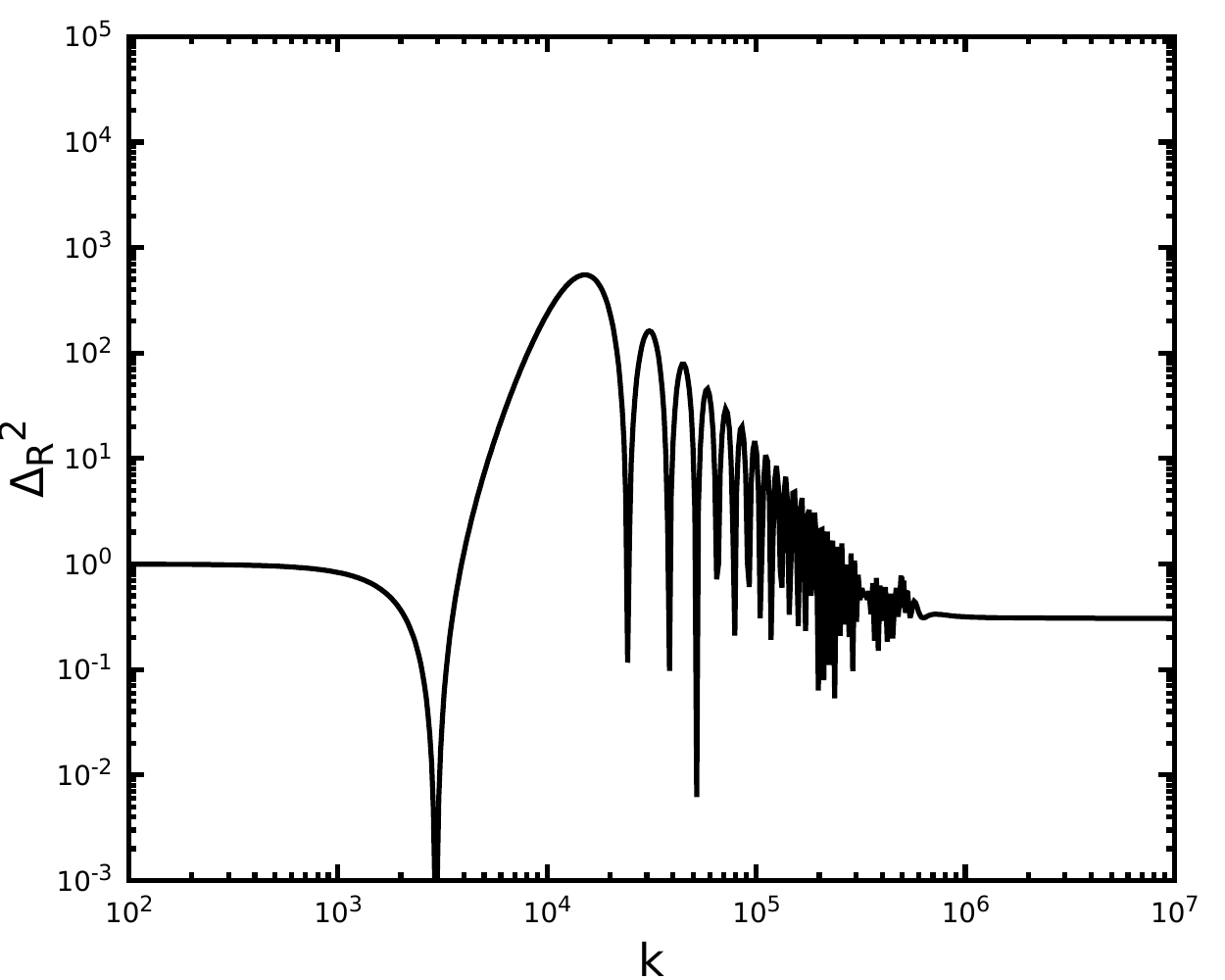}
\includegraphics[width=0.24\textwidth]{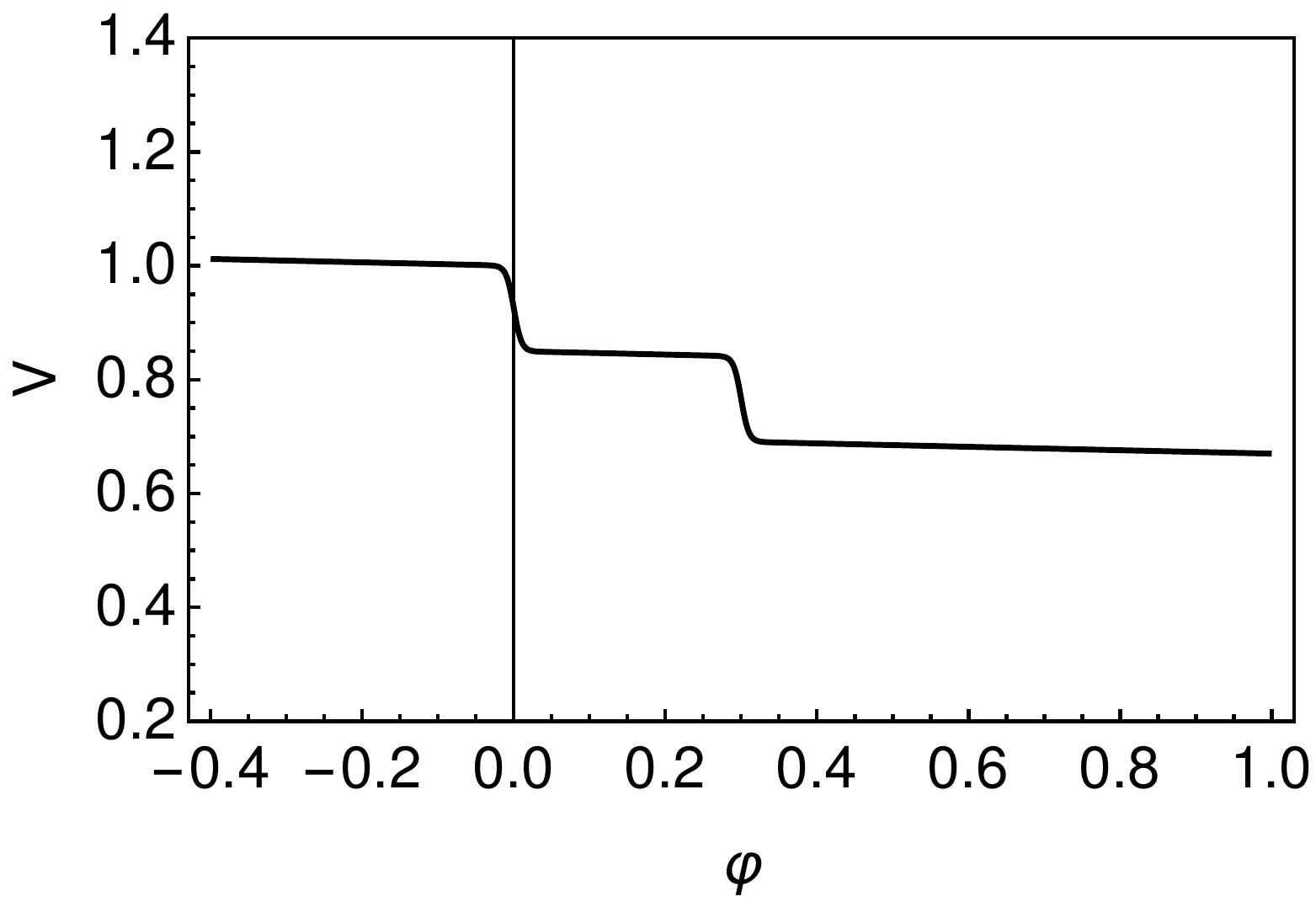} \includegraphics[width=0.24\textwidth]{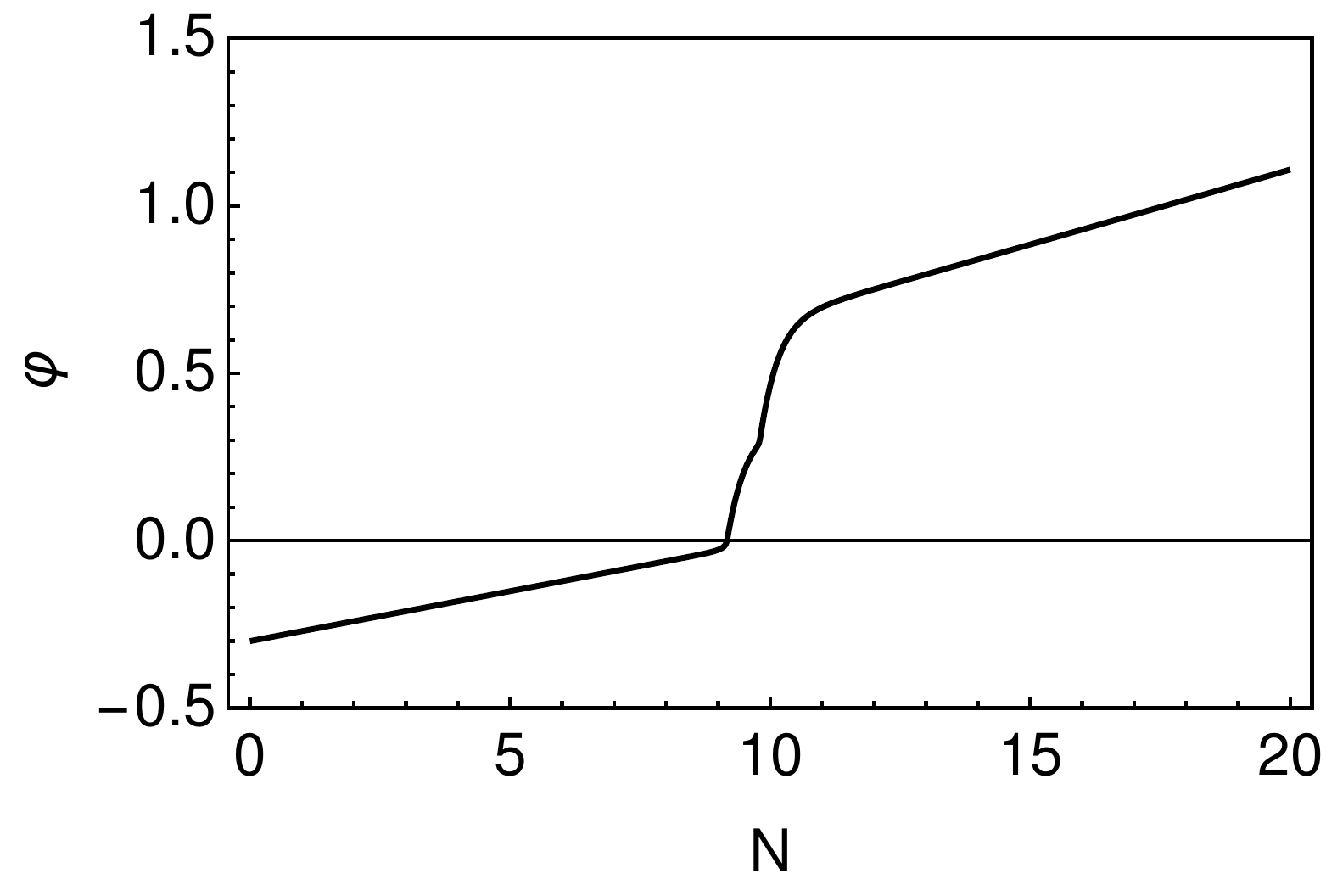}
\includegraphics[width=0.24\textwidth]{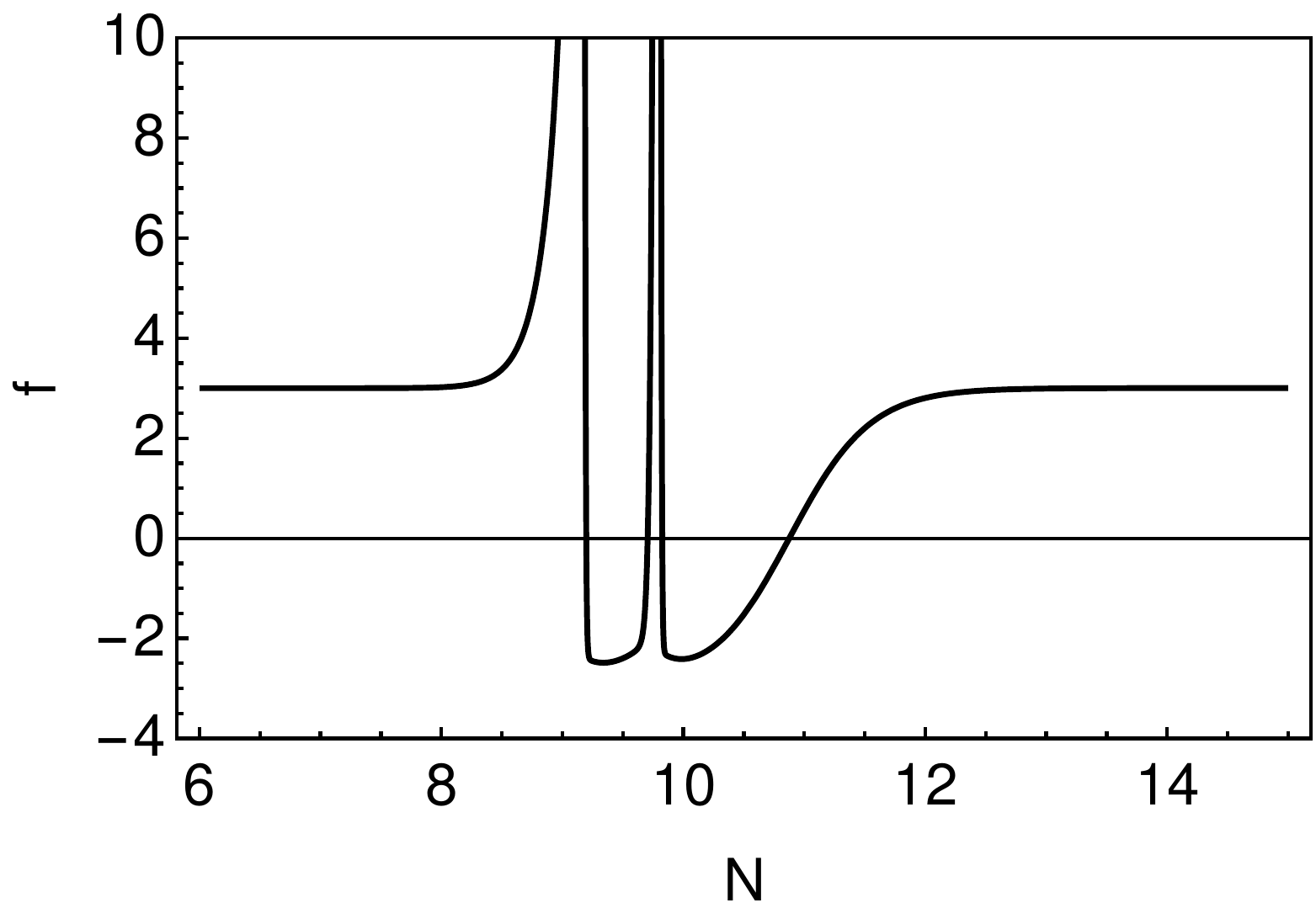} \includegraphics[width=0.25\textwidth,height= 28mm]{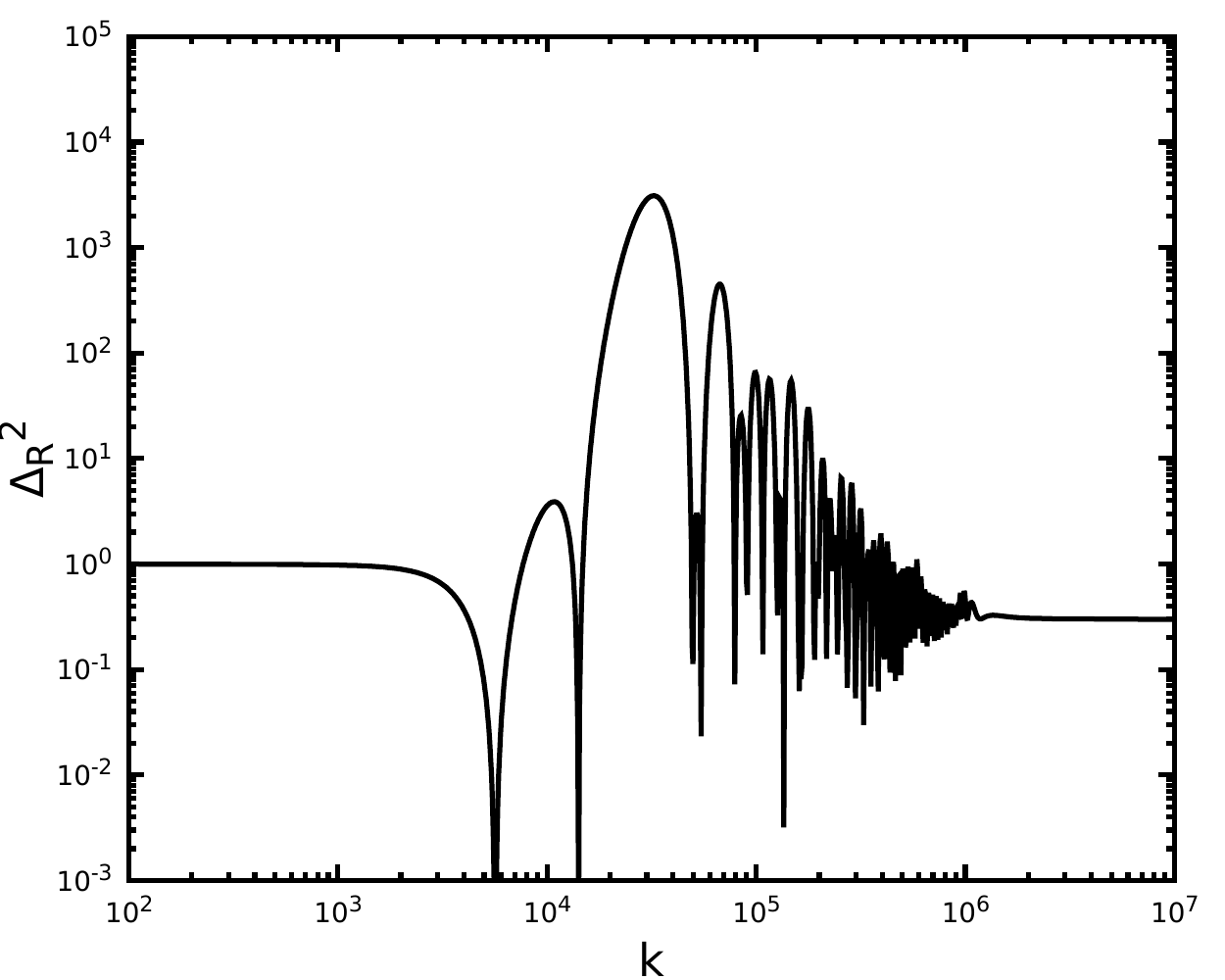}
\includegraphics[width=0.24\textwidth]{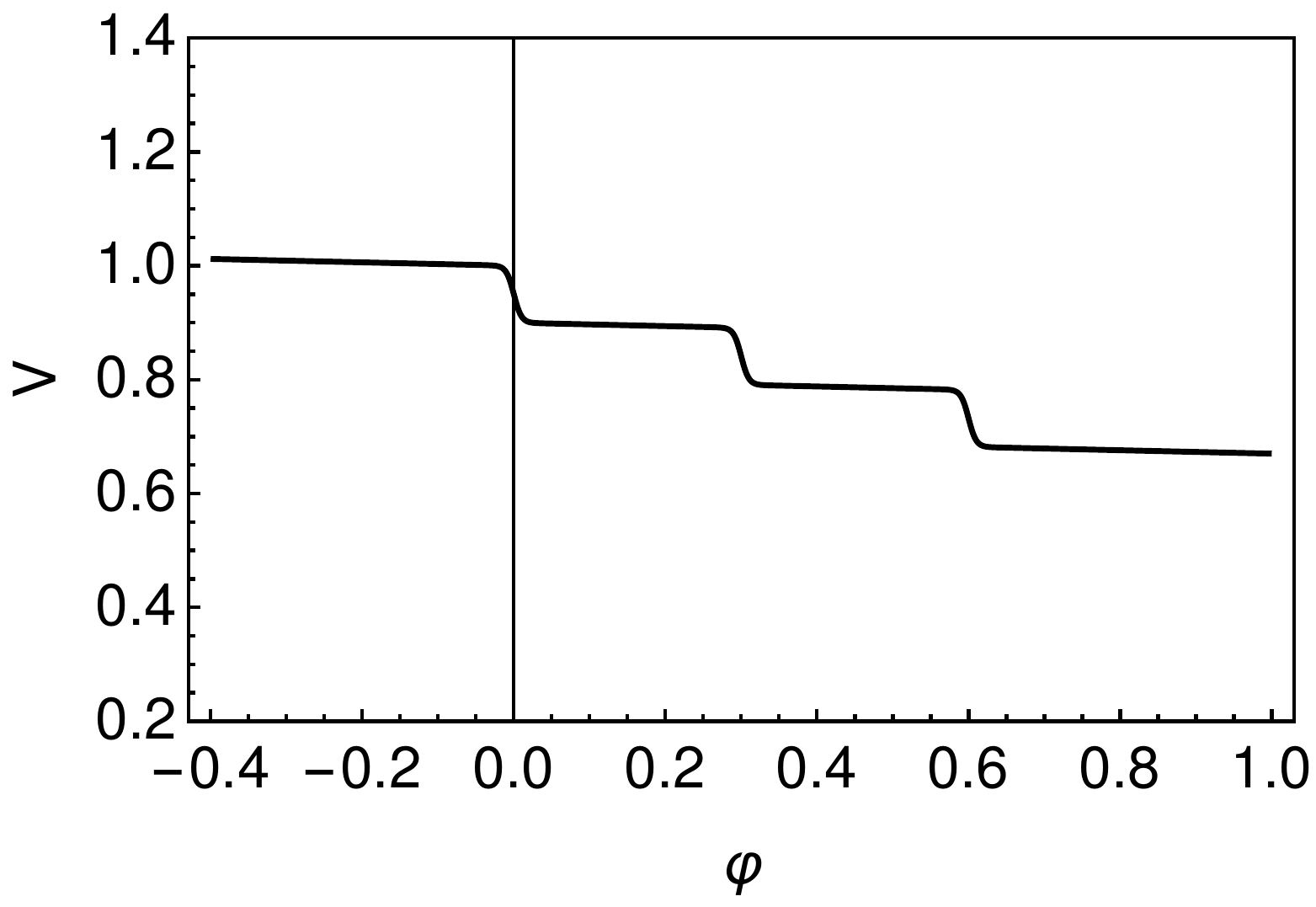} \includegraphics[width=0.24\textwidth]{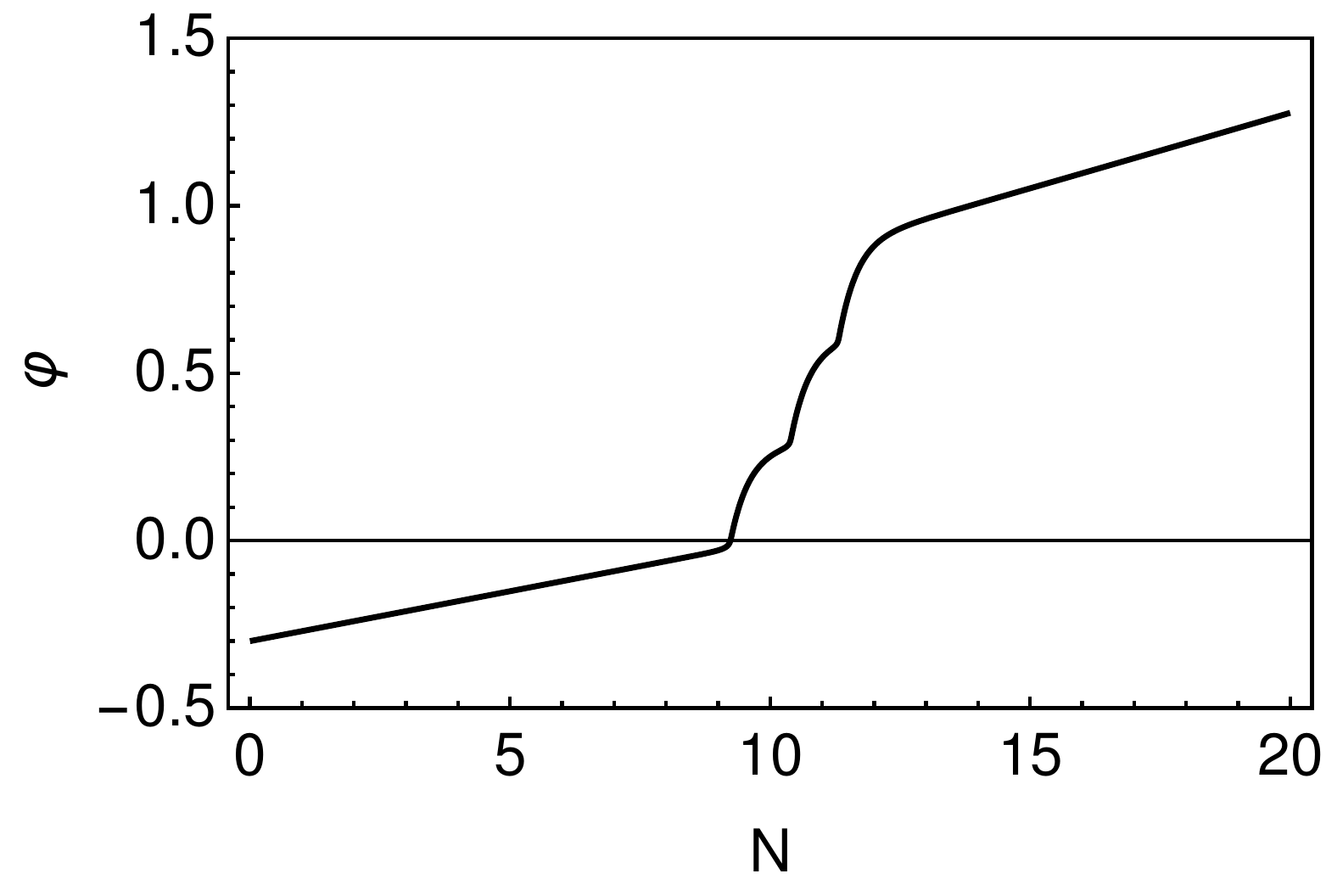}
\includegraphics[width=0.24\textwidth]{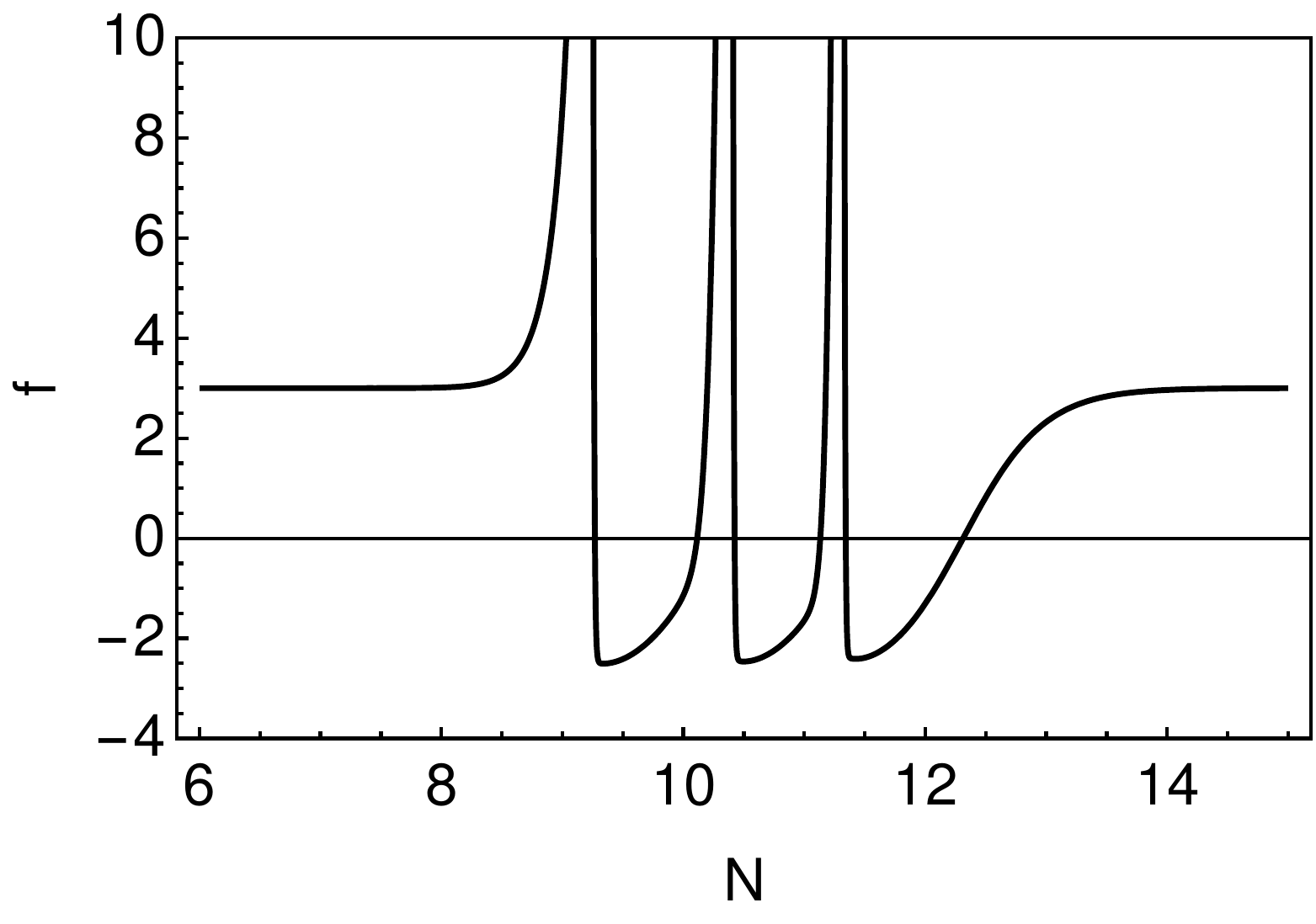} \includegraphics[width=0.25\textwidth,height= 28mm]{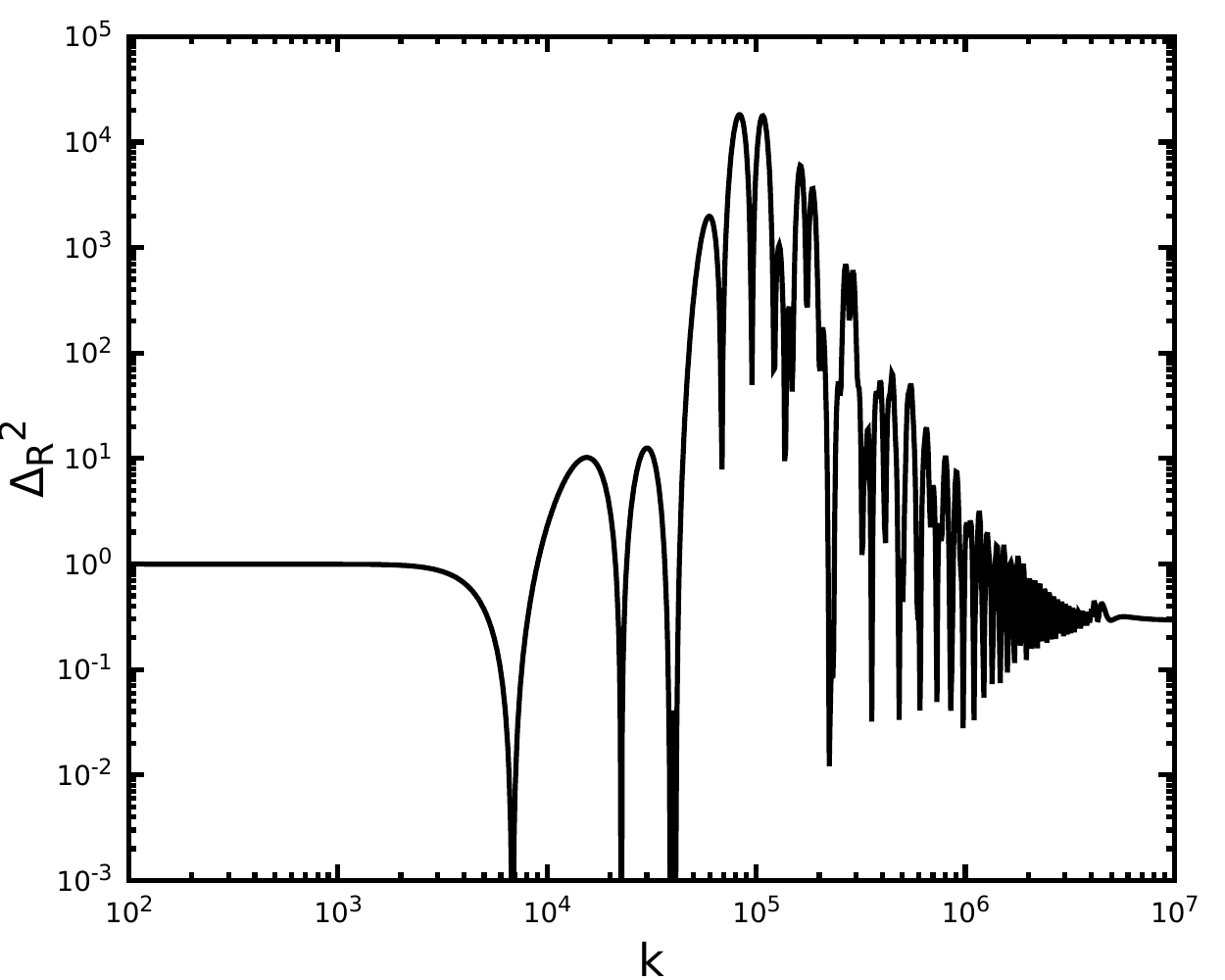}
\caption{
The inflaton potential $V(\varphi)$ of eq. (\ref{pote}), the evolution of the inflaton
$\varphi$, the function
$f(N)$ defined in eq. (\ref{AN}), and the power spectrum of curvature perturbations
with wavenumber $k$, for various choices of the parameters of the potential: 
First row: $A_1=-0.3$, $c_i=100$, $B=-0.03$.~
Second row: $A_1=-0.15$, $A_2=-0.15$, $\varphi_2=0.3$, $c_i=100$, $B=-0.03$.~
Third row:  $A_1=-0.1$, $A_2=-0.1$, $A_3=-0.1$, $\varphi_2=0.3$, $\varphi_3=0.6$, $c_i=100$, $B=-0.03$.
The scales of $k$ and $V$ are arbitrary.
}
\label{figtwo}
\end{figure}

A drawback of the potential (\ref{pote}) is that it is not possible to make
a connection with the range of the
spectrum that is relevant for the cosmic-microwave-background (CMB). 
The slope $B$ of the potential required for agreement with the 
measured spectral index is too steep for obtaining a large number of efoldings. 
As a result, contact with the observations is not possible
and we treat the pivot scale $k_*$, the amplitude $A_s$ and the 
spectral index $n_s$, introduced in the previous section, 
as free parameters. In particular, we assume that 
$k_*$ is located deep in the nonlinear part of the spectrum and the 
spectral index is sufficiently close to 1 for a large number of 
efoldings to be produced. We present our results for the spectrum in
units of $A_s$, which is equivalent to setting $A_s=1$. It is also obvious
from eq. (\ref{eomH}) that the absolute scale $V_0$ of the potential does not
play any role for our considerations. In practice, we set $V_0=1$ for
the numerical analysis. Finally, the inflaton field and the constants
$c_i$, $B$ can be given in units
of $\mpl$, which is equivalent to setting $\mpl=1$ in eq. (\ref{eomH}).

Before computing the spectrum, it is instructive to understand which
type of background evolution leads to its enhancement. 
The perusal of eq. (\ref{RcN}) leads to the conclusion that the 
sign of the function $f(N)$ defined in eq. (\ref{AN}) is crucial.
For $f(N)>0$ the second term of eq. (\ref{RcN}) acts as a friction 
term, suppressing the growth of the curvature perturbation. The opposite happens
for $f(N)<0$. It is known that the presence of an inflection point in
the potential enhances the spectrum. For this reason, we
examine first the form of $f(N)$ for such a case. Then we analyse the
conditions under which a similar enhancement of the spectrum can 
occur for a potential with a step-like structure.
It must be emphasized that the two cases are distinct. The rolling
of the inflaton through an inflection point does not
stop inflation, even though the standard slow-roll conditions
are not satisfied because of the large value of $\eta_H$. On the
other hand, the transition through a sharp drop in the potential
leads to a fast increase of the time-derivative of the inflaton, and in many cases to
a brief interruption of inflation. This is apparent from the effect of a
large value of $\exx_H$ on the effective equation-of-state
parameter $w=2\exx_H/3-1$.

\begin{figure}[t!]
\centering
\includegraphics[width=0.24\textwidth]{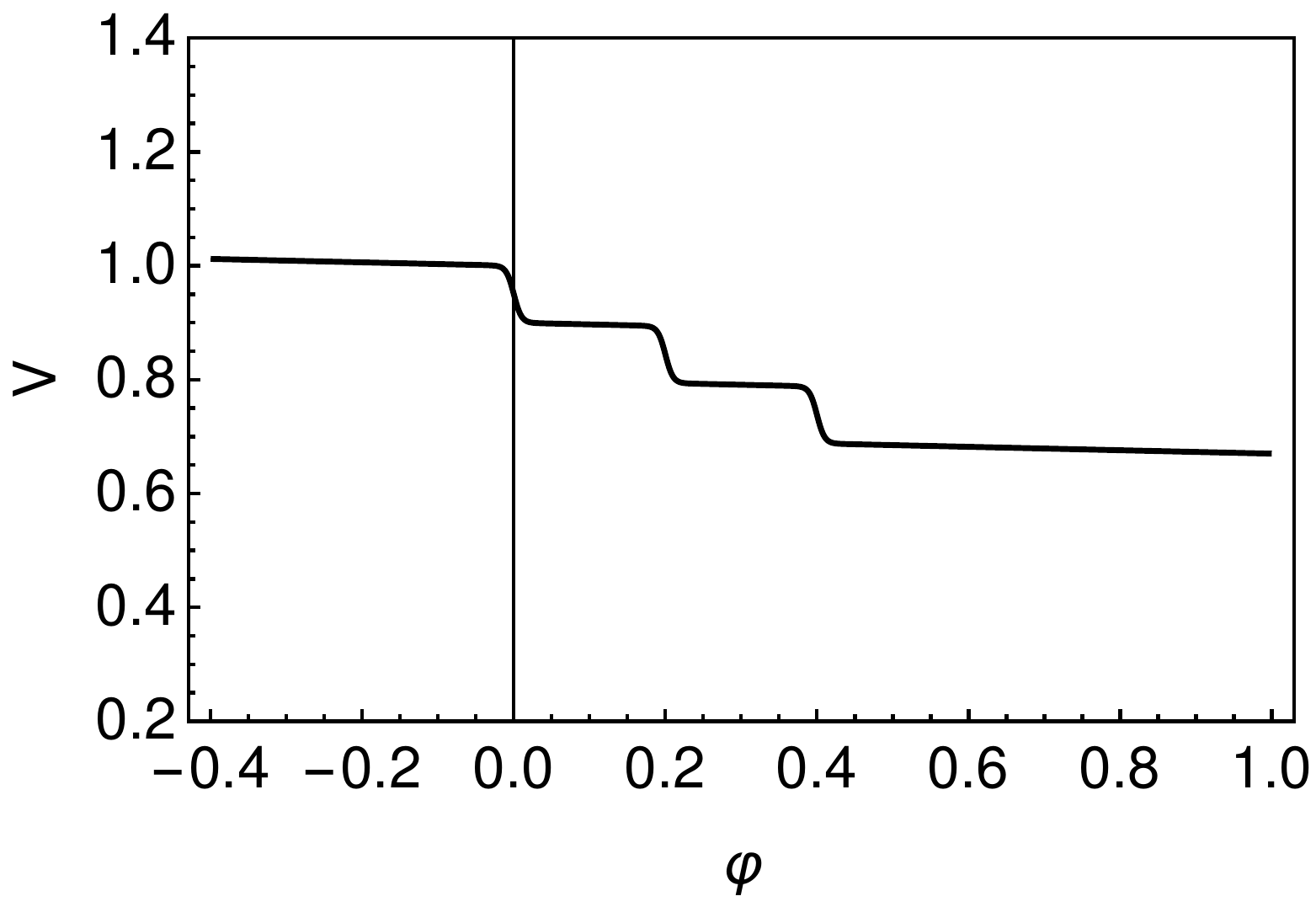} \includegraphics[width=0.24\textwidth]{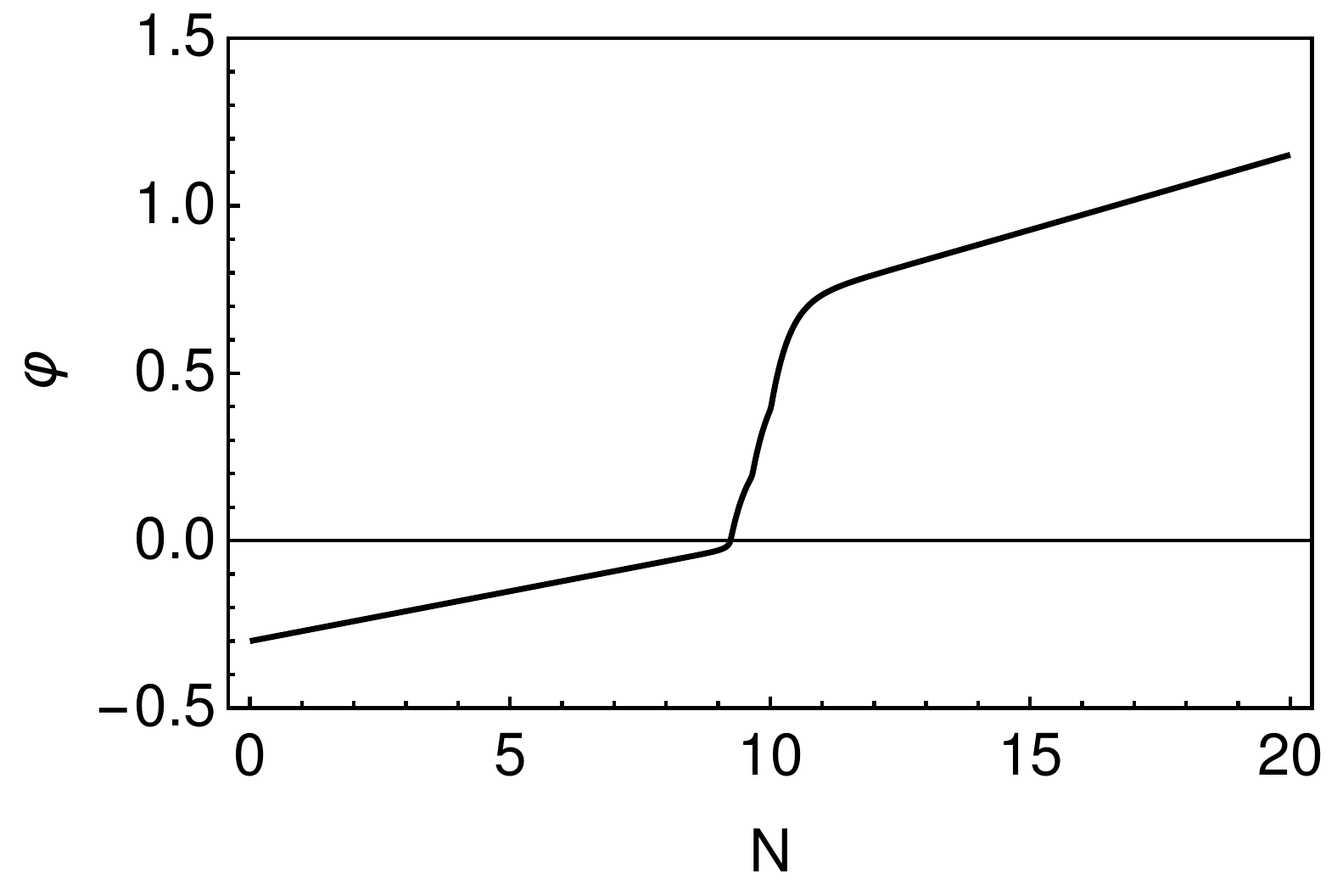} \includegraphics[width=0.24\textwidth]{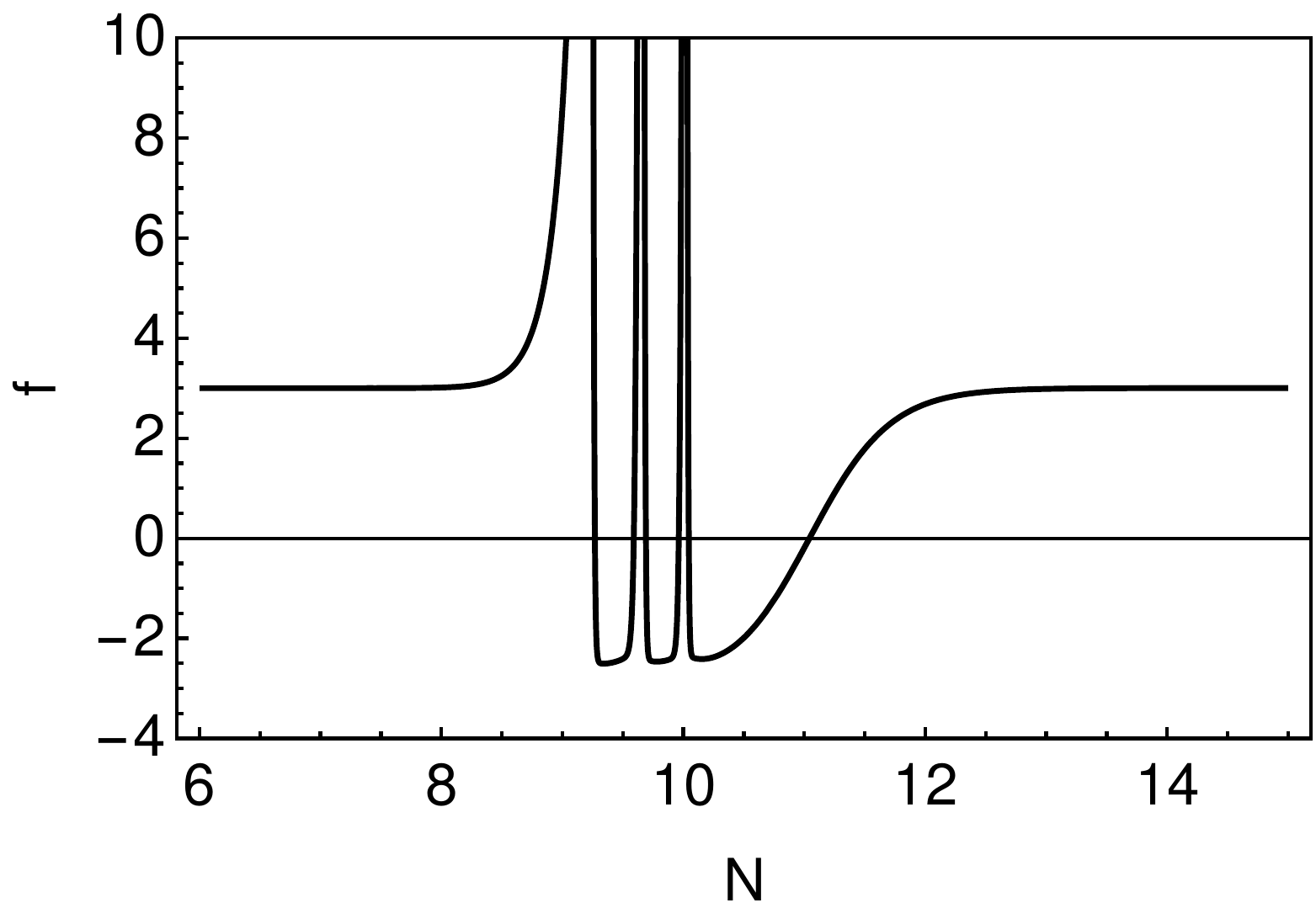} \includegraphics[width=0.25\textwidth,height= 28mm]{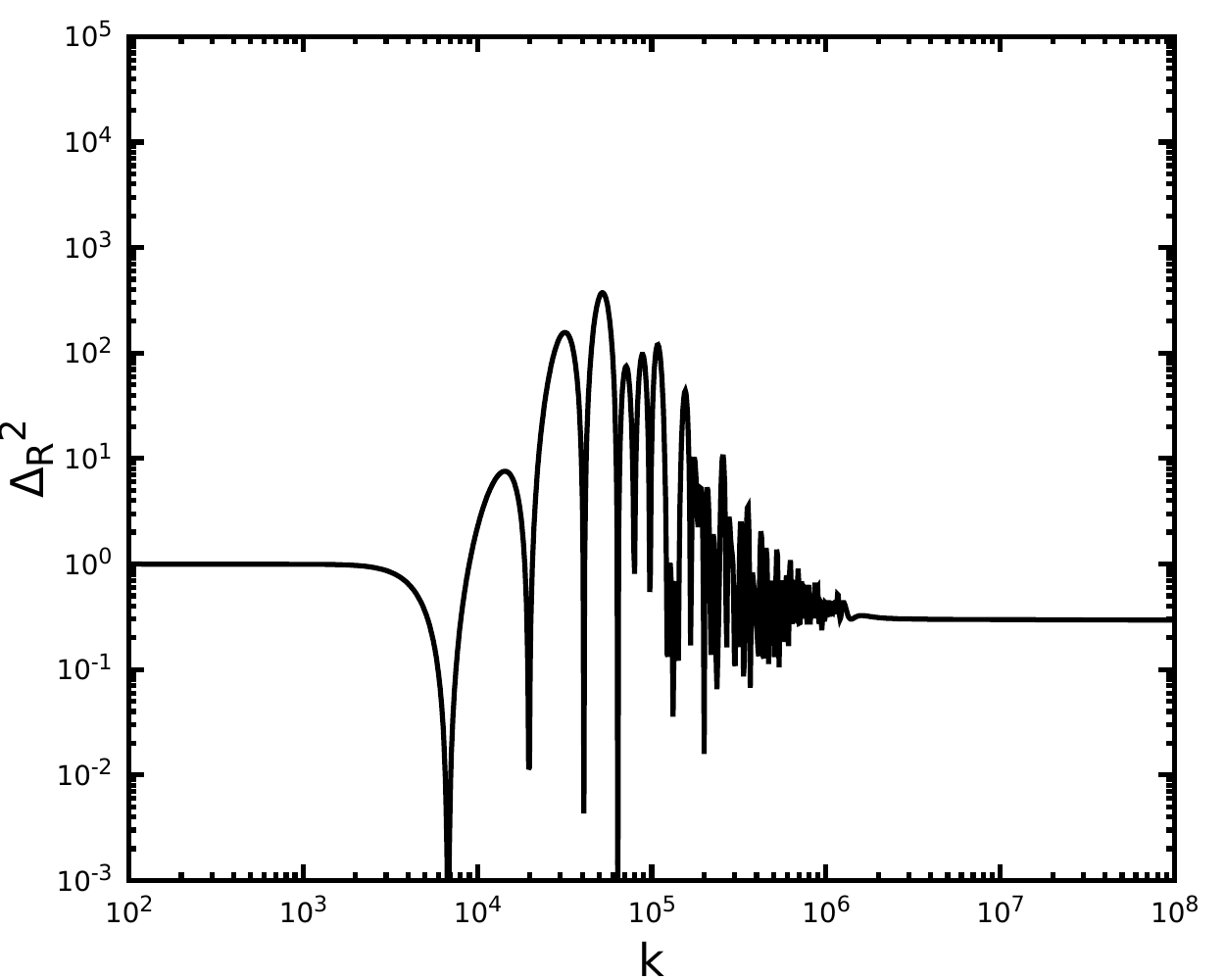}
\includegraphics[width=0.24\textwidth]{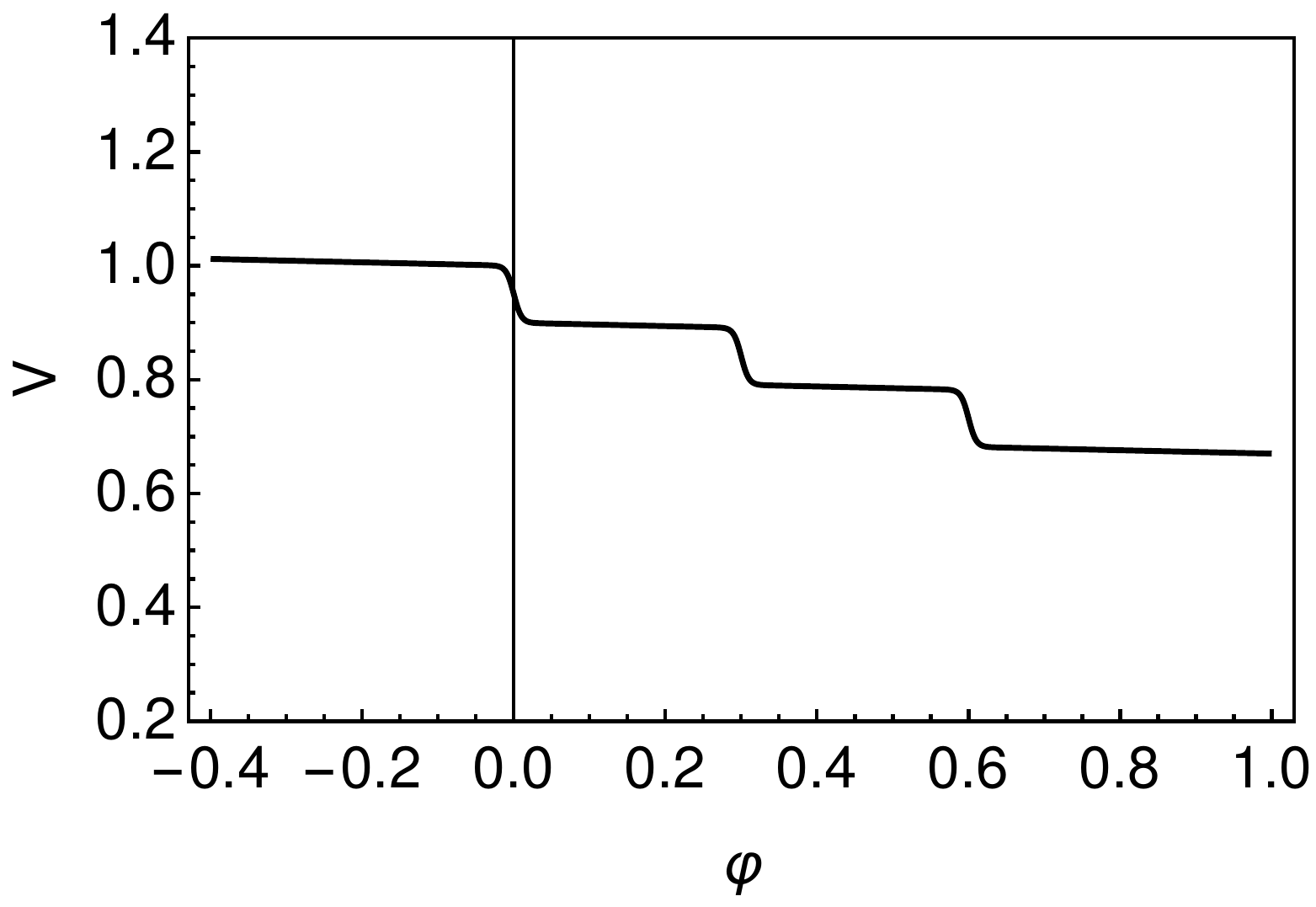} \includegraphics[width=0.24\textwidth]{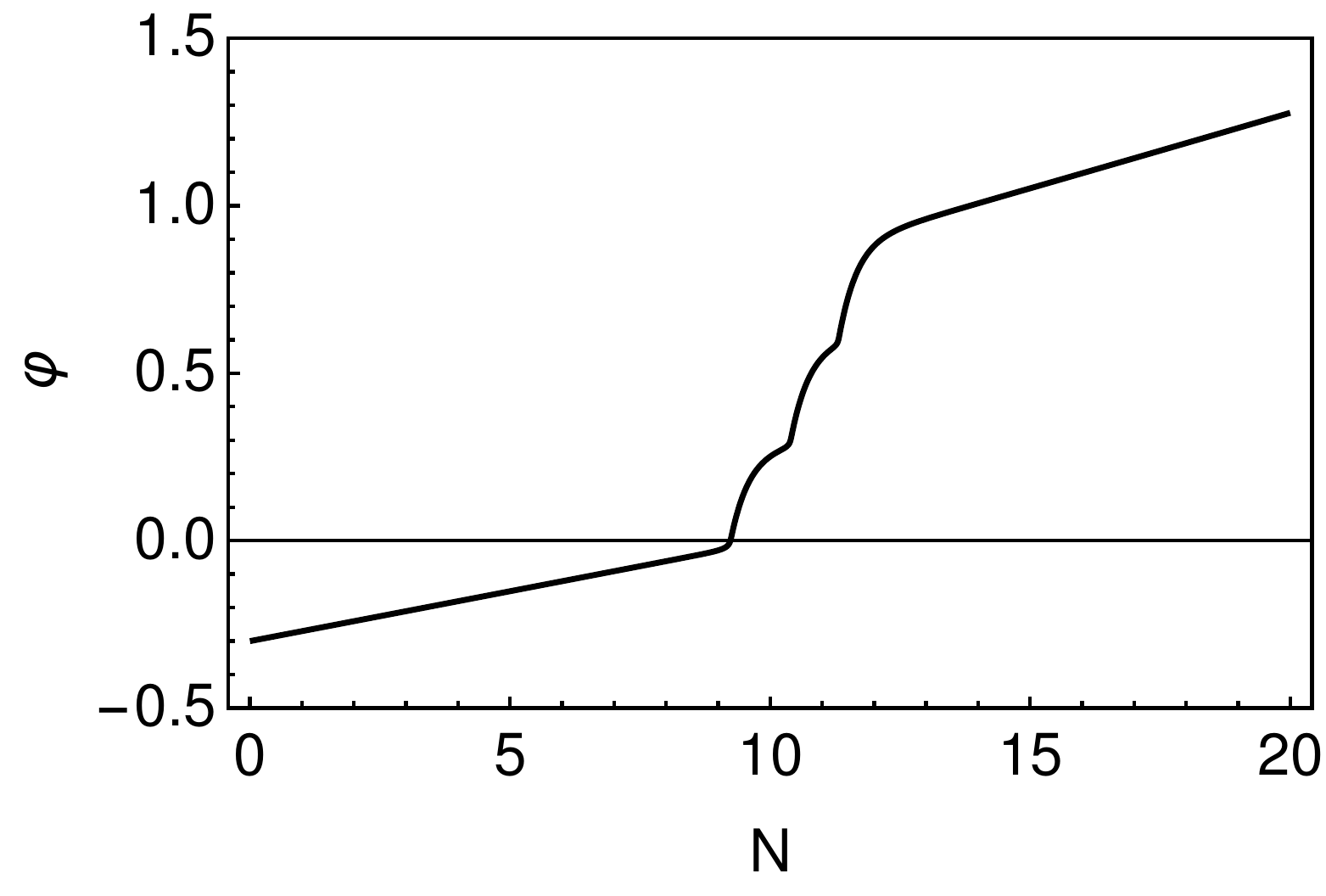}
\includegraphics[width=0.24\textwidth]{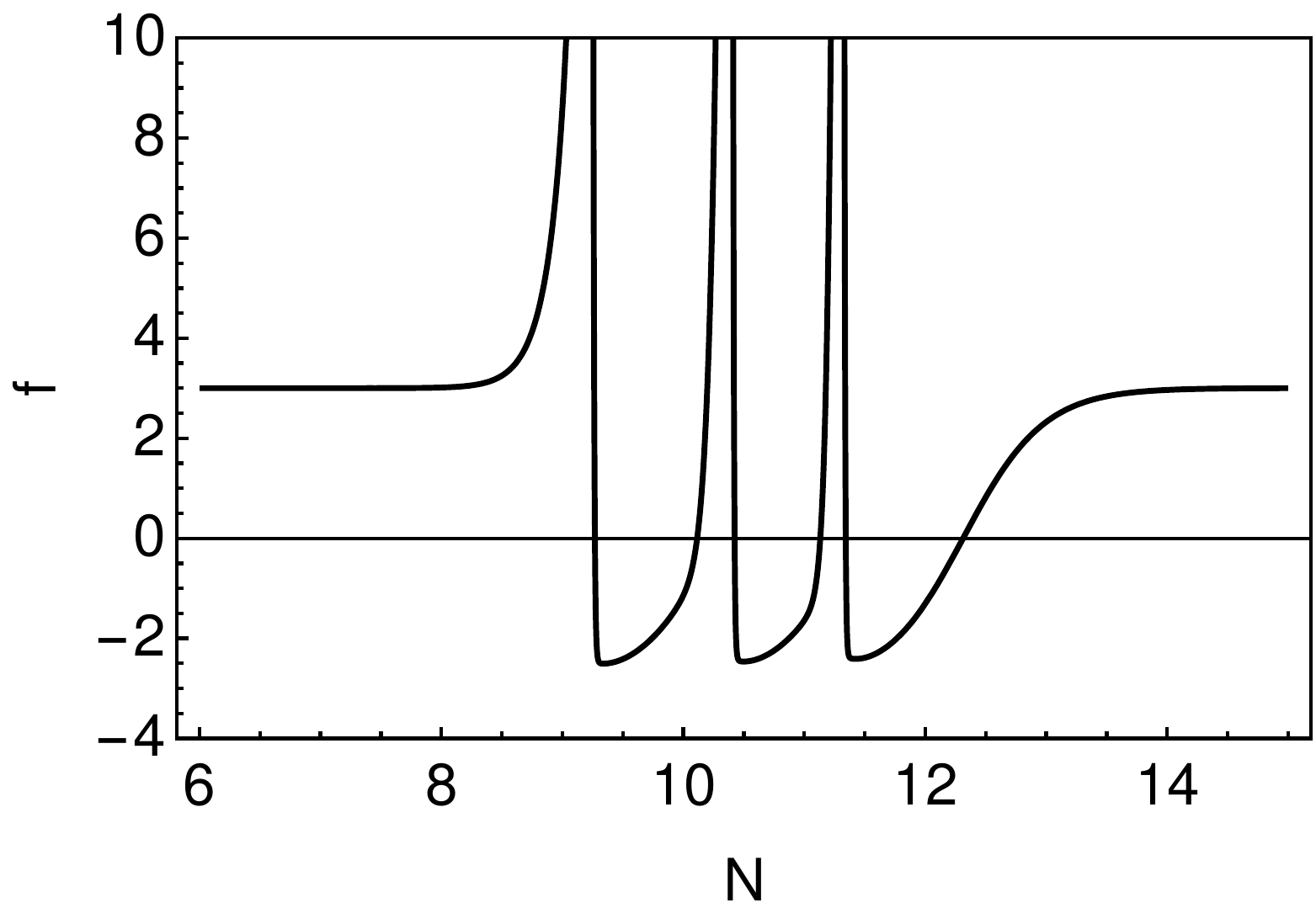} \includegraphics[width=0.25\textwidth,height= 28mm]{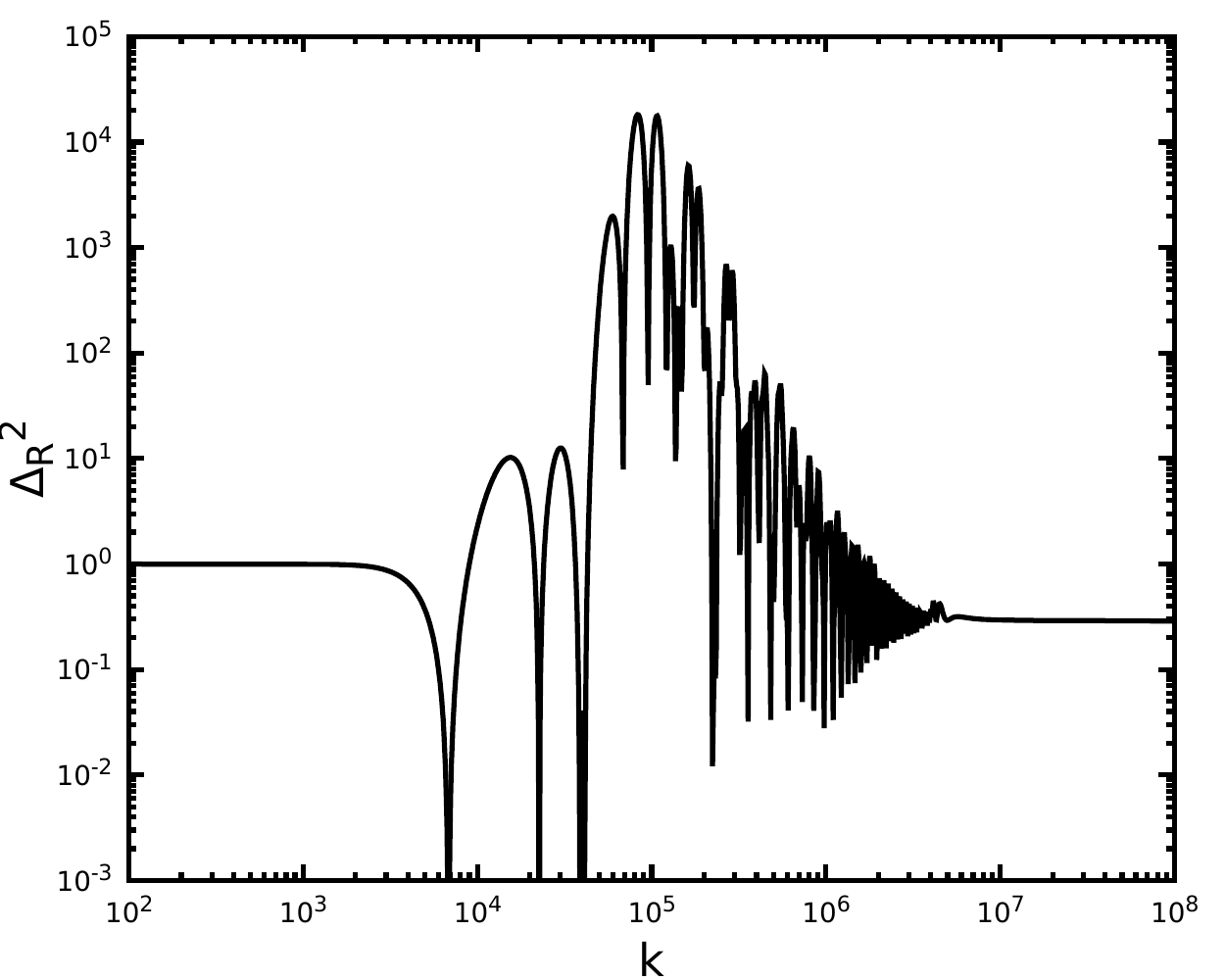}
\includegraphics[width=0.24\textwidth]{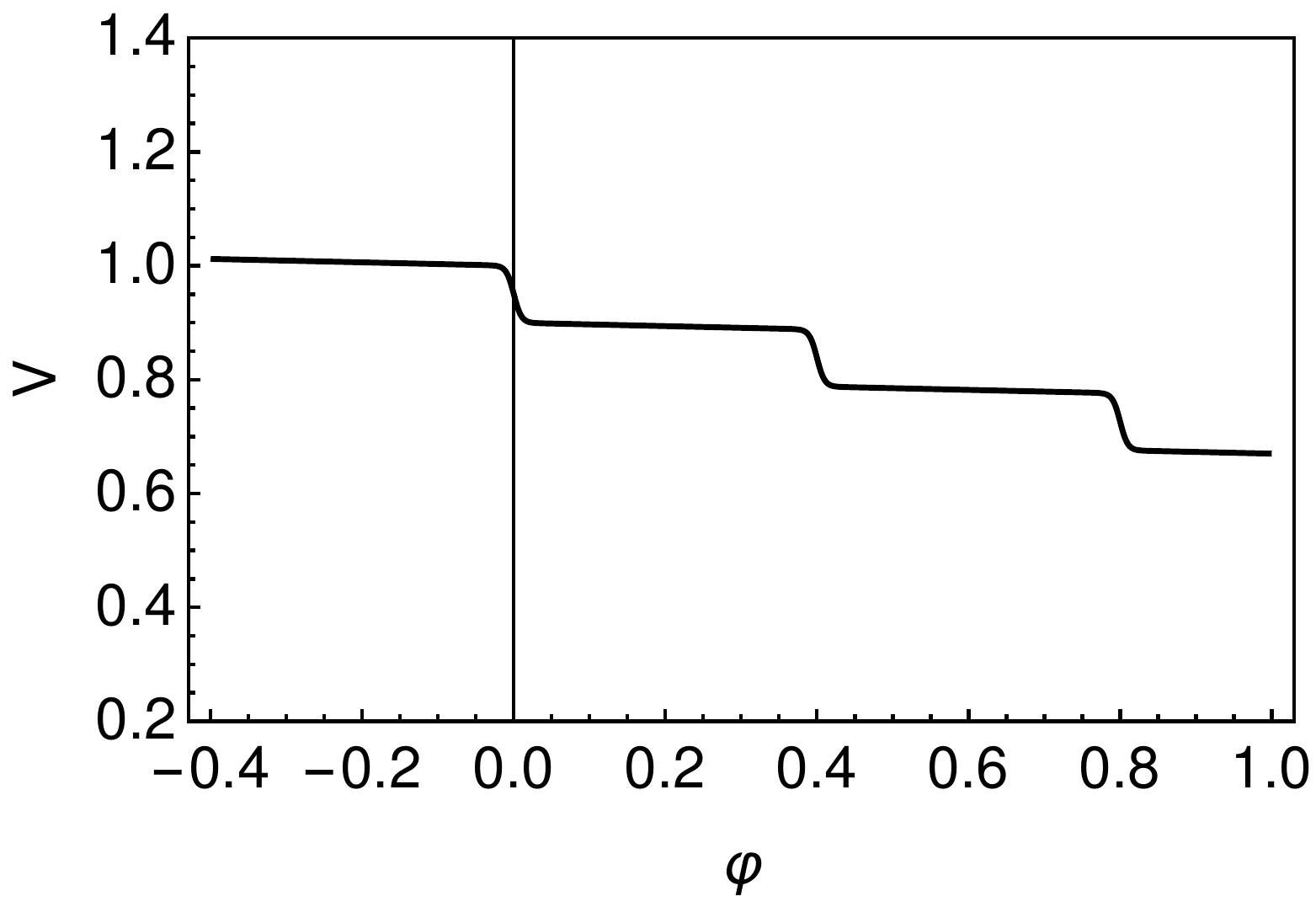} \includegraphics[width=0.24\textwidth]{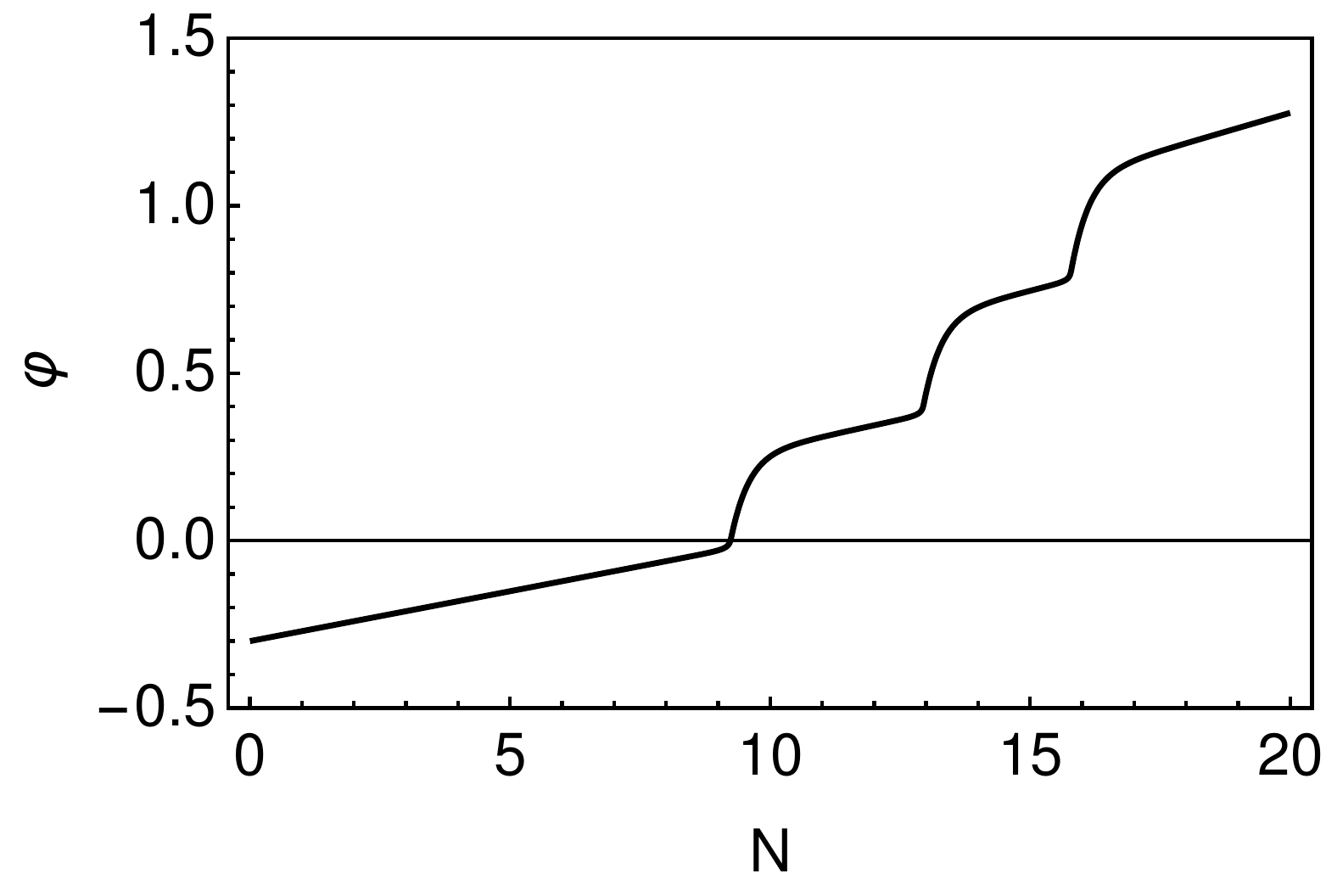}
\includegraphics[width=0.24\textwidth]{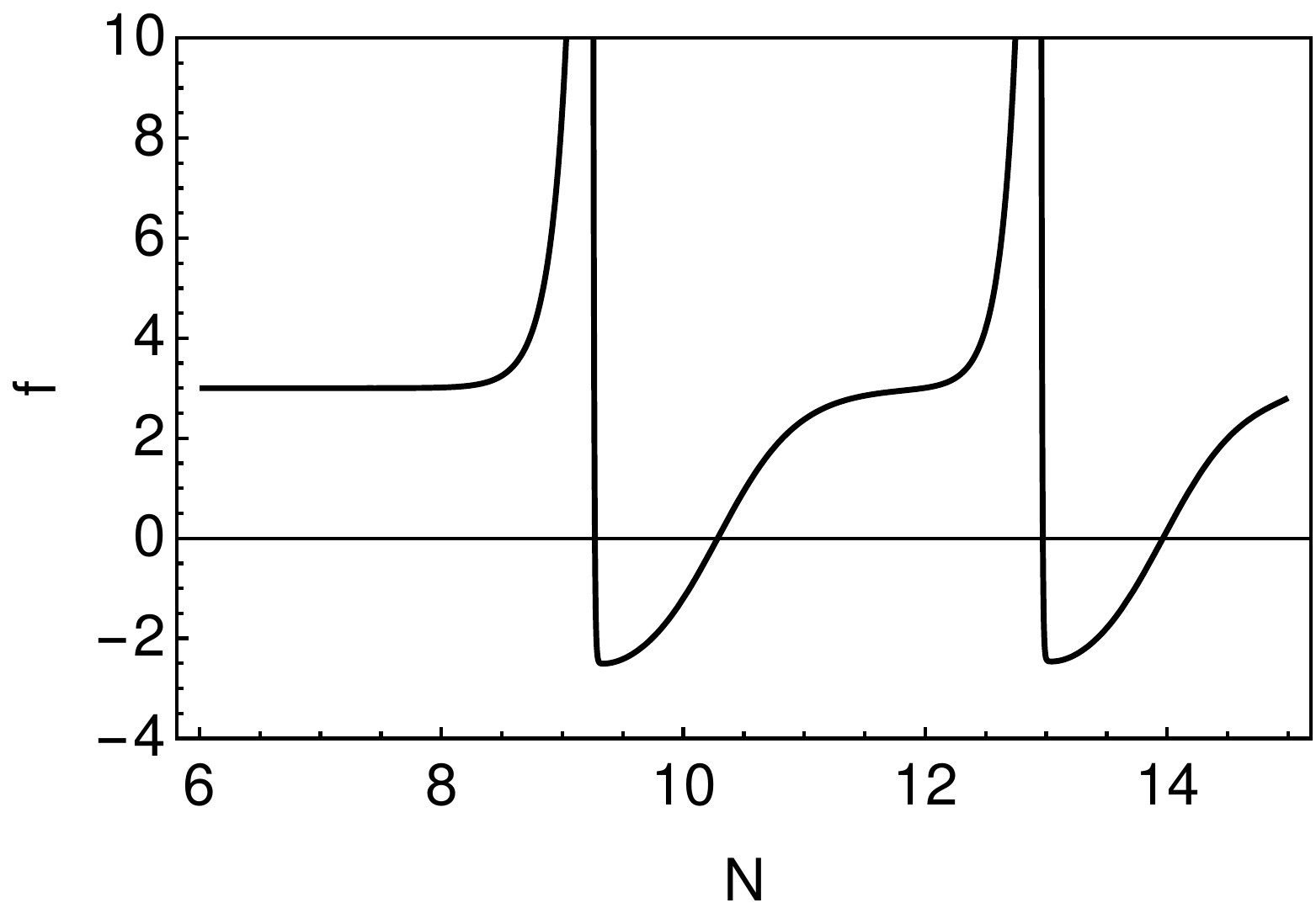} \includegraphics[width=0.25\textwidth,height= 28mm]{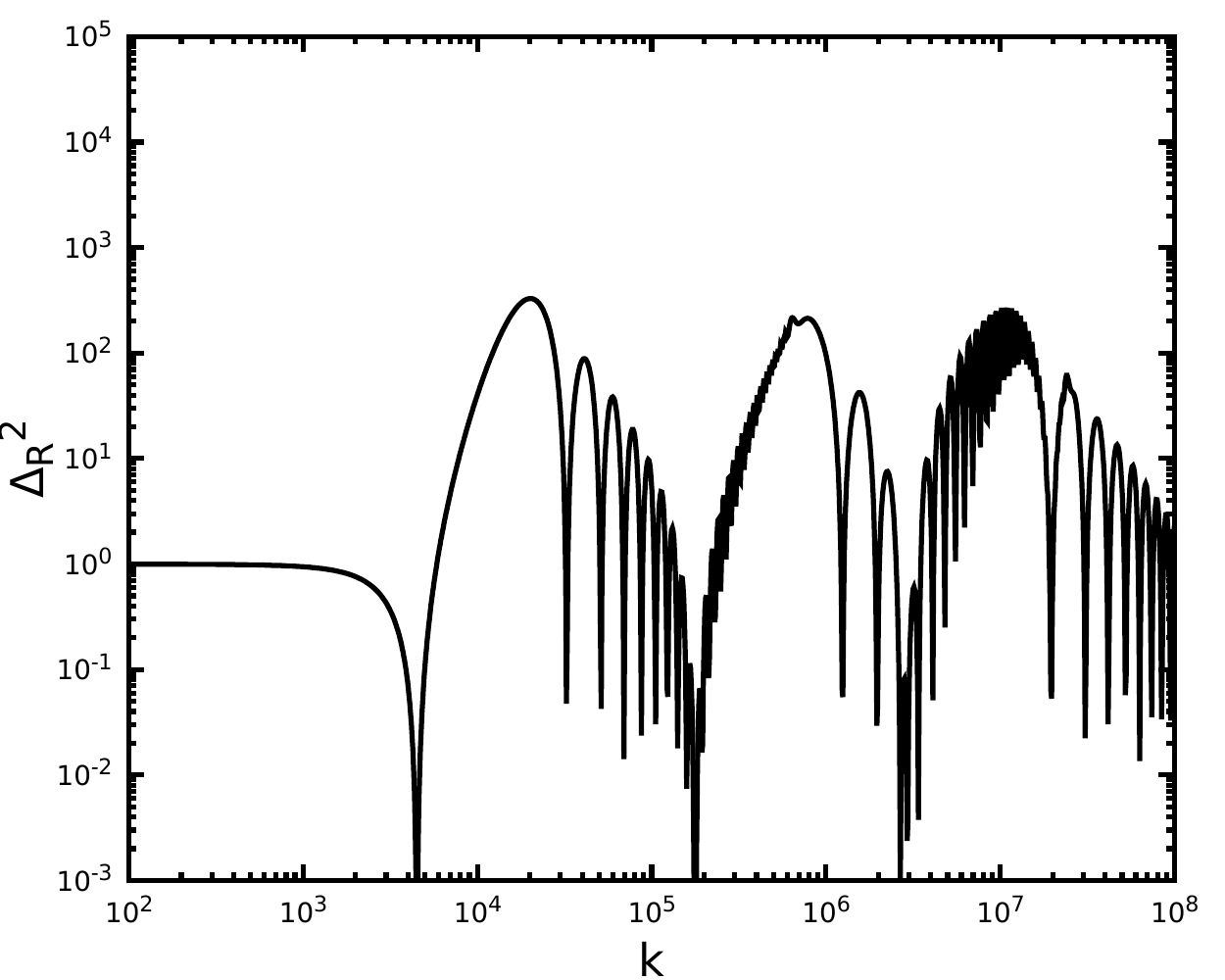}
\caption{
The inflaton potential $V(\varphi)$ of eq. (\ref{pote}), the evolution of the inflaton
$\varphi$, the function
$f(N)$ defined in eq. (\ref{AN}), and the power spectrum of curvature perturbations
with wavenumber $k$, for various choices of the parameters of the potential: 
First row: $A_1=-0.1$, $A_2=-0.1$, $A_3=-0.1$, $\varphi_2=0.2$, $\varphi_3=0.4$, $c_i=100$, $B=-0.03$.~
Second row: $A_1=-0.1$, $A_2=-0.1$, $A_3=-0.1$, $\varphi_2=0.3$, $\varphi_3=0.6$, $c_i=100$, $B=-0.03$.~
Third row:  $A_1=-0.1$, $A_2=-0.1$, $A_3=-0.1$, $\varphi_2=0.4$, $\varphi_3=0.8$, $c_i=100$, $B=-0.03$.
The scales of $k$ and $V$ are arbitrary.}
\label{figthree}
\end{figure}

In fig. \ref{figone} we present various elements of the 
calculation of the power spectrum for different potentials.
We have used the same scale for all related plots in order 
to make the comparison easy. 
The first plot in each row depicts the inflaton potential. 
The potential at the top has an inflection point at $\varphi=0$, even
though this is not clearly visible. The potentials in the next three 
rows display a step at $\varphi=0$, whose steepness is increased from 
top to bottom by choosing larger values of the parameter $c_1$. 
The form of the potential is reflected in the field evolution.
In the second plot of the first row, 
the field stays almost constant near zero for several
efoldings. In the following rows it evolves very quickly, within 2-3 
efoldings, from one plateau of the potential to the next. 
The third plot in each row depicts the ``effective friction"
$f(N)$. In all cases this function becomes negative during part of the
evolution, thus leading to a strong enhancement of the fluctuations.
For an inflection point it starts with the standard value 3, then becomes
negative, returns to positive values larger than 3, and eventually 
becomes equal to 3 again. For a step in the potential, there is a strong
increase to very large positive values before the function becomes
negative. This increase is confined within a period of efoldings that 
shrinks with increasing steepness (and $c_1$). 
On the other hand, the form of 
$f(N)$ in the interval where it is negative is largely independent 
of $c_1$, because it is determined by the approach of the field to
slow roll on the second plateau. It seems reasonable to expect 
that, for steeper steps, the suppression of
the perturbation during the strong increase of $f(N)$ is 
a subleading effect relative to the subsequent enhancement. 
This expectation is confirmed by the spectrum depicted in the last
plot of each row. In the first row we observe the strong and broad 
enhancement of the spectrum associated with an inflection point. 
After an initial dip, the spectrum grows rather steeply towards 
a maximum, beyond which it decays smoothly towards its almost scale-invariant form.
This behavior is consistent with the general analysis of ref. \cite{ozsoy}. 
The spectra of the next three rows display a strong oscillatory behaviour,
which will be discussed in the following. The largest enhancement is 
achieved for a band of wavenumbers during the first oscillation. 
It is clear that the magnitude of this enhancement increases with $c_1$.

The maxima of the spectra in fig. \ref{figone} are larger by up to
three orders of magnitude relative to the standard value for the
scale-invariant case. The enhancement is restricted by the fact that
the maximal ``velocity" achieved by the rolling field 
is limited by the size of the step.
It is possible, however, that the potential includes several
step-like features. We examine their effect in fig. \ref{figtwo}, 
where we compare potentials with one, two or three steps. The
total drop in the potential is the same in all three cases. 
It is apparent from the last column that the presence of
several features in the potential can lead to the increase of the
spectrum by several orders of magnitude. The reason can be 
traced to the ``effective friction" $f(N)$, displayed in the third column.
The presence of several steps increases the total number of
efoldings over which this function takes negative values. This is
reflected in the larger enhancement of the perturbations.

The field values at which the features of the potential appear 
play a crucial role for the form of the resulting spectrum. 
This feature is demonstrated in fig. \ref{figthree} in which
we consider potentials with three steps, at field values
with increasing distance from each other. It is apparent from the 
first row that when the steps are very close to each other the function
$f(N)$ stays negative for a small number of efoldings and the 
enhancement is comparable to the one-step case. Increasing the
distance leads to spectrum enhancement, as $f(N)$ stays negative 
longer. However, the enhancement persists up to a certain distance
between the features of the potential, beyond
which each step acts independently on the spectrum. This beaviour
is apparent in the second and third rows of fig. \ref{figthree}.

\begin{figure}[t!]
\centering
$$
\includegraphics[width=0.9\textwidth]{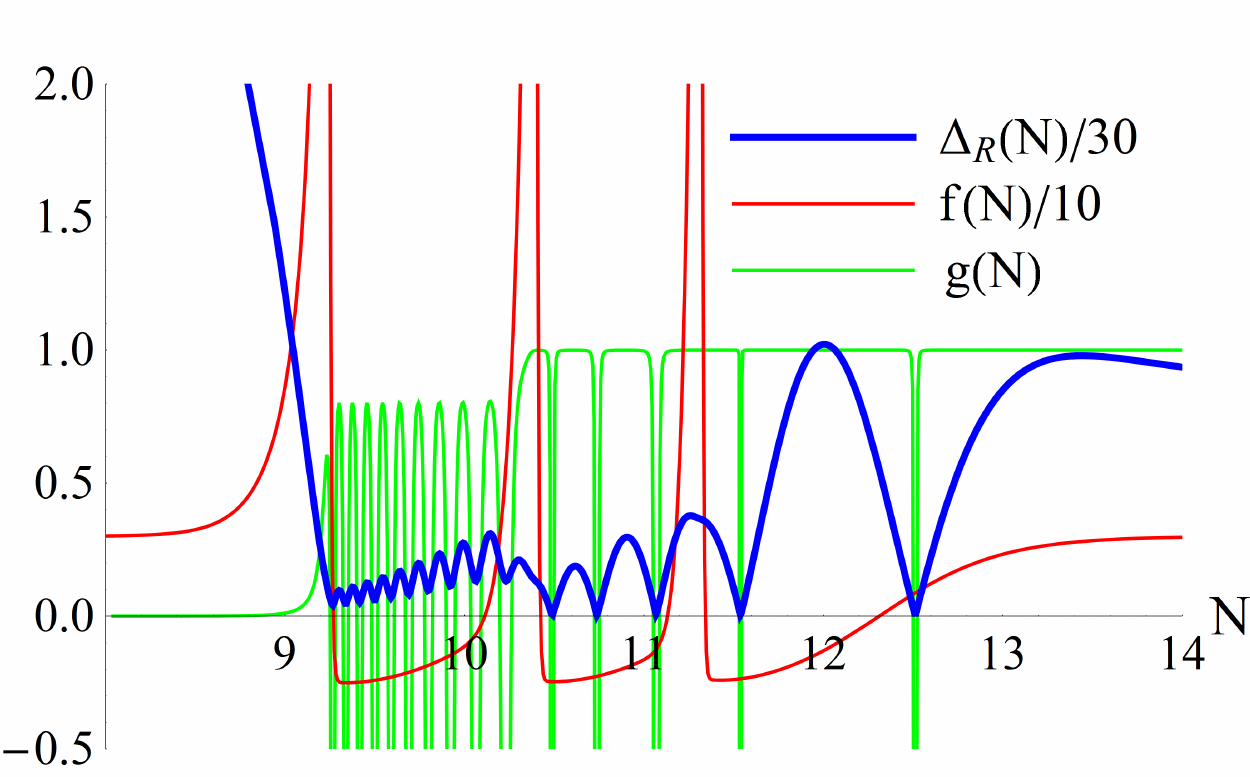}\hspace*{0.02\textwidth}
$$
\caption{The curvature perturbation as a function of the number of efoldings $N$, and
the functions $f(N)$, $g(N)$, 
for the potential (\ref{pote}) with $A_1=-0.1$, $A_2=-0.1$, $A_3=-0.1$, $\varphi_2=0.3$, $\varphi_3=0.6$, $c=100$, $B=-0.03$.~
Blue line: $\frac{1}{30} \sqrt{\Delta^2_R(k,N)}$ for $k=2.66 \times 10^5$.~
Red line: The function $\frac{1}{10} f(N)$ defined in (\ref{AN}).~
Green line: The function $g(N)$ defined in (\ref{BN}).~
}
\label{oscil}
\end{figure}

A prominent feature of the spectra resulting from sharp drops in
the inflaton potential is the appearance of strong oscillations, whose
origin we would like to understand. One can speculate that the oscillatory
pattern arises when modes within a wavenumber range exit the 
horizon, but then reenter during the period when inflation stops and the
comoving horizon grows. Upon reentry they start oscillating again, until
they exit for a second time during a subsequent period of inflation 
\cite{physrep,multcross}. However, the onset or freezing of the
oscillatory behaviour is not instantaneous, while the crossing of
the horizon is essentially a continuous process with a certain width.
An exact analytical treatment is difficult, 
and the evolution of each mode can be computed only numerically.
In fig. \ref{oscil} we present the evolution of the curvature perturbation
$\Rct_\kt(N)$ (blue line) for a given Fourier mode  $\kt=2.66 \times 10^5$ for an inflaton
background arising from a potential with three steps. 
The red and green lines depict the functions $f(N)$ and $g(N)$ 
defined by eqs. (\ref{AN}) and (\ref{BN}), respectively.
The enhancement of the curvature perturbation during the periods of
inflation with negative $f(N)$ is apparent. 
Similarly, the freezing of the perturbation during the periods with 
positive $f(N)$ is also
apparent, resulting in $\Rct_\kt(N)$ becoming asymptotically constant.

A striking feature is the series of oscillations for
the amplitude of perturbations, which approaches zero at several values
of $N$. At these points the function $g(N)$ becomes very negative, thus
preventing the amplitude from crossing zero. 
The origin of the oscillations
can be understood if one considers eq. (\ref{RN}) for constant
$f(N)=\kappa$. Its solution involves a linear combination of the 
Bessel functions 
$J_{\pm\kappa/2}$ and has
the form
\be
R_k(N)=A e^{-\frac{1}{2}\kappa\, N} \left(J_{\kappa/2}\left( e^{- N}\frac{k}{H} \right) + c\, J_{-\kappa/2}\left( e^{-N}\frac{k}{H} \right)  \right).
\label{solBessel} \ee
The initial subhorizon evolution of the perturbation during the 
slow-roll regime corresponds to the solution with $\kappa=3$
and $c=i$. This particular choice of $c$ eliminates the 
oscillatory behaviour in the amplitude of $R_k(N)$. However, the
nontrivial background evolution that we are considering 
corresponds to a varying $\kappa$, 
as well as a varying relative coefficient of the 
Bessel functions. As a result the zeros of the Bessel functions become
apparent in the amplitude $\Rc_k(N)$, which becomes very small 
for $e^{-N}k/H$ approaching one 
of these zeros. The asymptotic value of $\Rc_k$ for large $N$ depends
on the time of the transition of the background solution to 
positive values of $f(N)$. The freezing of $R_k$ can occur at any stage
of the oscillatory cycle, depending on the value of $k$. 
Eventually, this is reflected in the strong 
oscillatory behaviour of the spectrum as a function of $k$.

\begin{figure}[t!]
\centering
\includegraphics[width=0.46\textwidth]{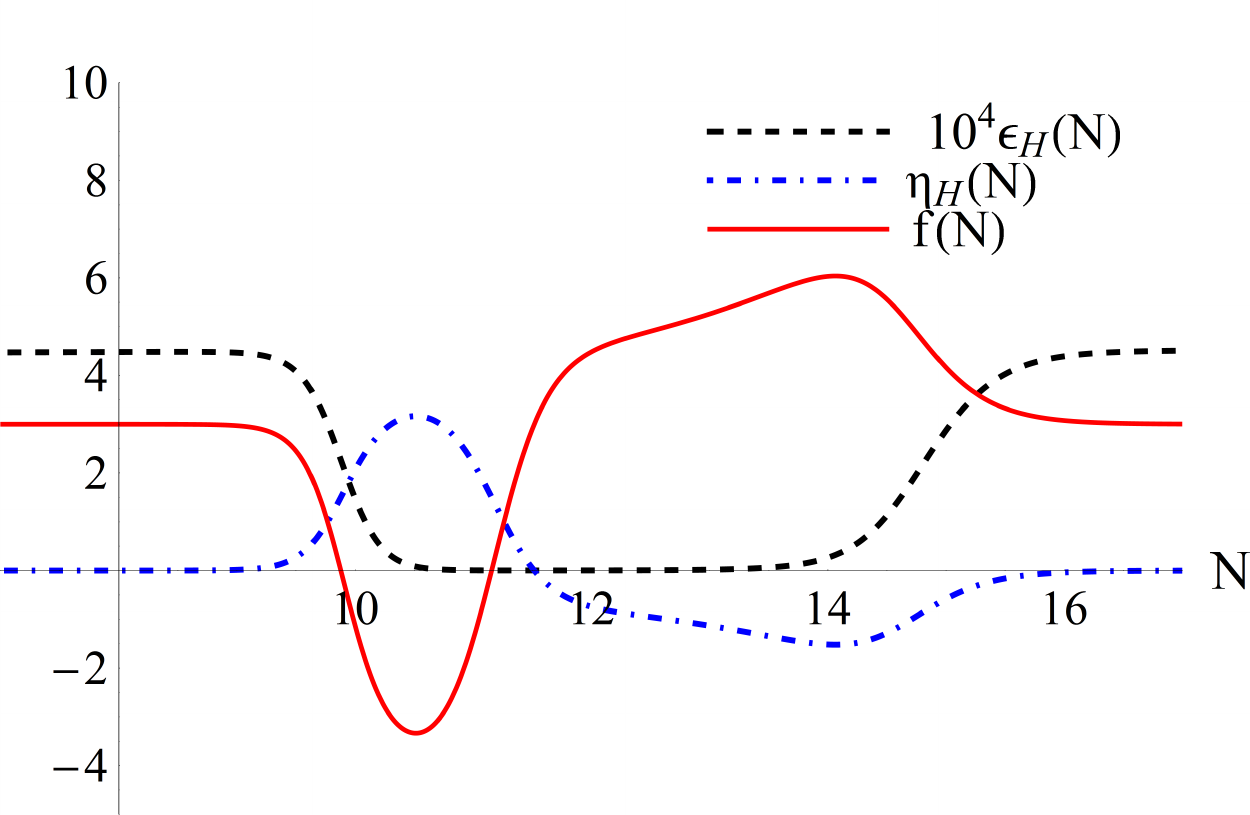} \includegraphics[width=0.46\textwidth]{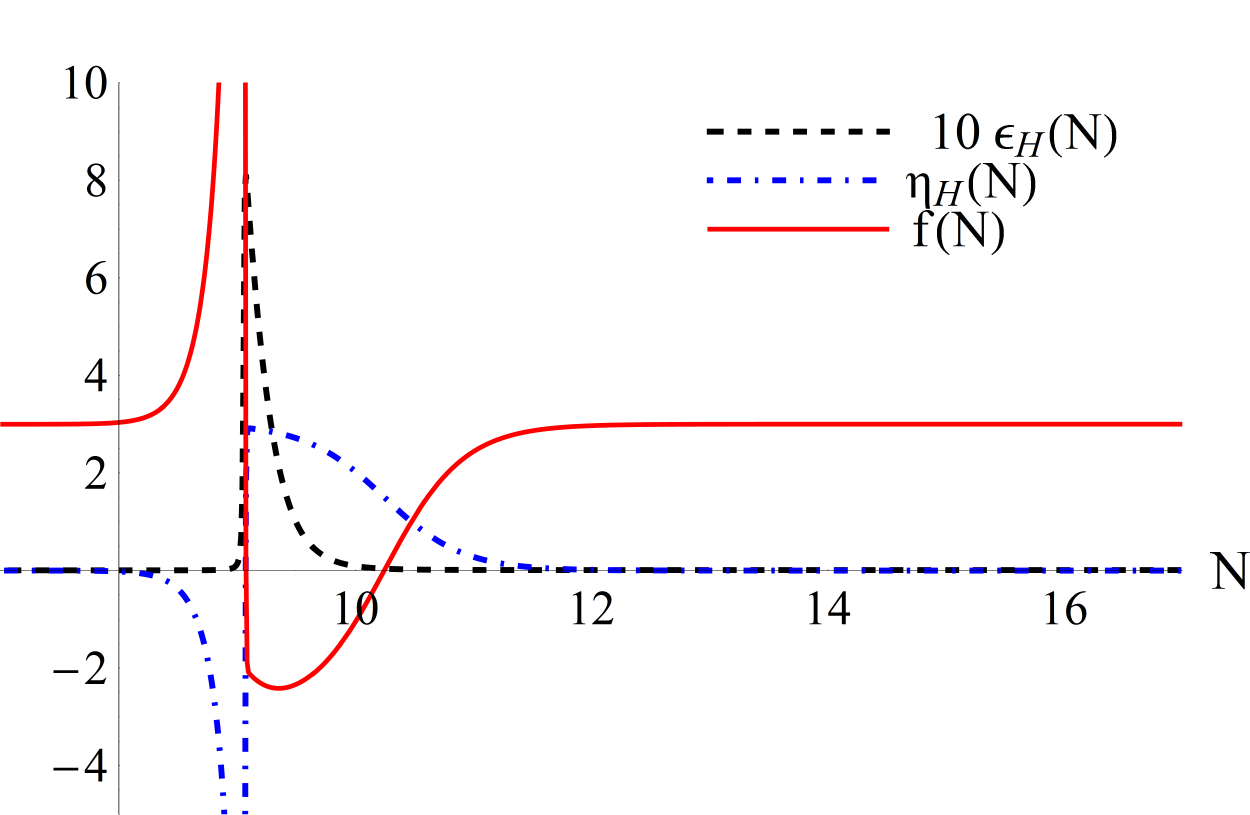} 
\caption{
The Hamilton-Jacobi slow-roll parameters $\epsilon_H$ and $\eta_H$, defined in eqs. (\ref{b1}) and 
(\ref{b2}) respectively, and the ``effective-friction" term $f(N)$ 
defined in eq. (\ref{AN}), as a function of the number of efoldings, for two 
choices of the parameters of the potential: 
Left plot: $A_1=0.000605$, $c_1=100$, $B=-0.03$.~
Right plot:  $A_1=-0.3$, $c_1=100$, $B=-0.03$. The two sets of parameters correspond to the first 
and third row 
of fig. \ref{figone}.
}
\label{ehg}
\end{figure}

In fig. \ref{ehg} we look in detail at the role of the slow-roll parameters in the enhancement of the
spectrum. We contrast the case of an inflection point in the potential (left plot) with that of a
step-like feature (right plot). In the first case, the solution remains inflationary during the
whole evolution. The Hamilton-Jacobi parameter $\epsilon_H$ has a constant value, apart from 
the part of the evolution near the inflection point, during which it approaches zero. 
The parameter $\eta_H$ starts from a value close to zero during the slow-roll regime, first turns positive
and subsequently negative, eventually returning close to zero during the second slow-roll regime.
The ``effective-friction" term is strongly influenced by $\eta_H$ and becomes negative during the
time that $\eta_H$ is significantly larger than zero.
In the case of a step-like feature, the parameter $\epsilon_H$ grows large during the interval that this
feature is transversed. For sharp steps or when the second plateau is sufficiently low, 
the solution ceases to be inflationary for a short time, 
as can be verified by computing the equation of state parameter
$w=-1+2\epsilon_H/3$. The parameter $\eta_H$ first turns negative, but then positive as the inflaton
``decelerates" while settling on a slow-roll regime on the second plateau. 
The ``effective friction" is again mainly influenced
by $\eta_H$ and becomes negative when
$\eta_H$ takes large positive values. 
The effect is sufficiently strong for the friction term to be negative even when 
$\epsilon_H$ is of order 1.

\subsection{A specific model}

\begin{figure}[t!]
\centering
\includegraphics[width=0.48\textwidth]{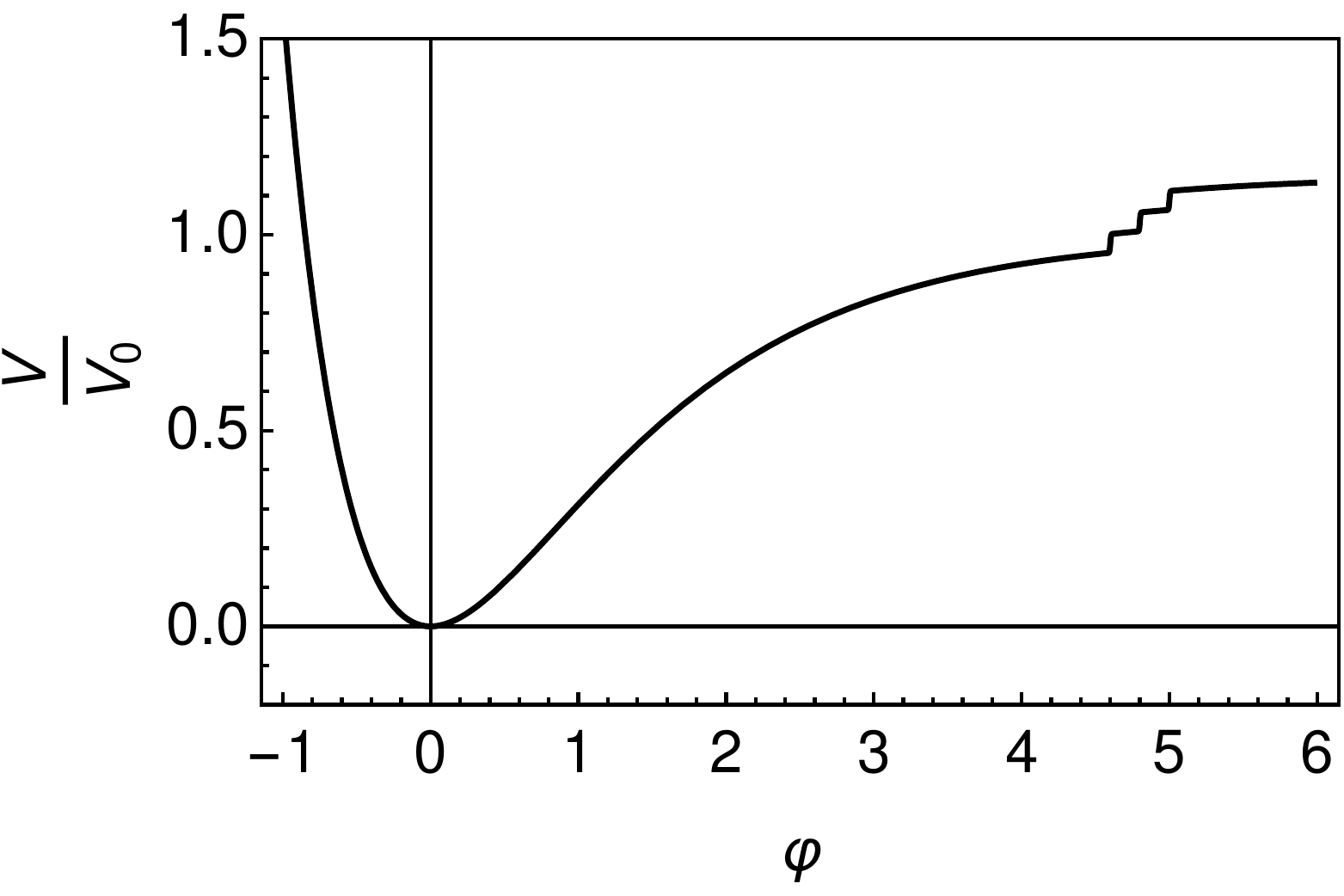} \includegraphics[width=0.48\textwidth]{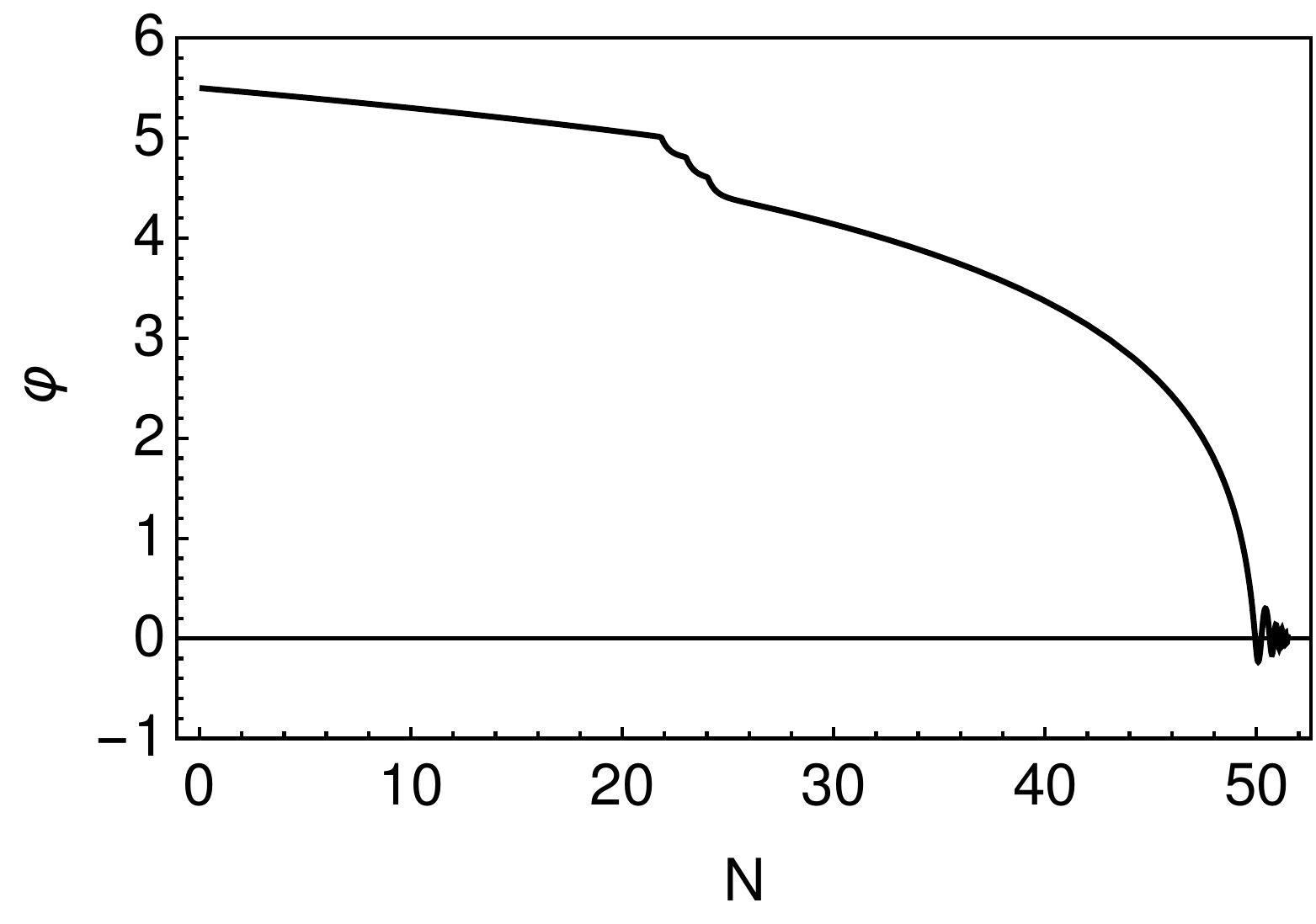} \includegraphics[width=0.48\textwidth]{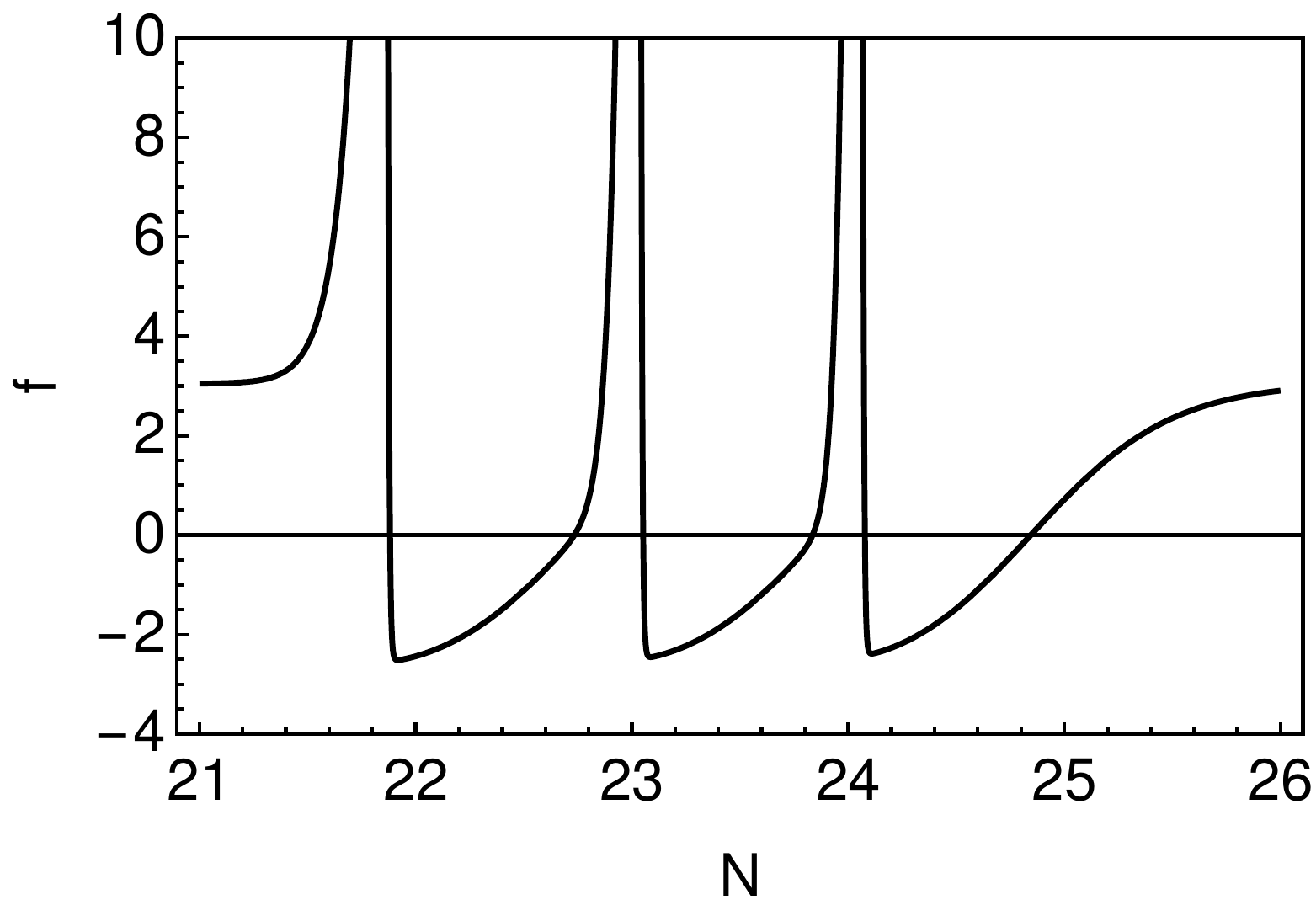} \includegraphics[width=0.50\textwidth,height= 50mm]{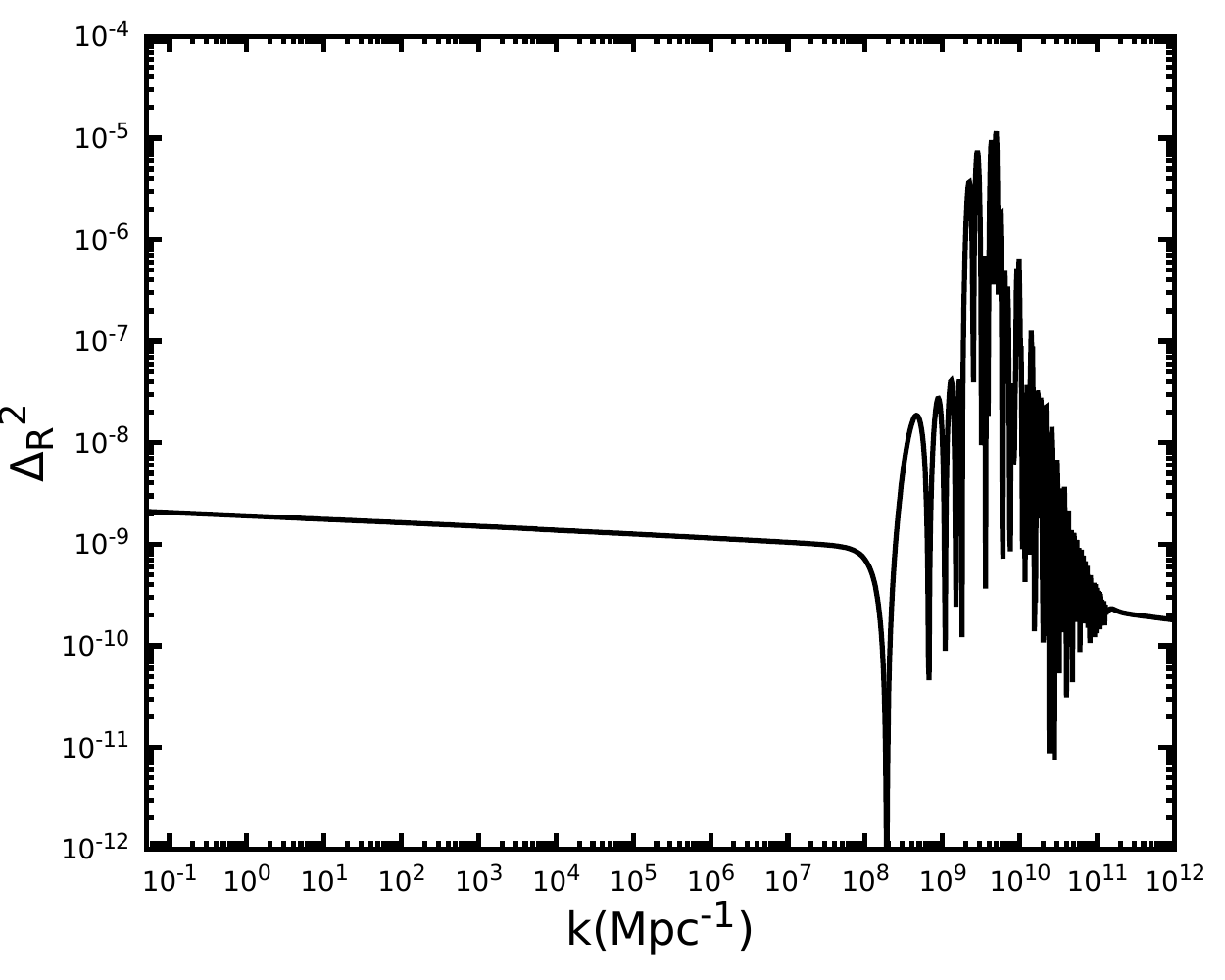}
\caption{
The inflaton potential $V(\varphi)$
defined in (\ref{potstaro}), the evolution of the inflaton $\varphi$ as a function of
the number of efoldings $N$, the function
$f(N)$ defined in eq. (\ref{AN}), and the power spectrum of curvature perturbations with
wavenumber $k$, for
$A_1=A_2=A_3=0.05$, $c_1=c_2=c_3=200$, $\varphi_1=5$, $\varphi_2=4.8$, $\varphi_3=4.6$, $B=-\sqrt{2/3}$. Dimensionful parameters in units of $\mpl$. 
The scale of the potential $V_0$ has been
adjusted in order to reproduce the amplitude of curvature perturbations in the CMB range. 
}
\label{figfive}
\end{figure}

The analysis of the previous subsection relied on a simplified potential
which did not allow us to make contact with the physical scales of the
power spectrum.
In order to obtain a more complete picture we study in this 
subsection a potential inspired by the Starobinsky model	
\cite{starobinsky}, to 
which we introduce step-like features. The potential is given by the 
expression
\be
V(\varphi)=V_0
\left(1-e^{B\varphi} \right)^2
\left( 1+\frac{1}{2}\sum_i A_i 
\left( 1+\tanh(c_i (\varphi-\varphi_i)) \right) \right).
\label{potstaro} \ee
As we do not engage in model building in this work, 
the above potential has not been derived from 
a more fundamental framework, such as supergravity. 
It is a phenomenological construction that has enough flexibility to 
allow for a sufficient number of efoldings, as well as power-spectrum 
scale and spectral
index compatible with the CMB observations. 

In fig. \ref{figfive} we present the various elements in the calculation of
the power spectrum of curvature perturbations for this model. The first
plot depicts the potential with the characteristic step-like feature. 
The values of the parameters are: 
$A_1=A_2=A_3=0.05$, $c_1=c_2=c_3=200$, $\varphi_1=5$, $\varphi_2=4.8$, $\varphi_3=4.6$, $B=-\sqrt{2/3}$. Dimensionful parameters are given in units of $\mpl$.  
The evolution of the inflaton $\varphi$ as a function of the number of efoldings $N$
is shown in the second plot. We count the number of efoldings from the moment that
the scale with wavenumber $k_*=0.05\,{\rm Mpc}^{-1}$, which we use as a pivot scale, 
exits the horizon. The above parameters result in a power spectrum in the 
CMB range with a spectral index $n_s\simeq 0.969$ and a tensor to scalar ratio
$r\simeq 0.0027.$
The third plot depicts the ``effective-friction" function $f(N)$
defined in eq. (\ref{AN}). It deviates from the standard value 3 during the period
in which the inflaton field takes values in the vicinity of the step-like
feature of the potential. When $f(N)$ is negative, it acts as negative friction,
leading to the enhancement of the curvature modes that cross the horizon during
this period. The enhancement for certain wavenumber bands can be significant. 
For this paticular choice of parameters the spectrum is enhanced by roughly 
four orders of magnitude. The enhancement can be made larger with an appropriate
choice of the potential, or with the inclusion of additional step-like features.
The curvature power spectrum is depicted in the last plot. It
has been normalized to the standard value $\simeq 2.1\times 10^{-9}$ for 
$k_*=0.05\,{\rm Mpc}^{-1}$ through an appropriate choice of the scale $V_0$ of
the potential.

The strong features in the spectrum appear deep in the nonlinear region, where
the phenomenological constraints are not strict because of the lack of 
analytical understanding of the evolution of the perturbations. 
The approximate wavenumber value $k_f$ 
for which these features appear can be estimated by 
noting that $k_f=\exp(N_f)H_f$ must hold at horizon crossing. For the 
pivot scale this relation is $k_*=\exp(N_*)H_*$, and we have set
$N_*=0$. If the Hubble parameter does not change substantially between
$N_*$ and $N_f$, we have $k_f/k_*\sim \exp(N_f)$. From the second plot of
fig. \ref{figfive} we obtain $N_f\sim 23$, which gives $k_f \sim 10^9\,{\rm Mpc}^{-1}$,
in agreement with the last plot.

\section{Conclusions}

We explored the possible enhancement of the power spectrum of 
curvature perturbations in single-field inflation when particular features 
appear in the inflaton potential. Our motivation stems from the
possibility that a strong enhancement of the spectrum within
a range of wavenumbers may have resulted in the
copious production of primordial black holes.  
One characteristic feature that is known to induce the enhancement of the 
spectrum is an inflection point of the potential at some value of the 
inflaton field \cite{inflection}. We analysed here the opposite case, i.e.
a sharp decrease of the potential, which may result even in the interruption
of inflation in certain cases, in contrast to what happens around an inflection point. 
Therefore, it comes as a surprise that 
the fast ``rolling" of the inflaton field through such a feature can have
as a consequence the enhancement of the fluctuations. 
Building on previous work \cite{stepinflation1}, we explored the conditions
under which the enhancement can be very large, by several orders of magnitude
relative to its magnitude within the almost scale-invariant range. 
It must be noted that it is always possible to enhance the spectrum by engineering
the transition to a second very flat plateau of the potential.
The time derivative of the inflaton under slow-roll conditions on the plateau
would be very small, resulting in an enhanced power spectrum.
In our analysis we exclude this rather trivial possibility by keeping the slope
rougly constant, apart from at the transition point or points, and focus on
the effect of the transition itself. 

We analysed in detail the simplified potential of eq. (\ref{pote}).
We found that sharp transitions lead
to the strong growth of the curvature perturbation. The reason can be traced 
to the ``effective-friction" term of eq. (\ref{RN}), which is given by the
function $f(N)$ defined in eq. (\ref{AN}). Even though this function is 
positive during the first part of the transition, thus suppressing the perturbation,
it can become negative during the second part, when the inflaton approaches
slow roll on the second plateau, and can lead to a dramatic enhancement.
The main effect comes from the slow-roll parameter $\eta_H$ 
taking large positive values, even when the parameter $\epsilon_H$ is large.
The effect is increased by the steepness of the potential, but is also limited 
by the size of the potential drop that bounds the maximal inflaton ``velocity".
However, successive nearby steps give an additive effect, leading to a
spectrum enhancement by several orders of magnitude. We discussed up to three steps, 
but increasing this number can increase the enhancement arbitrarily.

The second prominent feature of the spectrum is its strong oscillatory form
as a function of wavenumber. We analysed the origin of this feature during 
the discussion of fig. \ref{oscil} in the previous section.
The appearance of wavenumber bands in which the spectrum takes very large values
can lead to the creation of primordial black holes of characteristic sizes 
when the corresponding fluctuations enter the horizon. 
The suppression of the spectrum in other bands indicates the absence of 
black holes of other sizes. The combined effect can lead to a very 
distinctive pattern.

Another very exciting prospect is the possibility of detecting 
the oscillatory pattern in the spectrum of gravitational waves generated 
through the scalar perturbations at second order \cite{igws}. 
This scenario is independent of the creation of primordial black holes
and becomes possible even for a milder enhancement of the spectrum.
The detection of stochastic gravitational waves  
is a portal to the primordial spectrum of scalar perturbations at 
small scales and can be used in order to look for
strong features in the inflationary dynamics. The oscillatory patterns 
appearing in the scenario we discussed provide a prime example of a possibly
detectable feature. Because of a double integration over momenta in the 
expression for the spectrum of gravitational waves, 
the oscillatory pattern is expected to be superimposed on a
smooth underlying curve with one or two peaks
\cite{gwsmooth}.
However, a clear distinction is possible between smooth spectra resulting
from an inflection point in the inflaton potential
and the oscillatory spectra in our scenario. 
It must be noted that such oscillatory features have been considered recently
for the spectra resulting from
two-field inflationary models \cite{gwspectra}.

A realistic inflaton potential must generate a sufficient number of efoldings and 
result in a spectrum consistent with the CMB constraints. Even though our
aim here was not to engage in detailed model building, we 
discussed the potential of eq. (\ref{potstaro}), which is inspired by the 
Starobinsky model \cite{starobinsky}. The potential is constructed in 
a rather artificial manner and can serve only as a toy model. However, it is very
useful in order to establish that the type of spectrum enhancement that we
are suggesting can appear in realistic setups. One particular property of
the potentials that we are considering is their dependence on the 
hyperbolic tangent of the field. As we discussed in the introduction, this occurs 
in models associated with $\alpha$-attractors in supergravity \cite{alpha,alpha0},
in which the function $\tanh(\varphi/\sqrt{\alpha})$ becomes part of the
potential in the Einstein frame. 
The study of the spectra of density perturbations and induced gravitational waves
in specific models will be the subject of
future work.

\section*{Acknowledgments}

We would like to thank I. Dalianis and V. Spanos for useful discussions.
The work of G. Kodaxis, I. Stamou and N. Tetradis was supported by the Hellenic Foundation 
for Research and Innovation (H.F.R.I.) under the  “First Call for
H.F.R.I. Research Projects to support Faculty members and Researchers and the procurement of high-cost research equipment grant”  (Project Number: 824).



\begin{thebibliography}{99}




\bibitem{Virgo}
B.~Abbott \textit{et al.} [LIGO Scientific and Virgo],
Phys. Rev. Lett. \textbf{116} (2016) no.6, 061102
[arXiv:1602.03837 [gr-qc]];
\\
B.~P.~Abbott \textit{et al.} [LIGO Scientific and Virgo],
Phys. Rev. Lett. \textbf{116} (2016) no.24, 241103
[arXiv:1606.04855 [gr-qc]];
\\
B.~P.~Abbott \textit{et al.} [LIGO Scientific and Virgo],
Phys. Rev. Lett. \textbf{118} (2017) no.22, 221101
[arXiv:1706.01812 [gr-qc]];
\\
B.~P.~Abbott \textit{et al.} [LIGO Scientific and Virgo],
Astrophys. J. \textbf{851} (2017) no.2, L35
[arXiv:1711.05578 [astro-ph.HE]];
\\
B.~Abbott \textit{et al.} [LIGO Scientific and Virgo],
Phys. Rev. Lett. \textbf{119} (2017) no.14, 141101
[arXiv:1709.09660 [gr-qc]].

            
             \bibitem{pbh1}
             Ya.~B.~Zeldovich and I.~D.~Novikov, 
              Sov.\ Astron.\ -AJ   {\bf 10} (1967) 602;
  \\
               S.~Hawking,
               Mon.\ Not.\ Roy.\ Astron.\ Soc.\  {\bf 152} (1971) 75;
               \\
                B.~J.~Carr and S.~W.~Hawking,
                Mon.\ Not.\ Roy.\ Astron.\ Soc.\  {\bf 168} (1974) 399;
\\
                B.~J.~Carr,
                Astrophys. J. \textbf{201} (1975), 1-19.




\bibitem{reviews}
B.~Carr, F.~Kuhnel and M.~Sandstad,
Phys. Rev. D \textbf{94} (2016) no.8, 083504
[arXiv:1607.06077 [astro-ph.CO]];
\\
M.~Sasaki, T.~Suyama, T.~Tanaka and S.~Yokoyama,
Class. Quant. Grav. \textbf{35} (2018) no.6, 063001
[arXiv:1801.05235 [astro-ph.CO]];
\\
B.~Carr, K.~Kohri, Y.~Sendouda and J.~Yokoyama,
[arXiv:2002.12778 [astro-ph.CO]];
\\
B.~Carr and F.~Kuhnel,
[arXiv:2006.02838 [astro-ph.CO]];
\\
A.~M.~Green and B.~J.~Kavanagh,
[arXiv:2007.10722 [astro-ph.CO]].

\bibitem{slowroll}
C.~Germani and T.~Prokopec,
Phys. Dark Univ. \textbf{18} (2017), 6-10
[arXiv:1706.04226 [astro-ph.CO]];
\\
H.~Motohashi and W.~Hu,
Phys. Rev. D \textbf{96} (2017) no.6, 063503
[arXiv:1706.06784 [astro-ph.CO]].


\bibitem{inflectionorig}
P.~Ivanov, P.~Naselsky and I.~Novikov,
Phys. Rev. D \textbf{50} (1994), 7173-7178.


\bibitem{stepinflation1}
J.~A.~Adams, B.~Cresswell and R.~Easther,
Phys. Rev. D \textbf{64} (2001), 123514
[arXiv:astro-ph/0102236 [astro-ph]];
\\
S.~M.~Leach and A.~R.~Liddle,
Phys. Rev. D \textbf{63} (2001), 043508
[arXiv:astro-ph/0010082 [astro-ph]];
\\
 S.~M.~Leach, M.~Sasaki, D.~Wands and A.~R.~Liddle,
 Phys. Rev. D \textbf{64} (2001), 023512
 [arXiv:astro-ph/0101406 [astro-ph]].
 



\bibitem{pbhdarkmatter}
S.~Clesse and J.~García-Bellido,
Phys. Rev. D \textbf{92} (2015) no.2, 023524
[arXiv:1501.07565 [astro-ph.CO]];
\\
M.~Kawasaki, A.~Kusenko, Y.~Tada and T.~T.~Yanagida,
Phys. Rev. D \textbf{94} (2016) no.8, 083523
[arXiv:1606.07631 [astro-ph.CO]];
\\
K.~Inomata, M.~Kawasaki, K.~Mukaida, Y.~Tada and T.~T.~Yanagida,
Phys. Rev. D \textbf{96} (2017) no.4, 043504
[arXiv:1701.02544 [astro-ph.CO]].



\bibitem{inflection}
J.~Garcia-Bellido and E.~Ruiz Morales,
Phys. Dark Univ. \textbf{18} (2017), 47-54
[arXiv:1702.03901 [astro-ph.CO]];
\\
J.~M.~Ezquiaga, J.~Garcia-Bellido and E.~Ruiz Morales,
Phys. Lett. B \textbf{776} (2018), 345-349
[arXiv:1705.04861 [astro-ph.CO]];
\\
H.~Di and Y.~Gong,
JCAP \textbf{07} (2018), 007
[arXiv:1707.09578 [astro-ph.CO]];
\\
K.~Kannike, L.~Marzola, M.~Raidal and H.~Veermäe,
JCAP \textbf{09} (2017), 020
[arXiv:1705.06225 [astro-ph.CO]];
\\
G.~Ballesteros and M.~Taoso,
Phys. Rev. D \textbf{97} (2018) no.2, 023501
[arXiv:1709.05565 [hep-ph]];
\\
M.~P.~Hertzberg and M.~Yamada,
Phys. Rev. D \textbf{97} (2018) no.8, 083509
[arXiv:1712.09750 [astro-ph.CO]];
\\
J.~Espinosa, D.~Racco and A.~Riotto,
Phys. Rev. Lett. \textbf{120} (2018) no.12, 121301
[arXiv:1710.11196 [hep-ph]];
\\
S.~Cheng, W.~Lee and K.~Ng,
JCAP \textbf{07} (2018), 001
[arXiv:1801.09050 [astro-ph.CO]];
\\
O.~\"Ozsoy, S.~Parameswaran, G.~Tasinato and I.~Zavala,
JCAP \textbf{07} (2018), 005
[arXiv:1803.07626 [hep-th]];
\\
M.~Biagetti, G.~Franciolini, A.~Kehagias and A.~Riotto,
JCAP \textbf{07} (2018), 032
[arXiv:1804.07124 [astro-ph.CO]];
\\
G.~Franciolini, A.~Kehagias, S.~Matarrese and A.~Riotto,
JCAP \textbf{03} (2018), 016
[arXiv:1801.09415 [astro-ph.CO]];
\\
T.~Gao and Z.~Guo,
Phys. Rev. D \textbf{98} (2018) no.6, 063526
[arXiv:1806.09320 [hep-ph]];
\\
M.~Cicoli, V.~A.~Diaz and F.~G.~Pedro,
JCAP \textbf{06} (2018), 034
[arXiv:1803.02837 [hep-th]];
\\
I.~Dalianis, A.~Kehagias and G.~Tringas,
JCAP \textbf{01} (2019), 037
[arXiv:1805.09483 [astro-ph.CO]];
\\
R.~Mahbub,
Phys. Rev. D \textbf{101} (2020) no.2, 023533
[arXiv:1910.10602 [astro-ph.CO]];
\\
S.~S.~Mishra and V.~Sahni,
JCAP \textbf{04} (2020), 007
[arXiv:1911.00057 [gr-qc]];
\\
G.~Ballesteros, J.~Rey and F.~Rompineve,
JCAP \textbf{06} (2020), 014
[arXiv:1912.01638 [astro-ph.CO]];
\\
R.~G.~Cai, Z.~K.~Guo, J.~Liu, L.~Liu and X.~Y.~Yang,
JCAP \textbf{06} (2020), 013
[arXiv:1912.10437 [astro-ph.CO]];
\\
Y.~Aldabergenov, A.~Addazi and S.~V.~Ketov,
[arXiv:2006.16641 [hep-th]];
\\
S.~V.~Ketov and M.~Y.~Khlopov,
Symmetry \textbf{11} (2019) no.4, 511


\bibitem{multifield}
J.~Garcia-Bellido, A.~D.~Linde and D.~Wands,
Phys. Rev. D \textbf{54} (1996), 6040-6058
[arXiv:astro-ph/9605094 [astro-ph]];
\\
K.~Inomata, M.~Kawasaki, K.~Mukaida and T.~T.~Yanagida,
Phys. Rev. D \textbf{97} (2018) no.4, 043514
[arXiv:1711.06129 [astro-ph.CO]];
\\
S.~Pi, Y.~l.~Zhang, Q.~G.~Huang and M.~Sasaki,
JCAP \textbf{05} (2018), 042
[arXiv:1712.09896 [astro-ph.CO]];
\\
G.~A.~Palma, S.~Sypsas and C.~Zenteno,
Phys. Rev. Lett. \textbf{125} (2020) no.12, 121301
[arXiv:2004.06106 [astro-ph.CO]];
\\
J.~Fumagalli, S.~Renaux-Petel, J.~W.~Ronayne and L.~T.~Witkowski,
[arXiv:2004.08369 [hep-th]];
\\
M.~Braglia, D.~K.~Hazra, F.~Finelli, G.~F.~Smoot, L.~Sriramkumar and A.~A.~Starobinsky,
[arXiv:2005.02895 [astro-ph.CO]];
\\
Z.~Zhou, J.~Jiang, Y.~F.~Cai, M.~Sasaki and S.~Pi,
[arXiv:2010.03537 [astro-ph.CO]].



 
\bibitem{efstathiou}
S.~Chongchitnan and G.~Efstathiou,
JCAP \textbf{01} (2007), 011
[arXiv:astro-ph/0611818 [astro-ph]].




\bibitem{mukhsas}
V.~F.~Mukhanov,
Sov. Phys. JETP \textbf{67} (1988), 1297-1302;
\\
M.~Sasaki,
Prog. Theor. Phys. \textbf{76} (1986), 1036.


\bibitem{matterdom}
T.~Harada, C.~M.~Yoo, K.~Kohri and K.~I.~Nakao,
Phys. Rev. D \textbf{96} (2017) no.8, 083517
[erratum: Phys. Rev. D \textbf{99} (2019) no.6, 069904]
[arXiv:1707.03595 [gr-qc]];
\\
T.~Harada, C.~M.~Yoo, K.~Kohri, K.~i.~Nakao and S.~Jhingan,
Astrophys. J. \textbf{833} (2016) no.1, 61
[arXiv:1609.01588 [astro-ph.CO]].


\bibitem{stepinflation2}
A.~A.~Starobinsky,
JETP Lett. \textbf{55} (1992), 489-494;
\\
J.~A.~Adams, G.~G.~Ross and S.~Sarkar,
Nucl. Phys. B \textbf{503} (1997), 405-425
[arXiv:hep-ph/9704286 [hep-ph]];
\\
C.~P.~Burgess, R.~Easther, A.~Mazumdar, D.~F.~Mota and T.~Multamaki,
JHEP \textbf{05} (2005), 067
[arXiv:hep-th/0501125 [hep-th]];
\\
J.~Hamann, L.~Covi, A.~Melchiorri and A.~Slosar,
Phys. Rev. D \textbf{76} (2007), 023503
[arXiv:astro-ph/0701380 [astro-ph]];
\\
M.~Joy, A.~Shafieloo, V.~Sahni and A.~A.~Starobinsky,
JCAP \textbf{06} (2009), 028
[arXiv:0807.3334 [astro-ph]];
\\
M.~Joy, A.~Shafieloo, V.~Sahni and A.~A.~Starobinsky,
JCAP \textbf{06} (2009), 028
[arXiv:0807.3334 [astro-ph]];
\\
D.~K.~Hazra, M.~Aich, R.~K.~Jain, L.~Sriramkumar and T.~Souradeep,
JCAP \textbf{10} (2010), 008
[arXiv:1005.2175 [astro-ph.CO]];
\\
Z.~G.~Liu, J.~Zhang and Y.~S.~Piao,
Phys. Lett. B \textbf{697} (2011), 407-411
[arXiv:1012.0673 [gr-qc]];
\\
A.~Gallego Cadavid, A.~E.~Romano and S.~Gariazzo,
Eur. Phys. J. C \textbf{76} (2016) no.7, 385
[arXiv:1508.05687 [astro-ph.CO]];
Eur. Phys. J. C \textbf{77} (2017) no.4, 242
[arXiv:1612.03490 [astro-ph.CO]];
\\
M.~A.~Fard and S.~Baghram,
JCAP \textbf{01} (2018), 051
[arXiv:1709.05323 [astro-ph.CO]].


\bibitem{starobinsky}
A.~A.~Starobinsky,
Adv. Ser. Astrophys. Cosmol. \textbf{3} (1987), 130-133.



\bibitem{wetterich}
C.~Wetterich,
Phys. Lett. B \textbf{301} (1993), 90-94
[arXiv:1710.05815 [hep-th]].

\bibitem{review}
J.~Berges, N.~Tetradis and C.~Wetterich,
Phys. Rept. \textbf{363} (2002), 223-386
doi:10.1016/S0370-1573(01)00098-9
[arXiv:hep-ph/0005122 [hep-ph]].


\bibitem{alpha}
R.~Kallosh and A.~Linde,
JCAP \textbf{07} (2013), 002
[arXiv:1306.5220 [hep-th]];
\\
S.~Ferrara, R.~Kallosh, A.~Linde and M.~Porrati,
Phys. Rev. D \textbf{88} (2013) no.8, 085038
[arXiv:1307.7696 [hep-th]].

\bibitem{alpha0}
R.~Kallosh, A.~Linde and D.~Roest,
JHEP \textbf{08} (2014), 052
[arXiv:1405.3646 [hep-th]].




\bibitem{physrep}
V.~F.~Mukhanov, H.~Feldman and R.~H.~Brandenberger,
Phys. Rept. \textbf{215} (1992), 203-333.

\bibitem{multcross}
G.~Ballesteros, J.~Beltran Jimenez and M.~Pieroni,
JCAP \textbf{06} (2019), 016
[arXiv:1811.03065 [astro-ph.CO]].


\bibitem{ozsoy}
O.~\"Ozsoy and G.~Tasinato,
JCAP \textbf{04} (2020), 048
[arXiv:1912.01061 [astro-ph.CO]].

\bibitem{igws}
S.~Mollerach, D.~Harari and S.~Matarrese,
Phys. Rev. D \textbf{69} (2004), 063002
[arXiv:astro-ph/0310711 [astro-ph]];
\\
K.~N.~Ananda, C.~Clarkson and D.~Wands,
Phys. Rev. D \textbf{75} (2007), 123518
[arXiv:gr-qc/0612013 [gr-qc]];
\\
H.~Assadullahi and D.~Wands,
Phys. Rev. D \textbf{79} (2009), 083511
[arXiv:0901.0989 [astro-ph.CO]];
\\
D.~Baumann, P.~J.~Steinhardt, K.~Takahashi and K.~Ichiki,
Phys. Rev. D \textbf{76} (2007), 084019
[arXiv:hep-th/0703290 [hep-th]];
\\
R.~Saito and J.~Yokoyama,
Phys. Rev. Lett. \textbf{102} (2009), 161101
[erratum: Phys. Rev. Lett. \textbf{107} (2011), 069901]
[arXiv:0812.4339 [astro-ph]].


\bibitem{gwsmooth}
S.~Pi and M.~Sasaki,
JCAP \textbf{09} (2020), 037
[arXiv:2005.12306 [gr-qc]].

\bibitem{gwspectra}
J.~Fumagalli, S.~Renaux-Petel and L.~T.~Witkowski,
[arXiv:2012.02761 [astro-ph.CO]];
\\
M.~Braglia, X.~Chen and D.~K.~Hazra,
JCAP \textbf{03} (2021), 005
[arXiv:2012.05821 [astro-ph.CO]].




\end{thebibliography}
\end{document}